# Heterogeneous Choice Sets and Preferences[*]

Levon Barseghyan
Cornell University

Maura Coughlin
Rice University

Francesca Molinari
Cornell University

Joshua C. Teitelbaum
Georgetown University

February 9, 2021

**Abstract**

We propose a robust method of discrete choice analysis when agents' choice sets are unobserved. Our core model assumes nothing about agents' choice sets apart from their minimum size. Importantly, it leaves unrestricted the dependence, conditional on observables, between choice sets and preferences. We first characterize the sharp identification region of the model's parameters by a finite set of conditional moment inequalities. We then apply our theoretical findings to learn about households' risk preferences and choice sets from data on their deductible choices in auto collision insurance. We find that the data can be explained by expected utility theory with low levels of risk aversion and heterogeneous non-singleton choice sets, and that more than three in four households require limited choice sets to explain their deductible choices. We also provide simulation evidence on the computational tractability of our method in applications with larger feasible sets or higher-dimensional unobserved heterogeneity.

[*]We are grateful to the editor, the referees, Panle Jia Barwick, Aureo de Paula, Jean-Francois Houde, Chuck Manski, Ulrich Müller, Matthew Thirkettle, Lin Xu, and participants at numerous conferences and seminars. We acknowledge financial support from National Science Foundation grants SES-1031136 and SES-1824448 and from the Institute for Social Sciences at Cornell University. Part of the research for this paper was conducted while Barseghyan and Molinari were on sabbatical leave at the Department of Economics at Duke University, whose hospitality is gratefully acknowledged.

# 1 Introduction

The primitives of any discrete choice model include two sets: a known universal set of feasible alternatives—the feasible set—and the finite subset of the feasible set from which an agent makes her choice—her choice set. Discrete choice analysis in the tradition of McFadden (1974) rests on the assumption that agents' choice sets are observed. McFadden shows that when this assumption holds, one can apply the principle of revealed preference to learn about agents' unobserved preferences from data on their observed choices. Moreover, he shows that with additional restrictions on the structure and distribution of agents' preferences, one can achieve point identification of a parametric model of discrete choice.

In practice, however, choice sets are often unobserved (Manski 1977). Sometimes this is a missing data problem—agents' choice sets are observable in principle but are not recorded in the data. For example, one studying the college enrollment choices of high school students may not observe the colleges to which a student applied and was admitted (Kohn et al. 1976); one studying the travel mode choices of urban commuters may not observe if some modes normally available to a commuter were temporarily unavailable on a given day (Ben-Akiva and Boccara 1995); or one studying the hospital choices of English patients may not observe which alternatives were offered to a patient by her referring physician (Gaynor et al. 2016).

At other times the problem is that agents' choice sets are unobservable mental constructs. This is the case in models of limited attention or limited consideration, where an agent considers only a strict subset of the feasible set due to, for example, search costs, brand preferences, or cognitive limitations. For instance, one studying the personal computer choices of retail consumers can be sure that a consumer was not aware of all computers for sale but cannot observe the computers of which a consumer was aware (Goeree 2008); one studying the Medigap plan choices of Medicare insureds cannot observe which of the available plans an insured in fact considered (Starc 2014); or one studying the energy retailer choices of residential electricity customers cannot observe whether or to what extent a customer considered the alternatives to her default, incumbent retailer (Hortaçsu et al. 2017).

When choice sets are unobserved the econometrician is forced to make additional assumptions in order to achieve point identification (Ben-Akiva 1973). The most common approach is to assume, often implicitly, that all choice sets coincide with the feasible set or a known subset of the feasible set. More sophisticated approaches allow for heterogeneity in choice sets and obtain point identification by relying on auxiliary information about their composition or distribution, two-way exclusion restrictions (i.e., variables assumed to impact choice sets but not preferences and vice versa), and other restrictions on the choice set formation process (e.g., conditional independence between choice sets and preferences). In some applications these approaches seem reasonable or at least plausible. In many applications, however, they likely result in misspecified models, biased estimates, and incorrect inferences.

More fundamentally, the basic revealed preference argument breaks down when choice sets are unobserved. At one extreme, when an agent's choice set coincides with the feasible set, her choice reveals that she prefers the chosen alternative to all others. At the other extreme, when an agent's choice set comprises a single alternative, her choice is driven entirely by her choice set and reveals nothing about her preferences. In all other cases her choice is a function of both her preferences and her choice set. Learning about preferences from choices when choice sets are unobserved is the main challenge we address in this paper.

We propose a new, robust method of discrete choice analysis when agents' choice sets are unobserved. We lay out our core model in Section 2. We begin with the classic random utility model developed by McFadden (1974) and others, though we allow for a utility function that is neither linear in parameters nor additively separable in unobservables. Our key point of departure from the classic model, however, is that we relax the assumption that the agents' choice sets are observed. Instead, we assume only that the minimum size of the agents' choice sets is a known integer greater than one. Consequently, our model admits any choice set formation process (subject to the minimum size assumption) and allows for any dependence structure, without restriction, between agents' choice sets and their observables and, conditional on observables, between agents' choice sets and their preferences.

In Section 3 we first show that our model implies multiple optimal choices for an agent, resulting from the multiple possible realizations of her choice set. It is this multiplicity that, in the absence of additional restrictions on the choice set formation process, generally precludes point identification of the model's parameters. Because we avoid making such additional, unverifiable assumptions, our approach yields a robust method of statistical inference. We then present our main identification results, which leverage a result in random set theory, due to Artstein (1983), to define a finite set of conditional moment inequalities that characterizes the sharp identification region for the model's parameters. We also discuss the practicalities of computing the sharp identification region.[1]

In Sections 4 and 5 we demonstrate the usefulness of our theoretical findings by applying them to learn about households' risk preferences and choice sets from data on their deductible choices in auto collision insurance. In Section 4 we specify an empirical model that allows for unobserved heterogeneity in households' risk preferences and in their choice sets. Although we observe the feasible set of deductibles, we do not observe which deductibles enter a household's choice set. In our setting unobserved heterogeneity in choice sets may be due to missing data—e.g., if different households are quoted different subsets of deductibles—or to unobserved constraints—e.g., if some households disregard low deductibles due to budget constraints or high deductibles due to liquidity constraints.

---

[1] Sharpness means that the identification region comprises all and only those parameters for which there exists a choice set formation process such that the distribution of model-implied choices matches the distribution of observed choices. The recent econometrics literature uses the result in Artstein (1983), discussed in detail in Molchanov and Molinari (2018, Chapter 2), to conduct identification analysis in various partially identified models (e.g., Beresteanu and Molinari 2008; Beresteanu et al. 2011; Galichon and Henry 2011; Chesher et al. 2013; Chesher and Rosen 2017). For a review, see Molinari (2020).

We present our empirical findings in Section 5. Our key finding on preferences is that the data can be explained by expected utility theory with a distribution of risk aversion that has low mean and variance, with at least a quarter of households being effectively risk neutral. Our key finding on choice sets is that more than three in four households require limited choice sets (i.e., strict subsets of the feasible set) to explain their deductible choices, and we discuss two drivers of this result: suboptimal choices and violations of the law of demand.

Our empirical findings highlight the importance of using a robust method to conduct inference on discrete choice models when there may be unobserved heterogeneity in choice sets. The literature on risky choice, motivated in part by reported estimates of risk aversion that seem implausibly high in light of the Rabin (2000) critique (e.g., Cicchetti and Dubin 1994; Sydnor 2010), has focused on developing and estimating models that depart from expected utility theory in their specification of *how* agents evaluate risky alternatives. Our findings provide new evidence on the importance of developing models that differ in their specification of *which* alternatives agents evaluate, and of data collection efforts that seek to directly measure agents' heterogeneous choice sets (Caplin 2016).

In Section 6 we provide simulation evidence on the computational tractability of our method in applications that feature larger feasible sets or higher-dimensional unobserved heterogeneity. We also illustrate how the informational content of the data and the model varies with the relative values of the size of the feasible set and the minimum size of the agents' choice sets, and with the dependence between the agents' choice sets, on the one hand, and their preferences or observables, on the other.

We conclude the paper in Section 7 with a discussion in which we review the prior literature on discrete choice analysis with unobserved heterogeneity in choice sets and recap our contributions to the literature. Supplemental Material (Barseghyan et al. 2021) contains additional information and results, including on the computational aspects of our method.

# 2 A Random Utility Model with Unobserved Heterogeneity in Choice Sets

Our starting point is the random utility model developed by McFadden (1974). Let $\mathcal{I}$ denote a population of agents and $\mathcal{D}$ denote a finite set of alternatives, which we call the *feasible* *set.* Let $\mathcal{U}$ be a family of real-valued functions defined on $\mathcal{D}$. The model posits that for each agent $i \in \mathcal{I}$ there exists a function $U_i$ drawn from $\mathcal{U}$ according to some distribution such that

$$d \in^* C_i \iff U_i(d) \geq U_i(c) \text{ for all } c \in C_i, \tag{2.1}$$

where $\in^*$ denotes "is chosen from" and $C_i \subseteq \mathcal{D}$ denotes the agent's *choice set*.

We assume that each agent $i \in \mathcal{I}$ is characterized by a real-valued vector of observable attributes $\mathbf{x}_i = (\mathbf{s}_i, (\mathbf{z}_{ic}, c \in \mathcal{D}))$, where $\mathbf{s}_i$ is a subvector of attributes specific to agent $i$ that are constant across alternatives and $\mathbf{z}_{ic}$ is a subvector of attributes specific to alternative $c$ that may vary across agents. Let $\mathbf{x}_{ic} = (\mathbf{s}_i, \mathbf{z}_{ic})$ denote the vector of observable attributes relevant to alternative $c$. In addition, we assume that each agent $i \in \mathcal{I}$ is further characterized by a real-valued vector of unobservable attributes $\boldsymbol{\nu}_i$, which are idiosyncratic to the agent. Let $\mathcal{X}$ and $\mathcal{V}$ denote the supports of $\mathbf{x}_i$ and $\boldsymbol{\nu}_i$, respectively.

To operationalize $U_i$ as a random variable, we posit that it is a function of the agent's observable and unobservable attributes and we impose restrictions on its distribution.

ASSUMPTION 2.1 (Restrictions on Utility):

(I) *There exists a function $W : \mathcal{X} \times \mathcal{V} \mapsto \mathbb{R}$, known up to a finite-dimensional parameter vector $\boldsymbol{\delta} \in \Delta \subset \mathbb{R}^k$, where $\Delta$ is convex and compact, and continuous in each of its arguments such that $U_i(c) = W(\mathbf{x}_{ic}, \boldsymbol{\nu}_i; \boldsymbol{\delta})$ for all $c \in \mathcal{D}$, $(\mathbf{x}_{ic}, \boldsymbol{\nu}_i) - a.s.$*

(II) *The distribution of $\boldsymbol{\nu}_i$, denoted by $P$, is continuous, known up to a finite-dimensional parameter vector $\boldsymbol{\gamma} \in \Gamma \subset \mathbb{R}^l$, where $\Gamma$ is convex and compact, and independent of $\mathbf{x}_i$.*

Assumption 2.1 allows for nonadditive unobserved heterogeneity in $U_i$, indexed by $\boldsymbol{\nu}_i$. It is weaker than the standard assumption that $U_i$ is additively separable in unobservables. That said, one could let $\boldsymbol{\nu}_i = (\nu_{ic}, c \in \mathcal{D})$ and specify $W(\mathbf{x}_{ic}, \boldsymbol{\nu}_i; \boldsymbol{\delta}) = \omega(\mathbf{x}_{ic}; \boldsymbol{\delta}) + \nu_{ic}$ as in a conditional logit (McFadden 1974), or let $\boldsymbol{\nu}_i = (\boldsymbol{\upsilon}_i, (\epsilon_{ic}, c \in \mathcal{D}))$ and specify $W(\mathbf{x}_{ic}, \boldsymbol{\nu}_i; \boldsymbol{\delta}) = \omega(\mathbf{x}_{ic}, \boldsymbol{\upsilon}_i; \boldsymbol{\delta}) + \epsilon_{ic}$ as in a mixed logit (McFadden and Train 2000).

Assumption 2.1 also posits that the functional family of $U_i$ and the distributional family of $\boldsymbol{\nu}_i$ are known parametric classes, and that $\boldsymbol{\nu}_i$ is independent of $\mathbf{x}_i$. Though standard in discrete choice analysis, the parametric assumptions are not essential for our partial identification results (see Remark 3.1), and the independence assumption can be relaxed based on the structure of the empirical model (as we illustrate in our application). The assumption that $P$ is continuous, which ensures utility ties have probability zero, is also nonessential because our partial identification results allow for sets of model-implied optimal choices and thus can readily accommodate ties (see Section S1.2 of the Supplemental Material).

Our key point of departure from McFadden (1974) and the bulk of the discrete choice literature is the assumption regarding what is observed by the econometrician. It is standard to assume that (i) a random sample of choice sets $C_i$, choices $d_i$, and attributes $\mathbf{x}_i$, $\{(C_i, d_i, \mathbf{x}_i) : d_i \in^* C_i, i \in I \subset \mathcal{I}\}$, is observed, and that (ii) $|C_i| \geq 2$ for all $i \in \mathcal{I}$, where $|\cdot|$ denotes set cardinality (see, e.g., Manski 1975, Assumption 1). By contrast, we assume:

ASSUMPTION 2.2 (Random Sample and Minimum Choice Set Size):

*(I) A random sample of choices $d_i$ and attributes $\mathbf{x}_i$, $\{(d_i, \mathbf{x}_i) : i \in I \subset \mathcal{I}\}$, is observed.*

*(II) $\Pr(|C_i| \geq \kappa) = 1$ for all $i \in \mathcal{I}$, where $\kappa \geq 2$ is a known integer.*

Assumption 2.2(I) is weaker than the standard assumption as it omits the requirement that choice sets are observed. Given this difference, Assumption 2.2(II) is comparable to the standard assumption as it requires that choice sets have a known minimum size, $\kappa$, greater than one. The empirical content of the model increases with $\kappa$. Knowledge of $\kappa$ is immediate when choice sets are observed. We assume that $\kappa$ is known, either from information in the data or by assumption, even though choice sets are unobserved. In any event, Assumption 2.2(II) is weaker than the common assumption that every agent's choice set coincides with the feasible set or a known subset of the feasible set.

REMARK 2.1: Under Assumption 2.2(II) the model has no empirical content if $\kappa = 1$. However, Assumption 2.2(II) can be weakened to $\Pr(|C_i| = 1) \leq \bar{\pi}_1 < 1$ for all $i \in \mathcal{I}$, where $\bar{\pi}_1$ is known. In this case the empirical content of the model is decreasing in $\bar{\pi}_1$.

A key feature of our model is that it admits *any* choice set formation process, including any mixture process, subject only to Assumption 2.2(II). Choice sets may be formed by internal processes, such as simultaneous or sequential search (Stigler 1961; Weitzman 1979; Honka et al. 2019) or elimination-by-aspects or attention or attribute filters (Tversky 1972a,b; Masatlioglu et al. 2012; Kimya 2018; Cattaneo et al. 2020), or by external processes, such as advertising (Chamberlin 1933; Goeree 2008; Terui et al. 2011) or choice architecture (Thaler and Sunstein 2008; Johnson et al. 2012; Gaynor et al. 2016). Whether internal or external, the choice set formation process can admit any dependence structure, without restriction, between agents' choice sets and their observable attributes and, conditional on observables, between agents' choice sets and their unobservable attributes. That is, $C_i$ can be arbitrarily correlated with $\mathbf{x}_i$ and, conditional on $\mathbf{x}_i$, $C_i$ can be arbitrarily correlated with $\boldsymbol{\nu}_i$.

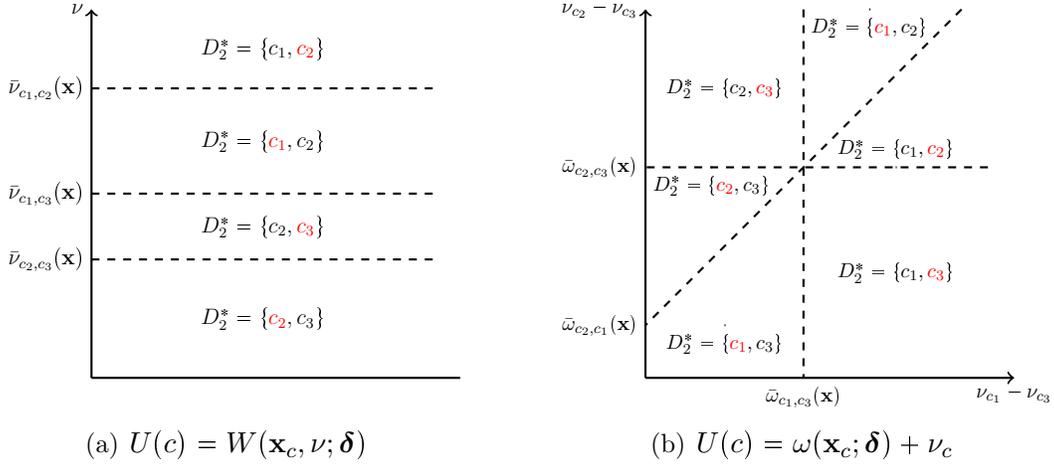

(a) $U(c) = W(\mathbf{x}_c, \nu; \boldsymbol{\delta})$

(b) $U(c) = \omega(\mathbf{x}_c; \boldsymbol{\delta}) + \nu_c$

Figure 3.1: Stylized depictions of $D_\kappa^*$ when $|\mathcal{D}| = 3$ and $\kappa = 2$.

Notes: In Panel (a), $\boldsymbol{\nu} \in \mathbb{R}$, $U(c) = W(\mathbf{x}_c, \nu; \boldsymbol{\delta})$, and the alternatives in $\mathcal{D}$ are vertically differentiated. The threshold $\bar{\nu}_{c_a,c_b}(\mathbf{x})$ is the value of $\nu$ above which $c_a$ has greater utility than $c_b$ and below which $c_b$ has greater utility than $c_a$. In Panel (b), $\boldsymbol{\nu} \in \mathbb{R}^3$ and $U(c) = \omega(\mathbf{x}_c; \boldsymbol{\delta}) + \nu_c$. The threshold $\bar{\omega}_{c_a,c_b}(\mathbf{x}) \equiv \omega(\mathbf{x}_{c_b}; \boldsymbol{\delta}) - \omega(\mathbf{x}_{c_a}; \boldsymbol{\delta})$ is the value of $\nu_{c_a} - \nu_{c_b}$ above which $c_a$ has greater utility than $c_b$ and below which $c_b$ has greater utility than $c_a$. Because $\kappa = 2$, either $|C| = 2$ or $|C| = 3$ and hence $D_2^*$ comprises the first and second best alternatives in $\mathcal{D}$. For a given $\boldsymbol{\nu}$, the first best appears in black and the second best in red. The agent's choice is determined by her realization $G$ of $C$. She chooses the first best if it is in $G$; otherwise she chooses the second best.

## 3 Partial Identification of the Model's Parameters

### 3.1 Preferences

The random utility model in Section 2 implies multiple optimal choices for the agent, due to the multiple possible realizations $G$ of her choice set $C_i$. Let $d_i^*(G, \mathbf{x}_i, \boldsymbol{\nu}_i; \boldsymbol{\delta})$ denote the model-implied optimal choice for agent $i$ with attributes $(\mathbf{x}_i, \boldsymbol{\nu}_i)$, choice set $C_i = G \subseteq \mathcal{D}$, and utility parameter $\boldsymbol{\delta}$. That is, $d_i^*(G, \mathbf{x}_i, \boldsymbol{\nu}_i; \boldsymbol{\delta}) \equiv \arg\max_{c \in G} W(\mathbf{x}_{ic}, \boldsymbol{\nu}_i; \boldsymbol{\delta})$.

The set of model-implied optimal choices given $(\mathbf{x}_i, \boldsymbol{\nu}_i)$ and $\boldsymbol{\delta}$ is

$$D_\kappa^*(\mathbf{x}_i, \boldsymbol{\nu}_i; \boldsymbol{\delta}) = \bigcup_{G \subseteq \mathcal{D}: |G| \geq \kappa} \left\{ d_i^*(G, \mathbf{x}_i, \boldsymbol{\nu}_i; \boldsymbol{\delta}) \right\} = \bigcup_{G \subseteq \mathcal{D}: |G| = \kappa} \left\{ d_i^*(G, \mathbf{x}_i, \boldsymbol{\nu}_i; \boldsymbol{\delta}) \right\}, \quad (3.1)$$

where the last equality follows from Sen's property $\alpha$: any alternative that is optimal for a given choice set $G' \subseteq \mathcal{D}$ is also optimal for every choice set $G \subset G'$ containing that alternative. The set $D_\kappa^*(\mathbf{x}_i, \boldsymbol{\nu}_i; \boldsymbol{\delta})$ is a *random closed set* with realizations in $\mathcal{D}$.[2] It contains the $|\mathcal{D}| - \kappa + 1$ best alternatives in $\mathcal{D}$, where "best" is defined with respect to $U_i$. Figure 3.1 contains stylized depictions of $D_\kappa^*(\mathbf{x}_i, \boldsymbol{\nu}_i; \boldsymbol{\delta})$ when $|\mathcal{D}| = 3$ and $\kappa = 2$.

---

[2]We formally define a random closed set in Definition A.1 in the Appendix. We formally establish that $D_\kappa^*(\mathbf{x}_i, \boldsymbol{\nu}_i; \boldsymbol{\delta})$ is a random closed set in Lemma A.1 in the Appendix.

When the information in the data and the economic model do not impose sufficiently strong restrictions on the distribution of $C_i$, the multiplicity of model-implied optimal choices generally precludes point identification of the model's parameters $\boldsymbol{\theta} = [\boldsymbol{\delta}; \boldsymbol{\gamma}]$. The reason is that the relationship between the data and the model is *incomplete* (Tamer 2003). To see this, let $\Pr(d_i^* = c|\mathbf{x}_i; \boldsymbol{\theta}, \mathsf{F})$ denote the model-implied conditional probability that alternative $c$ is chosen given $\mathbf{x}_i$ and $(\boldsymbol{\theta}, \mathsf{F})$, where $\mathsf{F} \equiv \mathsf{F}(\cdot; \mathbf{x}_i, \boldsymbol{\nu}_i)$ denotes the conditional probability mass function of $C_i$ given $(\mathbf{x}_i, \boldsymbol{\nu}_i)$. For all $c \in \mathcal{D}$,

$$\Pr(d_i^* = c|\mathbf{x}_i; \boldsymbol{\theta}, \mathsf{F}) = \int_{\boldsymbol{\tau} \in \mathcal{V}} \sum_{G \subseteq \mathcal{D}} \mathbf{1}(d_i^*(G, \mathbf{x}_i, \boldsymbol{\tau}; \boldsymbol{\delta}) = c) \mathsf{F}(G; \mathbf{x}_i, \boldsymbol{\tau}) dP(\boldsymbol{\tau}; \boldsymbol{\gamma}). \tag{3.2}$$

Because the only restriction we impose on $\mathsf{F}$ is that $\mathsf{F}(G; \mathbf{x}_i, \boldsymbol{\nu}_i) = 0$ for $G \subseteq \mathcal{D}$, $|G| < \kappa$, there may be multiple admissible values of $(\boldsymbol{\theta}, \mathsf{F})$ such that

$$\Pr(d_i^* = c|\mathbf{x}_i; \boldsymbol{\theta}, \mathsf{F}) = \Pr(d_i = c|\mathbf{x}_i), \ \forall c \in \mathcal{D}, \ \mathbf{x}_i - a.s., \tag{3.3}$$

where $d_i$ is the agent's observed choice.[3] Nonetheless, in general, it is not the case that for every $\boldsymbol{\theta}$ in a parameter space $\Theta$ there is an admissible $\mathsf{F}$ such that condition (3.3) holds. Hence, we can partially identify $\boldsymbol{\theta}$ from the information in the data and the model.

The set of values $\boldsymbol{\theta} \in \Theta$ for which there exists an admissible distribution $\mathsf{F}$ such that condition (3.3) holds forms the *sharp identification region* for $\boldsymbol{\theta}$. We denote this region by $\Theta_I$. The distribution $\mathsf{F}$, however, is an infinite-dimensional nuisance parameter, which creates difficulties for the computation of $\Theta_I$ and for statistical inference.[4] We circumvent these difficulties by working directly with the set $D_\kappa^*(\mathbf{x}_i, \boldsymbol{\nu}_i; \boldsymbol{\delta})$.

If the model is correctly specified, the agent's observed choice $d_i$ is maximal with respect to her preference among the alternatives in her choice set and it therefore satisfies

$$d_i \in D_\kappa^*(\mathbf{x}_i, \boldsymbol{\nu}_i; \boldsymbol{\delta}), \text{ almost surely}, \tag{3.4}$$

for the data generating value $\boldsymbol{\theta} \in \Theta$. To harness the empirical content of equation (3.4), we leverage a result in Artstein (1983), reported in Theorem A.1 in the Appendix. This result allows us to translate equation (3.4) into a finite number of conditional moment inequalities that fully characterize the sharp identification region $\Theta_I$.

---

[3] If $\mathsf{F}$ is known or sufficiently restricted (e.g., parametrically specified), then $\boldsymbol{\theta}$ can be point identified by condition (3.3) given sufficient variation in $\mathbf{x}_i$ and exclusion restrictions. For a discussion, see Section 7.

[4] Moreover, in (conditional or mixed) logit models, the fact that $\mathsf{F}$ may depend on $\boldsymbol{\nu}_i$ renders inapplicable the closed-from expressions for choice probabilities that are typical of these models. By contrast, as we show in Section S1.5 of the Supplemental Material, our method allows one to leverage such closed-form expressions to simplify computation.

THEOREM 3.1: *Let Assumptions 2.1 and 2.2 hold. In addition, let $\boldsymbol{\theta} = [\boldsymbol{\delta}; \boldsymbol{\gamma}]$, $\Theta = \Delta \times \Gamma$, and $\mathbb{K} = \{K \subset \mathcal{D} : |K| < \kappa\}$. Then*

$$\Theta_I = \left\{ \boldsymbol{\theta} \in \Theta : \Pr(d \in K | \mathbf{x}) \leqslant P(D^*_\kappa(\mathbf{x}, \boldsymbol{\nu}; \boldsymbol{\delta}) \cap K \neq \varnothing; \boldsymbol{\gamma}), \forall K \in \mathbb{K}, \mathbf{x} - a.s. \right\}. \qquad (3.5)$$

Our proof of Theorem 3.1, provided in Section A.2 of the Appendix, establishes that the characterization in equation (3.5) is sharp—all and only those values $\boldsymbol{\theta} \in \Theta$ for which the inequalities in equation (3.5) hold could have generated the observed data under the maintained assumptions.[5] These inequalities have a straightforward interpretation. At the data generating value $\boldsymbol{\theta} \in \Theta$, it must be the case that, for every subset $K \in \mathbb{K}$, the conditional probability that $K$ contains a model-implied optimal choice (right-hand side) is not less than the conditional probability of the observed choice (left-hand side), which itself is optimal. When $\boldsymbol{\nu}_i \in \mathbb{R}$ and the alternatives in $\mathcal{D}$ are vertically differentiated, the set $\mathbb{K}$ can be restricted to the subsets $\overrightarrow{K} = \{c_1\}, \{c_1, c_2\}, \ldots, \{c_1, c_2, \ldots, c_{\kappa-1}\}$ and $\overleftarrow{K} = \{c_{|\mathcal{D}|}\}, \{c_{|\mathcal{D}|}, c_{|\mathcal{D}|-1}\}, \ldots, \{c_{|\mathcal{D}|}, c_{|\mathcal{D}|-1}, \ldots, c_{|\mathcal{D}|-\kappa+2}\}$,[6] and the inequalities translate into statements about cumulative shares for higher (respectively lower) quality alternatives.

## 3.2 Choice Sets

Theorem 3.1 establishes that, under mild restrictions on the utility function (Assumption 2.1) and knowing only the minimum size of agents' choice sets (Assumption 2.2), one can learn features of the distribution of preferences without observing agents' choice sets or knowing how they are formed. We next show that, with an additional restriction on the choice set formation process, one can also learn features of the distribution of choice sets.

Let $\ell_i \equiv |C_i|$ denote the size of agent $i$'s choice set $C_i$. When $\ell_i = |\mathcal{D}|$ we say that $C_i$ has "full" size. When $\ell_i < |\mathcal{D}|$ we say that $C_i$ is "limited" or "restricted." More specifically, we say that $C_i$ is "full$-1$" when $\ell_i = |\mathcal{D}| - 1$, "full$-2$" when $\ell_i = |\mathcal{D}| - 2$, and so forth.

In addition to Assumptions 2.1 and 2.2, one could assume that:

ASSUMPTION 3.1 (Choice Set Size): *Agent $i$ draws the size $\ell_i$ of her choice set such that*

$$\Pr(\ell_i = q | \mathbf{x}_i, \boldsymbol{\nu}_i) = \Pr(\ell_i = q | \mathbf{x}_i) \equiv \pi_q(\mathbf{x}_i; \boldsymbol{\eta}), \ q = \kappa, \ldots, |\mathcal{D}|, \qquad (3.6)$$

*where $\pi_q(\mathbf{x}_i; \boldsymbol{\eta}) \geqslant 0$ for $q \geqslant \kappa$, $\sum_{q=\kappa}^{|\mathcal{D}|} \pi_q(\mathbf{x}_i; \boldsymbol{\eta}) = 1$, and the function $\pi$ is known up to a finite-dimensional parameter vector $\boldsymbol{\eta} \in H \subset \mathbb{R}^m$ where $H$ is convex and compact.*

---

[5]If per Remark 2.1 one weakens Assumption 2.2(II) to $\Pr(|C_i| = 1) \leqslant \bar{\pi}_1 < 1$ where $\bar{\pi}_1$ is known, then $\Theta_I = \{\boldsymbol{\theta} \in \Theta : \Pr(d \in K | \mathbf{x}) \leqslant \bar{\pi}_1 + (1 - \bar{\pi}_1) P(D^*_2(\mathbf{x}, \boldsymbol{\nu}; \boldsymbol{\delta}) \cap K \neq \varnothing; \boldsymbol{\gamma}), \forall K \in \mathbb{K}, \mathbf{x} - a.s.\}$.
[6]See Corollary S1.1 and Claim S1.1 in the Supplemental Material.

Assumption 3.1 posits that the size $\ell_i$ of agent $i$'s choice set is drawn from an unspecified distribution with support $\{\kappa, \ldots, |\mathcal{D}|\}$, which allows for the possibility that the agent's choice set has full size, $\ell_i = |\mathcal{D}|$, or is limited, $\ell_i < |\mathcal{D}|$. The only restrictions it imposes on the distribution of agents' choice sets are that the distributional family of $\ell_i$ is a known parametric class—though, as before, the parametric structure is not essential (see Remark 3.1)—and that $\ell_i$ is independent of $\boldsymbol{\nu}_i$. Conditional on $\ell_i$, however, the model with Assumption 3.1 continues to allow for any dependence structure, without restriction, between agents' choice sets and their observable attributes and, conditional on observables, between agents' choice sets and their unobservable attributes. Moreover, agents with choice sets of the same size need not have choice sets with the same composition.

Under Assumption 3.1, Theorem 3.1 specializes to the following corollary.[7]

COROLLARY 3.1: *Let Assumptions 2.1, 2.2, and 3.1 hold. In addition, let $\boldsymbol{\theta} = [\boldsymbol{\eta}; \boldsymbol{\delta}; \boldsymbol{\gamma}]$ and $\Theta = H \times \Delta \times \Gamma$. Then*

$$\Theta_I = \left\{ \boldsymbol{\theta} \in \Theta : \Pr(d \in K | \mathbf{x}) \leqslant \sum_{q=\kappa}^{|\mathcal{D}|} \pi_q(\mathbf{x}; \boldsymbol{\eta}) P(D_q^*(\mathbf{x}, \boldsymbol{\nu}; \boldsymbol{\delta}) \cap K \neq \varnothing; \boldsymbol{\gamma}), \forall K \subset \mathcal{D}, \mathbf{x} - a.s. \right\}. \quad (3.7)$$

The sharp identification region $\Theta_I$ in Corollary 3.1 has two noteworthy features. First, the projection of $\Theta_I$ on $[\boldsymbol{\delta}; \boldsymbol{\gamma}]$ is equal to the sharp identification region in Theorem 3.1. In other words, the information in $\Theta_I$ about the distribution of preferences is the same with or without Assumption 3.1. This is because $D_{q+1}^*(\mathbf{x}_i, \boldsymbol{\nu}_i; \boldsymbol{\delta}) \subset D_q^*(\mathbf{x}_i, \boldsymbol{\nu}_i; \boldsymbol{\delta})$ for all $q \geqslant \kappa$, and thus the projection of $\Theta_I$ on $[\boldsymbol{\delta}; \boldsymbol{\gamma}]$ is obtained with $\pi_\kappa(\mathbf{x}_i; \boldsymbol{\eta}) = 1$ and $\pi_q(\mathbf{x}_i; \boldsymbol{\eta}) = 0$ for $q > \kappa$. Second, $\Theta_I$ provides information about the distribution of choice set size, as well. It yields a lower bound on $\pi_\kappa(\mathbf{x}_i; \boldsymbol{\eta})$ (the upper bound is one provided $\kappa < |\mathcal{D}|$) and upper bounds on $\pi_q(\mathbf{x}_i; \boldsymbol{\eta})$ for $q = \kappa + 1, \ldots, |\mathcal{D}|$ (the lower bounds are zero provided $\kappa < |\mathcal{D}|$).

REMARK 3.1: Theorem 3.1 and Corollary 3.1 can be generalized for a *structure* $(W, P)$ or $(W, P, \pi)$, as the case may be, that is subject only to nonparametric restrictions. We focus on the case with parametric restrictions for computational reasons and because methods of statistical inference for moment inequality models focus on this case.

### 3.3 Illustration of the Inequalities Characterizing $\Theta_I$

Figure 3.2 contains stylized depictions of three inequalities in equation (3.7) when $|\mathcal{D}| = 5$, $\kappa = 4$, $\boldsymbol{\nu}_i = \nu_i \in \mathbb{R}$ with support $\mathcal{V} = [0, \bar{\nu}]$, and the alternatives in $\mathcal{D}$ are vertically differentiated. In this case $\Pr(\ell_i \in \{4, 5\}) = 1$. With probability $\pi_5$ the agent draws a

---
[7]The proof of Corollary 3.1 follows immediately from the proof of Theorem 3.1 and therefore is omitted.

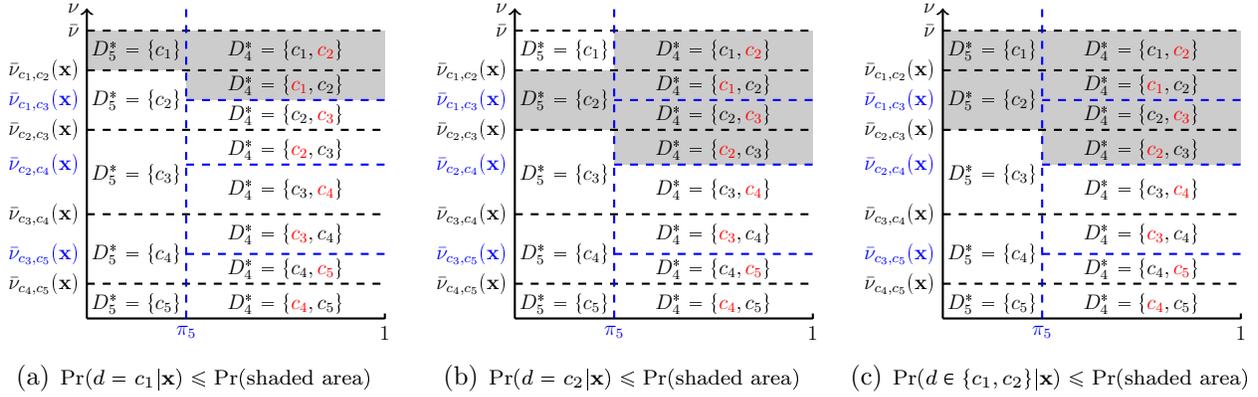

Figure 3.2: Stylized depictions of inequalities in $\Theta_I$ when $|\mathcal{D}| = 5$ and $\kappa = 4$.

Notes: Inequalities for three subsets $K \subset \mathcal{D}$ are depicted: (a) $K = \{c_1\}$; (b) $K = \{c_2\}$; and (c) $K = \{c_1, c_2\}$. For a given $\nu$, the first best alternative in $\mathcal{D}$ appears in black and the second best in red.

choice set of size 5, in which case $D_5^*$ comprises the first best alternative. With probability $\pi_4 = 1 - \pi_5$ she draws a choice set of size 4, in which case $D_4^*$ comprises the first and second best alternatives. In the former case the agent chooses the first best. In the latter case her choice is determined by her realization $G$ of $C_i$. She chooses the first best if it is contained in $G$; otherwise she chooses the second best.[8] The threshold $\bar{\nu}_{c_a,c_b}(\mathbf{x}_i)$ is the value of $\nu_i$ above which $c_a$ has greater utility than $c_b$ and below which $c_b$ has greater utility than $c_a$.

Panel (a) depicts the inequality for $K = \{c_1\}$. If $\ell_i = 5$ then $C_i = \mathcal{D}$ and $c_1$ is the optimal choice if $\nu_i > \bar{\nu}_{c_1,c_2}(\mathbf{x}_i)$. If $\ell_i = 4$ then $c_1$ is optimal if $\nu_i > \bar{\nu}_{c_1,c_2}(\mathbf{x}_i)$ and the realization $G$ of $C_i$ includes $c_1$ or if $\nu_i \in (\bar{\nu}_{c_1,c_3}(\mathbf{x}_i), \bar{\nu}_{c_1,c_2}(\mathbf{x}_i))$ and $G$ excludes $c_2$. It follows that

$$\Pr(d_i = c_1|\mathbf{x}_i) \leqslant \pi_5 P(\nu_i > \bar{\nu}_{c_1,c_2}(\mathbf{x}_i); \boldsymbol{\gamma}) + (1 - \pi_5) P(\nu_i > \bar{\nu}_{c_1,c_3}(\mathbf{x}_i); \boldsymbol{\gamma}).$$

Similar reasoning applies to the other singleton sets, with $K = \{c_2\}$ depicted in Panel (b).

The inequalities in equation (3.7) also include those for non-singleton sets. To see why, Panel (c) depicts the inequality for $K = \{c_1, c_2\}$. While the left-hand side is additive,

$$\Pr(d_i \in \{c_1, c_2\}|\mathbf{x}_i) = \Pr(d_i = c_1|\mathbf{x}_i) + \Pr(d_i = c_2|\mathbf{x}_i),$$

the right-hand side is subadditive: the shaded area in Panel (c) is smaller than the sum of the shaded areas in Panels (a) and (b). Hence, the values $\boldsymbol{\theta} \in \Theta$ that satisfy the inequalities for $K = \{c_1\}$ and $K = \{c_2\}$ may fail to satisfy the inequality for $K = \{c_1, c_2\}$.

---

[8]In general, the agent chooses the best alternative in the intersection of her realizations of $D_q^*$ and $C_i$.

Not all pairs of singleton sets, however, yield nonredundant inequalities. Consider, for example, $K = \{c_1\}$ and $K = \{c_5\}$. As is clear from Figure 3.2, there is no value $\nu_i \in \mathcal{V}$ for which $D_4^*$ contains both $c_1$ and $c_5$. It follows that the inequality for $K = \{c_1, c_5\}$ is redundant if the inequalities for $K = \{c_1\}$ and $K = \{c_5\}$ are satisfied. This type of reasoning can substantially reduce the number of inequalities that are needed to recover $\Theta_I$.[9]

Though not depicted in Figure 3.2, let us highlight the algebra that delivers an upper bound on $\pi_5$. Consider $K = \{c_1, c_2, c_3, c_4\}$. Given this $K$ we have

$$\Pr(d_i \in K | \mathbf{x}_i) \leqslant \pi_5 \Pr(D_5^* \cap K \neq \varnothing) + (1 - \pi_5) \Pr(D_4^* \cap K \neq \varnothing)$$
$$\Leftrightarrow \Pr(d_i = c_5 | \mathbf{x}_i) \geqslant \pi_5 \Pr(D_5^* = \{c_5\}) = \pi_5 P(\nu_i \leqslant \bar{\nu}_{c_4, c_5}(\mathbf{x}_i); \boldsymbol{\gamma}).$$

Given any $\boldsymbol{\gamma}$, this inequality yields an upper bound on $\pi_5$. In general, one obtains the upper bound on $\pi_q$, $q = \kappa + 1, \ldots, |\mathcal{D}|$, from a projection of $\Theta_I$ on the $\boldsymbol{\eta}$ component of $\boldsymbol{\theta}$.

## 3.4 Implementation of the Method

There are two challenges, both computational, in applying Theorem 3.1 and Corollary 3.1. First, given any $\kappa \geqslant 2$, the number of inequalities that characterize $\Theta_I$ grows superlinearly with $|\mathcal{D}|$. Second, computing the model-implied probabilities (the right-hand sides of the inequalities) may require evaluating a number of integrals equal to the dimension of $\boldsymbol{\nu}_i$.[10] In this section we discuss both challenges. For the sake of brevity we focus on Theorem 3.1.[11]

As the set $D_\kappa^*(\mathbf{x}_i, \boldsymbol{\nu}_i; \boldsymbol{\delta})$ comprises the $|\mathcal{D}| - \kappa + 1$ best alternatives in $\mathcal{D}$, it can have at most $h = \binom{|\mathcal{D}|}{|\mathcal{D}| - \kappa + 1}$ realizations, which we denote $D^1, \ldots, D^h$, with

$$\{D_\kappa^*(\mathbf{x}_i, \boldsymbol{\nu}_i; \boldsymbol{\delta}) = D^j\} = \{W(\mathbf{x}_{ic'}, \boldsymbol{\nu}_i; \boldsymbol{\delta}) > W(\mathbf{x}_{ic}, \boldsymbol{\nu}_i; \boldsymbol{\delta}) \ \forall c' \in D^j, \ \forall c \in \mathcal{D} \backslash D^j\}.$$

(In some models $P(D_\kappa^*(\mathbf{x}_i, \boldsymbol{\nu}_i; \boldsymbol{\delta}) = D^j; \boldsymbol{\gamma}) = 0$ for some $j \in \{1, \ldots, h\}$.[12]) It follows that

$$P(D_\kappa^*(\mathbf{x}_i, \boldsymbol{\nu}_i; \boldsymbol{\delta}) \cap K \neq \varnothing; \boldsymbol{\gamma}) = \sum_{j: D^j \cap K \neq \varnothing} P(D_\kappa^*(\mathbf{x}_i, \boldsymbol{\nu}_i; \boldsymbol{\delta}) = D^j; \boldsymbol{\gamma}). \qquad (3.8)$$

In some cases one can eliminate redundant inequalities through judicious use of set theory. For example, consider two disjoint subsets $K_1, K_2 \subset \mathcal{D}$ such that

$$P([D_\kappa^*(\mathbf{x}_i, \boldsymbol{\nu}_i; \boldsymbol{\delta}) \cap K_1 \neq \varnothing] \cap [D_\kappa^*(\mathbf{x}_i, \boldsymbol{\nu}_i; \boldsymbol{\delta}) \cap K_2 \neq \varnothing]; \boldsymbol{\gamma}) = 0.$$

---

[9]See Section 3.4 below.

[10]The left-hand sides can be estimated from the data.

[11]The same observations and results hold for Corollary 3.1 by replacing $\kappa$ with $q = \kappa + 1, \ldots, |\mathcal{D}|$.

[12]In the example presented in Section 3.3, this is the case for all $D^j$ comprised on non-adjacent elements, i.e., $D^j \in \{\{c_1, c_5\}, \{c_1, c_4\}, \{c_1, c_3\}, \{c_2, c_5\}, \{c_2, c_4\}, \{c_3, c_5\}\}$.

If the inequalities for $K = K_1$ and $K = K_2$ are satisfied, then the inequality for $K = \{K_1 \cup K_2\}$ is also satisfied, so the latter is redundant.[13] Now suppose

$$P(D_\kappa^*(\mathbf{x}_i, \boldsymbol{\nu}_i; \boldsymbol{\delta}) \cap K_1 \neq \varnothing; \boldsymbol{\gamma}) = P(D_\kappa^*(\mathbf{x}_i, \boldsymbol{\nu}_i; \boldsymbol{\delta}) \cap \{K_1 \cup K_2\} \neq \varnothing; \boldsymbol{\gamma}).$$

In this case, if the inequality for $K = \{K_1 \cup K_2\}$ holds, then the inequality for $K = K_1$ also holds, so the latter is redundant. In Theorem S1.1 in the Supplemental Material we provide an algorithm based on these considerations to eliminate redundant inequalities, and in Corollary S1.1 we provide sufficient conditions under which the number of inequalities is $2(\kappa - 1)$. One can check the conditions for the application of these results numerically or, in some cases, analytically (depending on the structure of the data and the model).

In some models, however, the predicate conditions in Theorem S1.1 and Corollary S1.1 do not apply, and the number of inequalities may be very large. This is the case, for instance, in a model where $\boldsymbol{\nu}_i = (\boldsymbol{v}_i, (\epsilon_{ic}, c \in \mathcal{D}))$, $U_i(c) = \omega(\mathbf{x}_{ic}, \boldsymbol{v}_i; \boldsymbol{\delta}) + \epsilon_{ic}$, and $(\epsilon_{ic}, c \in \mathcal{D})$ has full support on $\mathbb{R}^{|\mathcal{D}|}$ (e.g., a mixed logit). In Section S1.4 of the Supplemental Material we show that $\Theta_I$ can be equivalently characterized as the set of values $\boldsymbol{\theta} \in \Theta$ for which the optimal value of a convex program with $|\mathcal{D}|$ optimization variables is zero.[14] The convex program bypasses the need to enumerate all of the inequalities, and thanks to efficient algorithms for solving convex programs, the number of times that the objective function (which returns each inequality for specific choices of the optimization variables) is evaluated is typically less than the number of inequalities in equation (3.5).

The remaining challenge is computing $P(D_\kappa^*(\mathbf{x}_i, \boldsymbol{\nu}_i; \boldsymbol{\delta}) \cap K \neq \varnothing; \boldsymbol{\gamma})$ when the dimension of $\boldsymbol{\nu}_i$ is large. In light of equation (3.8) this amounts to computing $P(D_\kappa^*(\mathbf{x}_i, \boldsymbol{\nu}_i; \boldsymbol{\delta}) = D^j; \boldsymbol{\gamma})$ for all $j \in \{1, \ldots, h\}$. In Theorem S1.2 in the Supplemental Material we provide simplifications to compute these probabilities in a mixed logit model with unobserved heterogeneity in choice sets, where choice sets may be correlated with $\boldsymbol{\nu}_i$.[15] We show how one can exploit the logit closed-form choice probabilities and then numerically integrate over the random coefficients, thereby substantially reducing the computational burden.

---

[13]In the example in Section 3.3, this is the case, e.g., for $K_1 = \{c_1\}$ and $K_2 = \{c_5\}$. Another useful application of this result pertains to testing for full-size choice sets. With full-size choice sets, $D_\kappa^*(\mathbf{x}_i, \boldsymbol{\nu}_i; \boldsymbol{\delta})$ is a singleton and thus if $K_1$ and $K_2$ are disjoint it can intersect at most one of them. To test for full-size choice sets, therefore, one need only check the inequalities for the singleton subsets of $\mathcal{D}$ and their complements.

[14]This characterization can be used for inference with the method proposed by Andrews and Shi (2017).

[15]As in a typical mixed logit, we assume the random coefficients and additive errors are independent.

# 4 Deductible Choices in Auto Collision Insurance

In this section and the next we apply our theoretical findings to learn about the distributions of risk preferences and choice set size from data on households' deductible choices in auto collision insurance. In this section we specify a random expected utility model that allows for unobserved heterogeneity in risk aversion and choice sets and we describe our data.

## 4.1 Empirical Model

We model households' deductible choices in auto collision insurance. Each household $i$ (i) faces a menu of prices $\mathbf{p}_i = (p_{ic}, c \in \mathcal{D})$, where $p_{ic}$ is the household-specific premium associated with deductible $c$ and $\mathcal{D}$ is the feasible set of deductibles, (ii) has a probability $\mu_i$ of experiencing a claim during the policy period, and (iii) has an array of observed characteristics $\mathbf{t}_i$.[16] Following the related literature (e.g., Cohen and Einav 2007; Sydnor 2010; Barseghyan et al. 2011, 2013, 2016),[17] we make two simplifying assumptions about claims and their probabilities.

ASSUMPTION 4.1 (Claims and Claim Probabilities):

(I) *Households disregard the possibility of more than one claim during the policy period.*

(II) *Any claim exceeds the highest deductible in $\mathcal{D}$; payment of the deductible is the only cost associated with a claim; and deductible choices do not influence claim probabilities.*

Assumption 4.1(I) is motivated by the fact that claim rates are small, so the likelihood of two or more claims in the same policy period is very small. Assumption 4.1(II) abstracts from small claims, transaction costs, and moral hazard.

Under Assumption 4.1, household $i$'s choice of deductible involves a choice among binary lotteries, indexed by $c \in \mathcal{D}$, of the following form: $L_i(c) = (-p_{ic}, 1 - \mu_i; -p_{ic} - c, \mu_i)$. The household chooses among these lotteries based on the criterion in equation (2.1). We assume that household $i$'s preferences conform to expected utility theory,

$$U_i(c) = (1 - \mu_i)u_i(w_i - p_{ic}) + \mu_i u_i(w_i - p_{ic} - c), \tag{4.1}$$

where $w_i$ is the household's wealth and $u_i$ is its Bernoulli utility function.

We impose the following shape restriction on $u_i$.

ASSUMPTION 4.2 (CARA): *The function $u_i$ exhibits constant absolute risk aversion, i.e., $u_i(y) = \frac{1 - \exp(-\nu_i y)}{\nu_i}$ for $\nu_i \neq 0$ and $u_i(y) = y$ for $\nu_i = 0$.*

---

[16]As we explain in Section 4.2, we estimate $\mu_i$ and treat it as data.

[17]For a survey, see Barseghyan et al. (2018, Section 5.2).

Assuming CARA has two key virtues. First, $u_i$ is fully characterized by the coefficient of absolute risk aversion, $\nu_i \equiv -u_i''(y)/u_i'(y)$. Second, wealth does not affect utility comparisons. We note, however, that our approach can accommodate other shape restrictions (e.g., constant relative risk aversion) as well as non-expected utility models (e.g., the rank-dependent expected utility model in Barseghyan et al. 2013).

In terms of the core model developed in Section 2, household $i$'s observable attributes are $\mathbf{x}_i = (\mu_i, \mathbf{t}_i, \mathbf{p}_i)$, with $\mathbf{x}_{ic} = (\mu_i, \mathbf{t}_i, p_{ic})$, and its sole unobservable attribute is its coefficient of absolute risk aversion $\nu_i$.[18] Per Assumptions 2.1 and 4.2, we posit that $\nu_i \sim P(\boldsymbol{\gamma}(\mathbf{t}_i))$, where $P$ is specified below in Assumption 4.3(I), and that, $(\mathbf{x}_{ic}, \nu_i) - a.s.$,

$$U_i(c) = \frac{(1-\mu_i)(1-\exp(\nu_i p_{ic})) + \mu_i(1-\exp(\nu_i(p_{ic}+c)))}{\nu_i}. \tag{4.2}$$

Observe that, by equation (4.2), we assume that $\mu_i$ and $p_{ic}$ affect utility directly and we allow $\mathbf{t}_i$ to affect utility indirectly through $\nu_i$. To capture this indirect effect, we could specify $\boldsymbol{\gamma}(\mathbf{t}_i) = f(\mathbf{t}_i; \boldsymbol{\delta})$ where the functional form of $f$ is known up to $\boldsymbol{\delta} \in \Delta$. Instead, we account for (discrete) observed heterogeneity in preferences nonparametrically by conducting the analysis separately on population subgroups based on $\mathbf{t}_i$.

Per Assumption 2.2(I), we suppose that the deductible choices and observable attributes, $\{(d_i, \mathbf{x}_i) : i \in I\}$, for a random sample of households $I \subset \mathcal{I}$, $|I| = n$, are observed, but that the households' choice sets, $\{C_i : C_i \subseteq \mathcal{D}, i \in I\}$, are unobserved. Per Assumption 2.2(II), we assume that $\Pr(|C_i| \geq \kappa) = 1$ for every household $i \in \mathcal{I}$, where $\kappa \geq 2$.

We close the baseline empirical model with two final assumptions.

ASSUMPTION 4.3 (Heterogeneity Restrictions):

(I) *Conditional on $\mathbf{t}_i$, $\nu_i$ follows a Beta distribution on $[0, 0.03]$ with parameter vector $\boldsymbol{\gamma}(\mathbf{t}_i) = (\gamma_1(\mathbf{t}_i), \gamma_2(\mathbf{t}_i))$ and is independent of $(\mu_i, p_{ic})$. To simplify notation, we suppress below the dependence of $\boldsymbol{\gamma}$ on $\mathbf{t}_i$.*

(II) *The minimum choice set size is $\kappa = 3$.*

Assumption 4.3(I) specifies that $P$ is the Beta distribution with support $\mathcal{V} = [0, 0.03]$. The main attraction of the Beta distribution is its flexibility (e.g., Ghosal 2001). Its bounded support is a plus given our setting. A lower bound of zero rules out risk-loving preferences and seems appropriate for insurance markets that exist primarily because of risk aversion. Imposing an upper bound enables us to rule out absurd levels of risk aversion, and the choice of 0.03 is conservative both as a theoretical matter and in light of prior empirical estimates in similar settings (e.g., Cohen and Einav 2007; Sydnor 2010; Barseghyan et al. 2011, 2013, 2016). Assumption 4.3(II) posits that the size of every household's choice set is either full, full-1, or full-2. In our setting $|\mathcal{D}| = 5$. We set $\kappa = 3$ for reasons we explain in Section 4.2.

---

[18]In terms of the notation used in Section 2, $\mathbf{s}_i = (\mu_i, \mathbf{t}_i)$, $\mathbf{z}_{ic} = p_{ic}$, and $\boldsymbol{\nu}_i = \nu_i$.

REMARK 4.1: We also consider a mixed logit specification $U_i(c) = \omega(\mathbf{x}_{ic}, \nu_i) + \epsilon_{ic}$, where $\omega(\mathbf{x}_{ic}, \nu_i)$ is the certainty equivalent of the right-hand side of equation (4.2), $\nu_i$ is distributed per Assumption 4.3(I), and $\epsilon_{ic}$ is an i.i.d. disturbance that follows a Type 1 Extreme Value distribution and is independent of $(\mathbf{x}_{ic}, \nu_i)$; see Section 5.1.1.

In the baseline model we do not impose Assumption 3.1. Thus, conditional on $\mathbf{x}_i$, $C_i$ can be arbitrarily correlated with $\nu_i$. We impose Assumption 3.1 only in Section 5.2 when we apply Corollary 3.1 to learn about the distribution of choice set size. At that point, we could specify a functional form for $\pi_q(\mathbf{x}_i; \boldsymbol{\eta})$ known up to $\boldsymbol{\eta} \in H$. Instead, as with $\nu_i$, we assume $\pi_q$ is independent of $(\mu_i, p_{ic})$ conditional on $\mathbf{t}_i$, and we account for (discrete) observed heterogeneity nonparametrically by conducting the analysis separately on population subgroups based on $\mathbf{t}_i$. To simplify notation, we suppress below the dependence of $\pi_q$ on $\mathbf{t}_i$.

## 4.2 Data Description

We obtained the data from a large U.S. property and casualty insurance company. The data contain annual information on more than 100,000 households who first purchased auto policies from the company during the ten year period from 1998 to 2007. We focus on households' deductible choices in auto collision coverage. This coverage pays for damage to the insured vehicle, in excess of the deductible, caused by a collision with another vehicle or object, without regard to fault. The feasible set of auto collision deductibles is $\mathcal{D} = \{\$100, \$200, \$250, \$500, \$1000\}$ and thus $|\mathcal{D}| = 5$.

To construct our analysis sample, we initially include every household who first purchased auto collision coverage from the company between 1998 and 2007, retaining, at the time of first purchase, its deductible choice $d_i$, its pricing menu $\mathbf{p}_i$, its claim probability $\mu_i$, and an array $\mathbf{t}_i$ of three demographic characteristics: gender, age, and insurance score of the principal driver.[19] This yields an initial sample of 112,011 households. We then exclude households whose deductible choices cannot be rationalized by the model specified in Section 4.1 for any pair $(\nu_i, C_i)$ such that $\nu_i \in [0, 0.03]$ and $|C_i| \in \{3, 4, 5\}$. Importantly, our rationalizability check does *not* rely on the assumption that $P$ is the Beta distribution. This excludes 0.1 percent of the initial sample, yielding a final sample of 111,890 households.[20]

---

[19]Insurance score is a credit-based risk score.

[20]The data in this paper are not the same as the data in Barseghyan et al. (2013) and Barseghyan et al. (2016), though both data sets have the same source. In this paper, the data comprise 112,011 households who first purchased auto collision coverage between 1998 and 2007. In Barseghyan et al. (2013) and

Several comments are in order. First, we retain households' deductible choices at the time of first purchase to increase confidence that we are working with active choices. One might worry that households renew their policies without actively reassessing their deductibles.

Second, we require $\nu_i \in [0, 0.03]$ for the reasons stated in Section 4.1. However, the composition of our sample is robust to the upper bound of the support. If we decrease the upper bound to 0.02 the sample decreases by one household to 111,889 households. If we increase the upper bound to 0.04 the sample remains the same at 111,890 households.[21]

Third, we require $|C_i| \in \{3, 4, 5\}$—i.e., we assume $\kappa = 3$—to keep the model as close as possible to the standard approach that assumes full-size choice sets. As we explain in Section 5.2, $\kappa = 3$ is the highest value that is consistent with the data.

Fourth, the company generates each household's pricing menu, $\mathbf{p}_i = (p_{ic}, c \in \mathcal{D})$, according to the following pricing rule: $p_{ic} = g(c)\bar{p}_i + \zeta$, where $\bar{p}_i$ is the household's base price, $g$ is a decreasing positive function, and $\zeta > 0$. We observe $g$, $\zeta$, and the premium paid by each household given its chosen deductible. We thus can recover each household's base price. Given the company's pricing rule, the base price is a sufficient statistic for $\mathbf{p}_i$. Moreover, any $p_{ic} \in \mathbf{p}_i$ can be treated as the base price. We treat the premium associated with the $1000 deductible as the base price—i.e., $\bar{p}_i = p_{1000}$—and round it to the nearest five dollars. We use the rounded base prices and resulting pricing menus throughout our analysis.[22]

Fifth, we estimate the households' claim probabilities using the company's claims data. We assume that household $i$'s auto collision claims in year $t$ follow a Poisson distribution with mean $\lambda_{it}$. We also assume that deductible choices do not influence claim rates (Assumption 4.1(II)). We perform a Poisson panel regression with random effects and use the results to calculate a fitted claim rate $\widehat{\lambda}_i$ for each household.[23] In principle, a household may experience one or more claims during the policy period. We assume that households disregard the possibility of experiencing more than one claim (Assumption 4.1(I)). Given this, we transform

---

Barseghyan et al. (2016), the data comprise 4,170 households who first purchased auto collision coverage, auto comprehensive coverage, and home all perils coverage in the same year, in either 2005 or 2006.

[21]Moreover, our results are robust to increasing the upper bound from 0.03 to 0.04, as indicated by results available from the authors upon request.

[22]This includes our rationalizability check, though the final sample would be virtually identical if we used exact prices. Our use of rounded prices reduces the computational burden of recovering $\Theta_I$ and is supported by evidence that "people show a marked tendency to produce 0- and 5-ending numbers" in numerical cognition tasks, including price cognition (Schindler and Kirby 1997, p. 193). See also Schindler and Wiman (1989), Vanhuele and Drèze (2000), and Liang and Kanetkar (2006).

[23]To obtain the most precise estimates, we use the full set of auto collision claims data, which comprises 1,349,853 household-year records. We calculate $\widehat{\lambda}_i$ conditional on the household's observables at the time of first purchase and its subsequent claims experience; see Section S3.1 of the Supplemental Material.

Table 4.1: Descriptive Statistics

*Panel A. Summary Statistics*

|  | Mean | Std. dev. | 5th pctl. | Median | 95th pctl. |
|---|---|---|---|---|---|
| Deductible choice (dollars) | 439 | 178 | 200 | 500 | 500 |
| Pricing menus: | | | | | |
| $p_{500}$ | 217 | 137 | 77 | 181 | 480 |
| $p_{250} - p_{500}$ | 65 | 42 | 22 | 54 | 146 |
| $p_{500} - p_{1000}$ | 49 | 32 | 17 | 41 | 110 |
| Claim probability (annual) | 0.088 | 0.030 | 0.045 | 0.085 | 0.140 |
| Demographic characteristics: | | | | | |
| Female | 0.468 | 0.499 | 0 | 0 | 1 |
| Age (years) | 48.1 | 16.6 | 24.5 | 45.9 | 76.7 |
| Insurance score | 731 | 114 | 555 | 725 | 934 |

*Panel B. Deductible Choices*

|  | Obs. | Percent choosing deductible | | | | |
|---|---|---|---|---|---|---|
|  |  | $100 | $200 | $250 | $500 | $1000 |
| All households | 111,890 | 1.1 | 15.2 | 13.7 | 65.4 | 4.6 |
| Male | 59,476 | 1.0 | 14.9 | 12.9 | 65.9 | 5.4 |
| Female | 52,414 | 1.1 | 15.5 | 14.7 | 64.8 | 3.8 |
| Young | 36,932 | 0.1 | 6.9 | 10.7 | 77.1 | 5.2 |
| Old | 38,046 | 2.5 | 26.2 | 16.7 | 51.0 | 3.6 |
| Low Insurance Score | 37,087 | 0.4 | 10.1 | 12.7 | 72.2 | 4.6 |
| High Insurance Score | 38,371 | 1.8 | 20.9 | 14.6 | 58.1 | 4.6 |

Notes: Analysis sample (111,890 households). Pricing statistics are annual amounts in nominal dollars. Demographic statistics are for the principal driver.

$\widehat{\lambda}_i$ into a claim probability $\mu_i \equiv 1 - \exp(-\widehat{\lambda}_i)$, which follows from the Poisson probability mass function, and round it to the nearest half percentage point.[24] We treat $\mu_i$ as data.

Table 4.1 presents descriptive statistics for the analysis sample. Panel A summarizes the households' deductible choices, pricing menus, claim probabilities, and demographic characteristics. Panel B reports the sample distribution of deductible choices for the full sample and for subsamples based on gender, age, and insurance score.[25] In Table 4.1 and throughout the paper, young/old and low/high insurance scores are defined as bottom/top third based on the age and insurance score, respectively, of the principal driver.

---

[24]Our use of rounded claim probabilities reduces the computational burden of recovering $\Theta_I$ and is supported by evidence that people report rounded probabilities (Manski and Molinari 2010).

[25]In addition, Table S3.1 in the Supplemental Material reports the sample distribution of deductible choices by octiles of base price $\bar{p}_i$ and claim probability $\mu_i$.

# 5 Empirical Method and Findings

Our empirical application is motivated in part by the fact that, although we observe the feasible set of deductibles, we do not observe which deductibles enter a household's choice set. There are many plausible sources of unobserved heterogeneity in choice sets. It may be due to missing data—e.g., different sales agents may quote different subsets of deductibles to different households—or to unobserved constraints—e.g., some households may disregard low deductibles due to budget constraints or high deductibles due to liquidity constraints.

Our application is also motivated by a persistent finding in prior empirical studies of risk preferences which assume full-size choice sets. These studies tend to find that average risk aversion is quite high—arguably implausibly high. Two recent examples that utilize similar data are Cohen and Einav (2007) and Barseghyan et al. (2013). It is plausible that the assumption of full-size choice sets may be driving this finding, and that allowing for unobserved heterogeneity in choice sets may yield more credible estimates of risk preferences.

In what follows we first apply Theorem 3.1, which does *not* assume independence between preferences and choice sets, to learn about the distribution of risk aversion (Section 5.1). In this case $\boldsymbol{\theta} = (\gamma_1, \gamma_2)$. We then apply Corollary 3.1, which assumes that choice set size is independent of preferences (Assumption 3.1), to learn about the distribution of choice set size (Section 5.2). In this case $\boldsymbol{\theta} = (\gamma_1, \gamma_2, \pi_3, \pi_4, \pi_5)$. In the text we present results for the population (all households). As indicated in Section 4.1, we also conduct our analysis separately for population subgroups based on observed characteristics $\mathbf{t}_i$. The subgroup results are reported in Section S3.3 of the Supplemental Material.

The sample moments that we use to implement equation (3.5) are

$$\bar{m}_{n,K,j}(\boldsymbol{\theta}) = \frac{1}{n} \sum_{i=1}^{n} m_{K,j}(d_i, \mu_i, \bar{p}_i; \boldsymbol{\theta})$$
$$= \frac{1}{n} \sum_{i=1}^{n} \left[ \mathbf{1}(d_i \in K, (\mu_i, \bar{p}_i) \in B_j) - P(D_\kappa^*(\mu_i, \bar{p}_i) \cap K \neq \varnothing; \boldsymbol{\gamma}) \mathbf{1}((\mu_i, \bar{p}_i) \in B_j) \right], \quad (5.1)$$

and similarly for equation (3.7). In equation (5.1), $P(D_\kappa^*(\mu_i, \bar{p}_i) \cap K \neq \varnothing; \boldsymbol{\gamma})$ is a function known up to $\boldsymbol{\theta}$ that can be evaluated using the Beta cumulative distribution function, and $B_j$, $j = 1, \ldots, J$, are "hypercubes" as defined in Andrews and Shi (2013, Example 1) [hereafter, AS] that are used to transform the conditional moment inequalities into unconditional ones.[26]

---

[26] We follow AS and transform $(\mu_i, \bar{p}_i)$ using the upper-triangular Cholesky decomposition of their sample covariance matrix, so that the transformed variables $(\tilde{\mu}_i, \tilde{p}_i)$ have a sample covariance matrix equal to the identity matrix. We then let the side lengths of the hypercubes $B_j$ be determined by the octiles of the distributions of $\tilde{\mu}_i$ and $\tilde{p}_i$, and we also include a hypercube containing all values of $(\mu_i, \bar{p}_i)$, so $J = 65$. Each hypercube contains between 660 and 2,901 households, except for one that contains all households.

We apply the method proposed by AS to compute bootstrap-based critical values, denoted $\hat{c}_{n,\cdot}(\cdot)$ below, that define a confidence set which covers each $\boldsymbol{\theta} \in \Theta_I$ with asymptotic probability $1 - \alpha$ uniformly over a large class of probability distributions $\mathcal{P}$.[27] We first compute a confidence set for $(\gamma_1, \gamma_2)$, from which we obtain a confidence set for $(\mathrm{E}(\nu_i), \mathrm{Var}(\nu_i))$ leveraging the fact that for $\nu_i \sim Beta(\gamma_1, \gamma_2)$ a unique pair $(\mathrm{E}(\nu_i), \mathrm{Var}(\nu_i))$ corresponds to each value of $(\gamma_1, \gamma_2)$. Formally, the AS confidence set is

$$CS = \{\boldsymbol{\theta} \in \Theta : T_n(\boldsymbol{\theta}) \leq \hat{c}_{n,1-\alpha+\xi}(\boldsymbol{\theta}) + \xi\}. \tag{5.2}$$

In equation (5.2), $\xi > 0$ is an arbitrarily small constant (AS suggest setting $\xi = 10^{-6}$) and $T_n(\boldsymbol{\theta})$ is a Kolmogorov-Smirnov test statistic that aggregates sample moment violations:

$$T_n(\boldsymbol{\theta}) = n \max_{j=1,\ldots,J; K \in \mathbb{K}} \max\{\bar{m}_{n,K,j}(\boldsymbol{\theta})/\hat{\sigma}_{n,K,j}(\boldsymbol{\theta}), 0\}^2,$$

where $\hat{\sigma}_{n,K,j}(\boldsymbol{\theta})$ is the sample analog of the population standard deviation of $m_{K,j}(d_i, \mu_i, \bar{p}_i; \boldsymbol{\theta})$ and the set $\mathbb{K}$ is determined using Theorem S1.1 in the Supplemental Material.[28]

We obtain confidence intervals for $\mathrm{E}(\nu_i)$, $\pi_3$, $\pi_4$, and $\pi_5$ using the method proposed by Kaido et al. (2019) [hereafter, KMS]. The first is a smooth function of $\boldsymbol{\theta} = (\gamma_1, \gamma_2)$ with a gradient that satisfies the assumptions in KMS, while the latter three are linear projections of $\boldsymbol{\theta} = (\gamma_1, \gamma_2, \pi_3, \pi_4, \pi_5)$. Let $f(\boldsymbol{\theta})$ denote any of the aforementioned functions of $\boldsymbol{\theta}$. The lower and upper bounds of the KMS confidence interval for $f(\boldsymbol{\theta})$ are obtained by solving

$$\min_{\boldsymbol{\theta} \in \Theta} / \max_{\boldsymbol{\theta} \in \Theta} f(\boldsymbol{\theta}) \text{ s.t. } \sqrt{n} \bar{m}_{n,K,j}(\boldsymbol{\theta})/\hat{\sigma}_{n,K,j}(\boldsymbol{\theta}) \leq \hat{c}_n^f(\boldsymbol{\theta}), \ j = 1, \ldots, J, \ K \in \mathbb{K},$$

where $\hat{c}_n^f(\boldsymbol{\theta})$ is a bootstrap-based critical level calibrated such that $f(\boldsymbol{\theta})$ is uniformly asymptotically covered with probability $1 - \alpha$ over a large class of probability distributions $\mathcal{P}$.[29] For $f(\boldsymbol{\theta}) = \mathrm{E}(\nu_i)$ the set $\mathbb{K}$ is determined using Theorem S1.1, while for $f(\boldsymbol{\theta}) \in \{\pi_3, \pi_4, \pi_5\}$ the set $\mathbb{K}$ is determined using Corollaries S1.2 and S1.3 in the Supplemental Material.

---

[27] See AS (Theorem 2) for a formal statement. The AS confidence set asymptotically exploits all the information in the conditional moments, in the sense that as the sample size grows to infinity the number of inequalities used for inference increases and the confidence set shrinks to $\Theta_I$.

[28] We note that there are values $\boldsymbol{\theta} \in \Theta$ for which $T_n(\boldsymbol{\theta}) = 0$. This implies that we fail to reject the hypothesis that our empirical model is correctly specified.

[29] See KMS (Theorem 3.1) for a formal statement. Although they do not asymptotically exploit all the information in the conditional moments because they are based on a fixed number of inequalities, the KMS confidence intervals (implemented on the same sample with the same inequalities and tuning parameters) are shorter than the confidence intervals obtained by projecting the AS confidence set.

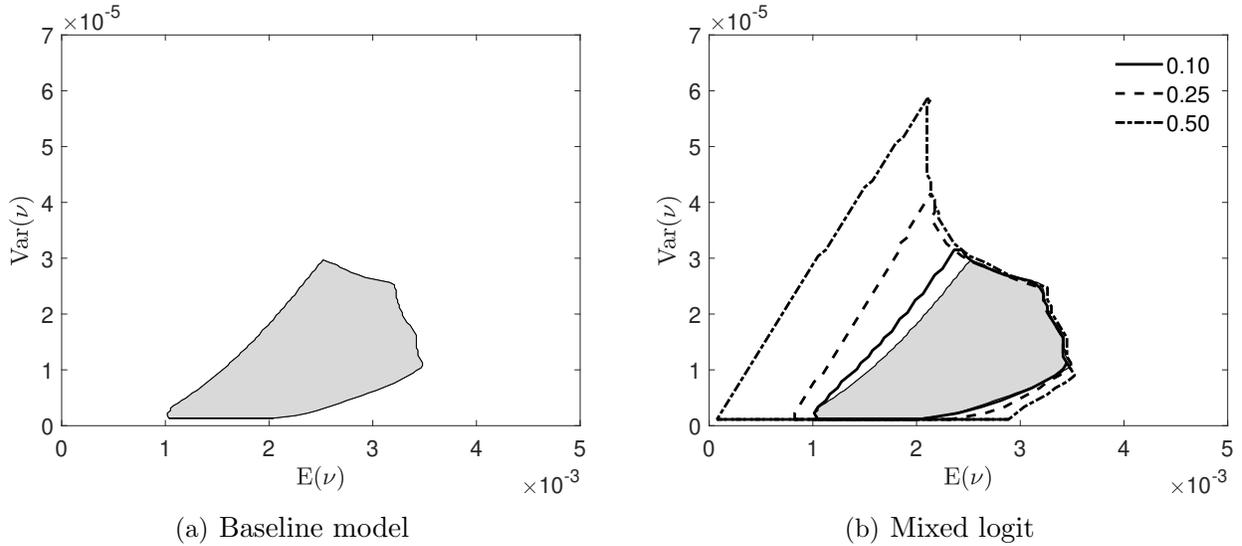

Figure 5.1: AS 95 percent confidence set for $(\mathrm{E}(\nu), \mathrm{Var}(\nu))$.

In Section S2 of the Supplemental Material we provide further details on implementation of the AS and KMS methods.[30] We refer to the original papers for a thorough discussion of the methods, and to Canay and Shaikh (2017) for a comprehensive presentation of the literature on inference in moment inequality models.

## 5.1 Risk Preferences

Panel (a) of Figure 5.1 depicts the AS 95 percent confidence set for $(\mathrm{E}(\nu_i), \mathrm{Var}(\nu_i))$ for all households.[31] In addition, Table 5.1 reports (i) the KMS 95 percent confidence interval for the mean of $\nu_i$ and (ii) 95 percent confidence intervals for the 25th and 75th percentiles of $\nu_i$ based on projections of the AS confidence set. For the mean, we report the actual confidence interval as well as the risk premium, for a lottery that yields a loss of $1000 with probability 10 percent, implied by each bound. For the percentiles, we report only the implied risk premia. Focusing on the lower bounds, the main takeaway is that the households' deductible choices can be explained by a distribution of absolute risk aversion that has a low mean, on the order of $10^{-3}$, and low variance, on the order of $10^{-6}$. Strikingly, the lower bound on the 25th percentile of $\nu_i$ corresponds to a risk premium of less than half a cent, implying that the data are consistent with at least a quarter of households being effectively risk neutral.

---

[30]Both the AS and KMS methods entail the selection of tuning parameters. We find that our results are robust to the choice of tuning parameters, as indicated by results available from the authors upon request.

[31]In Figure S3.2 in the Supplemental Material we also report a 95 percent confidence set for an outer region of admissible probability density functions of $\nu_i$.

Table 5.1: Distribution of Absolute Risk Aversion

|  | Mean | | Implied risk premium | | | | | |
|---|---|---|---|---|---|---|---|---|
|  | | | Mean | | 25th pctl. | | 75th pctl. | |
|  | LB | UB | LB | UB | LB | UB | LB | UB |
| Baseline model | 0.00105 | 0.00347 | $ 62 | $307 | $ 0 | $ 78 | $ 79 | $454 |
| UR | 0.00167 | 0.00170 | $115 | $117 | $ 86 | $ 88 | $142 | $145 |
| ASR | 0.00260 | 0.00264 | $211 | $216 | $ 40 | $ 43 | $333 | $340 |
| Cohen and Einav (2007) | 0.00310 | | $267 | | Not reported | | Not reported | |
| Barseghyan et al. (2013) | 0.00113 | | $ 68 | | Not reported | | Not reported | |

Notes: 95 percent confidence intervals for baseline, UR, and ASR models. LB = lower bound. UB = upper bound. Implied risk premia for a lottery that yields a loss of $1000 with probability 10 percent.

To provide context for these results, Table 5.1 also reports 95 percent confidence intervals for the mean, 25th percentile, and 75th percentile of $\nu_i$ obtained under two point-identified expected utility models that fully specify the choice set formation process. They are:

**Uniform random (UR):** Utility is given by equation (4.2). Choice sets are drawn uniformly at random from $\mathcal{D}$, conditional on $|C_i| = q$ for $q \geq \kappa$ and independent of $\nu_i$. Specifically, $\Pr(C_i = G||G| = q) = \binom{|\mathcal{D}|}{q}^{-1}$ for all $G \subseteq \mathcal{D}$, $|G| = q$, $q \geq \kappa$; and $C_i \perp \nu_i$.

**Alternative-specific random (ASR):** Utility is given by equation (4.2). Alternatives in $\mathcal{D}$ enter choice sets with alternative-specific probabilities, independent of one another and $\nu_i$, conditional on $|C_i| \geq \kappa$ (Manski 1977; Manzini and Mariotti 2014). Specifically, $\Pr(C_i = G||G| \geq \kappa) = \Pr(C_i = G)/(1 - \sum_{G \subseteq \mathcal{D}:|G|<\kappa} \Pr(C_i = G))$ for all $G \subseteq \mathcal{D}$, where $\Pr(C_i = G) = \prod_{c \in G} \varphi(c) \prod_{c \in \mathcal{D} \setminus G}(1 - \varphi(c))$ and $\varphi(c) = \Pr(c \in C_i)$; and $C_i \perp \nu_i$.

UR and ASR are "reduced form" models that can capture a wide range of choice set formation processes. For example, UR is consistent with a simultaneous search process with a uniform prior (cf. Stigler 1961),[32] and ASR may describe an advertising process in which alternatives are marketed with different intensities in independent, non-targeted campaigns. With dependence between $\varphi(c)$ and $\nu_i$, ASR can capture an even wider range of choice set formation processes, including, for instance, a sequential search process with free recall (e.g., Weitzman 1979) or an advertising process with correlated, targeted campaigns.

For additional context, Table 5.1 includes point estimates for the mean of $\nu_i$ reported by Cohen and Einav (2007) and Barseghyan et al. (2013) for their CARA models. Cohen and Einav (2007) estimate the distribution of $\nu_i$ in a parametric expected utility model using data on deductible choices in Israeli auto insurance. Barseghyan et al. (2013) estimate the distributions of $\nu_i$ and probability distortions in a parametric rank-dependent expected utility model using data on deductible choices in U.S. auto and home insurance.

---

[32]With a uniform prior, the simultaneous search problem reduces to choosing the optimal number of alternatives to search and, given this number, randomly choosing the alternatives to be searched.

The main takeaway is that the baseline lower bounds are substantially smaller than the lower bounds obtained under UR and ASR and the point estimate reported by Cohen and Einav (2007).[33] This suggests that if one properly allows for heterogeneity in choice sets, the data can be explained by expected utility theory with substantially lower levels of risk aversion than many familiar models—including some that allow for choice set heterogeneity but perhaps misspecify the choice set formation process—would imply. A second takeaway comes from results in Barseghyan et al. (2013). Their point estimate for the mean of $\nu_i$ is only slightly larger than the baseline lower bound ($68 versus $62 in terms of implied risk premium). However, because they allow for probability distortions, $\nu_i$ neither fully captures the level of risk aversion nor solely drives risk-averse behavior in their model. Taking into account their point estimate for probability distortions, the implied risk premium is $91. This suggests that failing to allow for heterogeneity in choice sets may affect inferences not only about the level of risk aversion, but also about the sources of risk-averse behavior.

### 5.1.1 Mixed Logit with Unobserved Heterogeneity in Choice Sets

We also compute the AS 95 percent confidence set for $(\mathrm{E}(\nu_i), \mathrm{Var}(\nu_i))$ for a mixed logit specification $U_i(c) = \omega(\mathbf{x}_{ic}, \nu_i) + \epsilon_{ic}$, where $\omega(\mathbf{x}_{ic}, \nu_i)$ is the certainty equivalent of the right hand side of equation (4.2), $\nu_i$ is distributed per Assumption 4.3(I), and $\epsilon_{ic}$ is an i.i.d. disturbance that follows a Type 1 Extreme Value distribution with scale parameter $\sigma$ and is independent of $(\mathbf{x}_{ic}, \nu_i)$. We define utility in terms of its certainty equivalent so that $\epsilon_{ic}$ is measured in dollars (which allows for a clear economic interpretation). Panel (b) of Figure 5.1 depicts the confidence set for three values of $\sigma$ chosen so that the standard deviation of $\epsilon_{ic}$ is equal to 10 percent, 25 percent, and 50 percent of the average price difference among adjacent deductibles in $\mathcal{D}$. (At zero percent, of course, the mixed logit specification reduces to the baseline model.) As the "noise factor" increases, the confidence set expands mainly to the "northwest," admitting higher values of $\mathrm{Var}(\nu_i)$ and lower values of $\mathrm{E}(\nu_i)$. Focusing on the latter, the projection of the confidence set on $\mathrm{E}(\nu_i)$ is essentially unchanged at a noise factor of 10 percent. At 25 percent the lower bound is smaller but still informative. By 50 percent, however, the confidence set effectively admits $(\mathrm{E}(\nu_i), \mathrm{Var}(\nu_i)) = (0, 0)$ (i.e., all households are risk neutral) and overall is quite large. The bottom line is that the confidence set remains informative at reasonable levels of noise. Not surprisingly, however, as the magnitude of the noise approaches that of the variation in observable covariates, the data loses much of its informational content about households' preferences.

---

[33]Moreover, the UR estimates lie outside $\Theta_I$ and hence this model is rejected in our application; see, e.g., the bounds on the 25th percentile reported in Table 5.1 and Figure S3.2 in the Supplemental Material.

Table 5.2: Distribution of Choice Set Size

|  | $\pi_5$ (full) | | $\pi_4$ (full-1) | | $\pi_3$ (full-2) | |
| --- | --- | --- | --- | --- | --- | --- |
|  | LB | UB | LB | UB | LB | UB |
| All households | 0.00 | 0.24 | 0.00 | 0.89 | 0.11 | 1.00 |

Notes: KMS 95 percent confidence intervals. LB = lower bound. UB = upper bound.

## 5.2 Choice Set Size

Table 5.2 reports KMS 95 percent confidence intervals for $\pi_5$, $\pi_4$, and $\pi_3$. The interesting quantities are the upper bounds on $\pi_5$ and $\pi_4$. The former is the maximum fraction of households whose deductible choices can be rationalized with full-size choice sets, while the latter is the maximum fraction of households whose deductible choices can be rationalized with full-1 choice sets.[34] The main result is that a large majority of households require limited choice sets (full-1 or full-2) to explain their deductible choices. Specifically, we find that at least 76 percent of households require limited choice sets, including at least 11 percent who require full-2 choice sets. In the remainder of this section we discuss two drivers of this result: suboptimal choices and violations of the law of demand.[35]

### 5.2.1 Suboptimal Choices

The first driver is the existence and frequency of suboptimal choices. In total, 16.7 percent of households in our sample choose a deductible that is suboptimal (i.e., not first best in $\mathcal{D}$) under our empirical model at all $\nu \in [0, 0.03]$. The vast majority of these households choose $200, which is a suboptimal alternative under the model for virtually every household in our sample.[36] In particular, $200 is dominated by $100 or $250, depending on $\mu$. Suboptimal alternatives, sometimes called dominated alternatives, are not uncommon in discrete choice settings, including insurance settings (see, e.g., Handel 2013; Bhargava et al. 2017).

To see why $200 is a suboptimal alternative under the model, consider a risk-neutral household with claim probability $\mu$. The household prefers $200 to $100 if and only if $\mu < \frac{p_{100}-p_{200}}{200-100}$, and prefers $200 to $250 if and only if $\mu > \frac{p_{200}-p_{250}}{250-200}$. In our data $p_{100} - p_{200} = p_{200} - p_{250}$ for all households. For the risk-neutral household, therefore, at most one of the foregoing inequalities holds and thus $200 is dominated by $100 or $250, depending on the value of $\mu$. A similar logic applies for risk-averse households with reasonable levels of risk aversion—under our model or any other model in which lotteries are evaluated by expectations over functions of final wealth (see Barseghyan et al. 2016)—and indeed for virtually every household in our sample $200 is suboptimal at all $\nu \in [0, 0.03]$.[37]

---

[34]With $\kappa = 3$, the lower bounds on $\pi_5$ and $\pi_4$ are zero, the lower bound on $\pi_3$ is one minus the upper bound on $\pi_4$, and the upper bound on $\pi_3$ is one.

[35]In other applications, additional or different data features may reveal the presence of heterogeneous choice sets. One example is zero shares for alternatives that are not suboptimal.

[36]The remainder of these households choose $1000 or $500 when $250 is optimal.

Yet 15.2 percent of households in our sample choose $200. At the same time, only 1.1 percent choose $100 and 13.7 percent choose $250. Hence, the combined demand for $100 and $250 is less than the demand for $200. This pattern is even more pronounced within certain subgroups, including households with old principal drivers and households with high insurance scores; see Table 4.1.

Heterogeneous choice sets can readily explain these choice patterns. In our model all that is required to rationalize a household's choice of $200 is the absence of $100 or $250, as the case may be, from the household's choice set. Moreover, all that is required to explain $\Pr(d = 100|\mathbf{x}) + \Pr(d = 250|\mathbf{x}) < \Pr(d = 200|\mathbf{x})$ is a choice set distribution in which the frequencies of $100 and $250 are sufficiently less than the frequency of $200.[38]

With full-size choice sets, however, our model cannot explain these choice patterns. The reason is that, with full-size choice sets, our model satisfies a rank order property which implies $\Pr(d = a|\mathbf{x}) + \Pr(d = b|\mathbf{x}) > \Pr(d = c|\mathbf{x})$ when $c$ is dominated by $a$ or $b$. Indeed, *any* model that satisfies an analogous rank order property is incapable of explaining the relative frequency of $200 in the distribution of observed deductible choices. This includes, inter alia, the conditional logit model (McFadden 1974), the mixed logit model (McFadden 1974; McFadden and Train 2000), the multinomial probit model (e.g., Hausman and Wise 1978), and semiparametric models such as the one in Manski (1975).[39] For a more complete discussion, see Section S3.5 of the Supplemental Material.

### 5.2.2 Law of Demand

Violations of the law of demand are also driving our main result on choice sets. With full-size choice sets, households' demand for high deductibles should increase as base price increases and should decrease as claim risk increases. If follows that, with full-size choice sets, we should observe for all $K \in \overleftarrow{K} \equiv \big\{\{\$1000\}, \{\$1000, 500\}, \{\$1000, \$500, \$250\}\big\}$,

$$\Pr(d \in K|\mu, \bar{p}) > \Pr(d \in K|\mu', \bar{p}') \text{ if } \mu < \mu' \text{ and } \bar{p} > \bar{p}'. \tag{5.3}$$

---

[37]Evaluating equation (4.2) for all 111,890 households over a fine grid of $\nu$, we find that the $200 deductible is optimal in 0.001 percent of cases, all of which entail $\nu \geqslant 0.0115$.

[38]That said, not all heterogeneous choice set formation processes can explain these choice patterns. For instance, UR cannot but ASR can; see Claim S3.1 in the Supplemental Material.

[39]This also includes the model in Barseghyan et al. (2016), which explains why they find that 13.0 percent of the households in their data cannot be rationalized by their model.

In our data, however, we observe multiple violations. In particular, when we compare all pairs of hypercubes, where one hypercube has a lower average $\mu$ and a higher average $\bar{p}$ than the other, over all subsets $K \in \overleftarrow{K}$, we find 61 violations (3 percent) of equation (5.3).[40]

The requirement in equation (5.3) holds generically for models in which $\frac{\partial[U(c)-U(c')]}{\partial \bar{p}} < 0$ and $\frac{\partial[U(c)-U(c')]}{\partial \mu} > 0$ for all $c, c' \in \mathcal{D}$, $c < c'$. Given the assumptions of our empirical model, the law of demand implies a second, stronger requirement (cf. Barseghyan et al. 2020). Observe that for any $\mathbf{x} = (\mu, \bar{p})$ and any subset $K \subset \mathcal{D}$ of adjacent deductibles, there exists an interval $\mathcal{S}_K(\mathbf{x}) \subseteq \mathcal{V}$ such that $d^*(\mathcal{D}, \mathbf{x}, \nu) \in K$ if $\nu \in \mathcal{S}_K(\mathbf{x})$ and $d^*(\mathcal{D}, \mathbf{x}, \nu) \in \mathcal{D} \backslash K$ if $\nu \in \mathcal{V} \backslash \mathcal{S}_K(\mathbf{x})$, where $d^*(\mathcal{D}, \mathbf{x}, \nu)$ denotes the model implied-optimal choice when the choice set has full size. It follows that, with full-size choice sets,

$$\Pr(d \in K | \mathbf{x}) \leqslant \Pr(d \in K' | \mathbf{x}') \text{ if } \mathcal{S}_K(\mathbf{x}) \subset \mathcal{S}_{K'}(\mathbf{x}') \tag{5.4}$$

for any subsets $K, K' \subset \mathcal{D}$ of adjacent deductibles and any $\mathbf{x} \neq \mathbf{x}'$. In our data, however, we observe numerous violations of equation (5.4). In particular, when we compare all pairs of hypercubes, where $\mathbf{x}$ denotes the average $(\mu, \bar{p})$ in one hypercube and $\mathbf{x}'$ denotes the average $(\mu', \bar{p}')$ in the other, over all subsets $K, K' \subset \mathcal{D}$ of adjacent deductibles where each subset contains either one, two, or three deductibles, we find 44,847 instances (15 percent) in which $\mathcal{S}_K(\mathbf{x}) \subset \mathcal{S}_{K'}(\mathbf{x}')$ but $\Pr(d \in K | \mathbf{x}) > \Pr(d \in K' | \mathbf{x}')$.[41]

We conclude by highlighting how equation (5.4) relates to the characterization of $\Theta_I$ in Corollary 3.1. Consider whether any parameter vector with $\pi_{|\mathcal{D}|} = 1$ belongs to $\Theta_I$. At that value $D^*_\kappa(\mathbf{x}, \nu) = \{d^*(\mathcal{D}, \mathbf{x}, \nu)\}$, a singleton, and hence the inequality in equation (3.7), evaluated at any subset $K \subset \mathcal{D}$ of adjacent deductibles and its complement $\mathcal{D} \backslash K$, implies

$$\Pr(d \in K | \mathbf{x}) = \sum_{c \in K} \int \mathbf{1}(d^*(\mathcal{D}, \mathbf{x}, \tau) = c) dP(\tau; \boldsymbol{\gamma}) = \int_{\mathcal{S}_K(\mathbf{x})} dP(\tau; \boldsymbol{\gamma}),$$

which in turn implies equation (5.4). Thus, a violation of equation (5.4) implies that no parameter vector with $\pi_{|\mathcal{D}|} = 1$ belongs to $\Theta_I$. A similar logic applies to the choice probabilities of suboptimal alternatives. In general, our method—through the inequalities in equation (3.7)—takes into account all restrictions implied by the data and the economic model, while accounting for finite sample uncertainty.

---

[40]We do not count violations where $K$ contains a suboptimal alternative under the model given the average $(\mu, \bar{p})$ in either hypercube.

[41]Again, we do not count violations where $K$ or $K'$ contains a suboptimal alternative under the model given $\mathbf{x}$ or $\mathbf{x}'$, respectively.

# 6  Computational Tractability of Our Method

As we note in Section 3.4, there are two computational challenges in applying Theorem 3.1. The first is that, given any $\kappa \geq 2$, the number of inequalities in equation (3.5) grows superlinearly with the number of feasible alternatives (i.e., cardinality of $\mathcal{D}$). The second challenge is computing the inequalities, the difficulty of which increases with the dimensionality of unobserved heterogeneity (i.e., dimension of $\boldsymbol{\nu}_i$). In our empirical application these challenges are mitigated by the fact that $|\mathcal{D}| = 5$ and $\boldsymbol{\nu}_i \in \mathbb{R}$. Other applications, however, may feature larger feasible sets or higher-dimensional unobserved heterogeneity.

In this section we provide simulation evidence on the computational tractability of our method when the cardinality of $\mathcal{D}$ or the dimension of $\boldsymbol{\nu}$ is large. Our simulations also illustrate how the informational content of the confidence set is impacted by the size of $\kappa$ (the minimum choice set size) relative to $|\mathcal{D}|$ and by the dependence between the agents' choice sets, on the one hand, and their preferences or observables, on the other.

## 6.1  Data Generating Processes

### 6.1.1  Large Feasible Set

A first set of simulations probes the computational tractability of our method when the feasible set is large. Specifically, we assume $\mathcal{D} = \{\$10, \$20, \ldots, \$1010\}$, so that $|\mathcal{D}| = 101$. These simulations otherwise parallel our empirical application and maintain the assumptions of our baseline empirical model in Section 4.1, except as follows.

For each household $i$, we fix the probability of experiencing a claim equal to $\mu_i = 0.10$ and we set prices to be proportional to the amount of coverage (hence, there are no suboptimal alternatives in the simulations): $p_{ic} = g(c)\bar{p}_i + \zeta$, where $g(c) = [(\$1010 - c)/\$1010] + \$1$, $\bar{p}_i$ is the household's base price, and $\zeta > 0$. We assume that the household draws $\bar{p}_i$ from a discrete Uniform distribution with support $\{\$10, \$20, ..., \$1000\}$. (The value of $\zeta$ is immaterial because, with CARA utility, $\zeta$ cancels out in utility comparisons.) We further assume that the household draws its coefficient of absolute risk aversion $\nu_i$ from a Uniform distribution with support $[0, 0.01]$—equivalently, $\nu_i \sim 0.01 \times Beta(1, 1)$.

We consider two formation processes for households' choice sets $C_i \subseteq \mathcal{D}$.

**FP1:** Full-size choice sets: $\Pr(C_i = \mathcal{D}) = 1$.

**FP2:** Limited choice sets: $\Pr(|C_i| = q) = 1$ where $q = 30$ or $q = 70$. For each $q$, we run simulations with three forms of dependence:

**No correlation:** $C_i$ is drawn uniformly at random from $\mathcal{D}$, independent of $\nu_i$ and $\bar{p}_i$ (hence, following the UR model).

**Correlation with $\nu$:** If $\nu_i < 0.005$ (hence, below the median), $C_i$ comprises the $q$ riskiest alternatives; if $\nu_i \geqslant 0.005$, $C_i$ comprises the $q$ safest alternatives.

**Correlation with $\bar{p}$:** $C_i$ comprises $q$ adjacent alternatives, the index of the first of which increases linearly (subject to rounding) from 1 to $|\mathcal{D}| - q + 1$ as $\bar{p}_i$ increases from \$10 to \$1000.

In obtaining $\Theta_I$, we assume only that $\Pr(|C_i| \geqslant \kappa) = 1$ where $\kappa = 10, 30, 50, 70, 90$ in the case of FP1 and $\kappa = q$ in the case of FP2. We use a sample size of $n = 100,000$ households.

Given this structure, all possible realizations of the set $D_\kappa^*(\mathbf{x}_i, \boldsymbol{\nu}_i; \boldsymbol{\delta})$ are given by adjacent elements of $\mathcal{D}$, as $\{c_j, c_{j+1}, \ldots, c_{j+|\mathcal{D}|-\kappa}\}$, for $j = 1, \ldots, \kappa$; see Claim S1.1 in the Supplemental Material. Accordingly, we can leverage the results set forth in Theorem S1.1 and Corollary S1.1 in the Supplemental Material to reduce the number of inequalities that are needed to obtain $\Theta_I$. Specifically, the number of inequalities needed here is $2(\kappa - 1)$.

### 6.1.2 High-dimensional Unobserved Heterogeneity

A second set of simulations probes the computational tractability of our method when unobserved heterogeneity is high-dimensional. These simulations are based on a mixed logit model, $U_i(c) = \omega(\mathbf{x}_{ic}, \nu_i) + \epsilon_{ic}$, as in Section 5.1.1. We maintain the assumptions of the model in Section 5.1.1, except that (i) we assume $\nu_i \sim 0.01 \times Beta(1, 1)$ (as in the first set of simulations) and (ii) for all $c \in \mathcal{D}$, we set the standard deviation of $\epsilon_{ic}$ equal to 10 percent of the average price difference among adjacent alternatives in $\mathcal{D}$. We also maintain the assumptions on $\mu_i$ and $p_{ic}$ in Section 6.1.1. We consider three feasible sets, with $|\mathcal{D}| = 7, 12, 17$, and we assume that choice sets are formed according to FP1. In obtaining $\Theta_I$, we assume only that $\Pr(|C_i| \geqslant \kappa) = 1$ where $\kappa = 5$. We use a sample size of $n = 1,000,000$.

Our choice of $|\mathcal{D}| = 17$ is motivated by the recent paper of Abaluck and Adams (2020), who estimate a model of limited consideration using data on Medicare Part D prescription drug plan choices. (We also consider $|\mathcal{D}| = 7, 12$ to illustrate the decrease in informational content as $|\mathcal{D}|$ becomes larger relative to $\kappa$.) Abaluck and Adams (2020) restrict the feasible set to 17 plans to manage the computational burden of estimating their model's alternative-specific attention parameters. They find this restriction necessary even though they make assumptions (discussed below in Section 7) to achieve point identification and their approach involves neither moment inequalities nor random coefficients.

As we note in Section 3.4, the predicate conditions in Theorem S1.1 and Corollary S1.1 do not hold for this model, and hence we cannot apply these results to reduce the number of inequalities characterizing $\Theta_I$ (though we can and do apply Theorem S1.2 to simplify the evaluation of the inequalities). Computationally, therefore, the problem here is harder than in the first set of simulations. Indeed, computational difficulties in conditional and mixed logit

Table 6.1: Computational Performance of Simulations

| Choice sets | $|\mathcal{D}|$ | $\kappa$ | Correlation | Number of inequalities | Evaluation points | Execution time per evaluation point (seconds) | Total execution time (hours) |
|---|---|---|---|---|---|---|---|
| *Panel A. Large $\mathcal{D}$ Simulations* | | | | | | | |
| FP1 | 101 | 10 | None | 1,818 | 32,936 | 0.005 | 0.07 |
| | | 30 | | 5,858 | 18,593 | 0.017 | 0.12 |
| | | 50 | | 9,898 | 11,177 | 0.030 | 0.13 |
| | | 70 | | 13,938 | 6,640 | 0.044 | 0.12 |
| | | 90 | | 17,978 | 3,667 | 0.058 | 0.10 |
| FP2 ($q = 30$) | 101 | 30 | None | 5,858 | 17,784 | 0.017 | 0.12 |
| | | | With $\nu$ | | 47,030 | 0.017 | 0.25 |
| | | | With $\bar{p}$ | | 72,416 | 0.012 | 0.28 |
| FP2 ($q = 70$) | 101 | 70 | None | 13,938 | 6,482 | 0.044 | 0.12 |
| | | | With $\nu$ | | 6,809 | 0.044 | 0.12 |
| | | | With $\bar{p}$ | | 7,872 | 0.040 | 0.13 |
| *Panel B. High-dimension $\nu$ Simulations* | | | | | | | |
| FP1 | 7 | 5 | None | 9,898 | 8,958 | 0.03 | 0.12 |
| | 12 | | | 80,093 | 24,880 | 0.28 | 1.98 |
| | 17 | | | 324,513 | 37,185 | 1.73 | 19.41 |

Notes: Total execution time includes pre-processing time.

models with unobserved heterogeneity in choice sets are pervasive even in fully parametric, point-identified models due to the need to enumerate all subsets of the feasible set. As doing so is often too costly in practice, it is common to resort to simulating choice sets (see, e.g. Goeree 2008). This approach, however, requires numerous strong assumptions (including, in particular, independence between preferences and choice sets) to obtain consistent estimates. Under our approach, if necessary, one can resort to computing an outer region for $\Theta_I$ that is based on fewer, and possibly cheaper to evaluate, inequalities. This outer region is guaranteed to contain all parameter values in $\Theta_I$, though it may also include some that lie outside $\Theta_I$.[42]

## 6.2 Number of Inequalities and Computation Time

Table 6.1 reports the number of inequalities needed and the total execution time (in hours) spent to compute the AS 95 percent confidence set for $(\mathrm{E}(\nu_i), \mathrm{Var}(\nu_i))$. It also reports the number of evaluation points used (through an adaptive grid construction that we built) and the execution time (in seconds) per evaluation point. Panel A covers the large $\mathcal{D}$ simulations, while Panel B covers the high-dimension $\nu$ simulations. To assess the tractability of our method from the perspective of a researcher who has access to run-of-the-mill computing power, we run all our simulations on a single Dell Precision Tower 7910 (Dual CPU E5-2687W v4 @ 3.00GHz with 128GB RAM). Of course, with more a powerful workstation, or with access to computer clusters or cloud computing, one can handle larger problems (more inequalities), reduce execution times, or both.

---

[42]It is not uncommon in the partial identification literature to resort to outer regions to preserve computational tractability (see, e.g., Ciliberto and Tamer 2009).

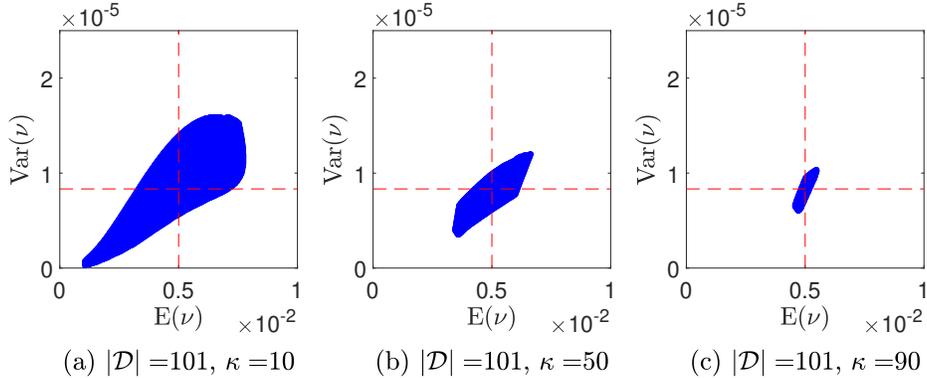

(a) $|\mathcal{D}| =101$, $\kappa =10$  (b) $|\mathcal{D}| =101$, $\kappa =50$  (c) $|\mathcal{D}| =101$, $\kappa =90$

Figure 6.1: AS 95 percent confidence set – large $\mathcal{D}$ simulations, FP1.

Panel A illustrates the power of Theorem S1.1 and Corollary S1.1. With $|\mathcal{D}| = 101$ and $\kappa$ ranging as high as 90, the number of inequalities never exceeds 18,000 and the total execution time never exceeds 20 minutes. Counterintuitively, the second fastest execution time is achieved in the case where $|\mathcal{D}| = 101$ and $\kappa = 90$, despite the fact that this case has the largest number of inequalities and hence the slowest execution time per evaluation point. The reason is our adaptive grid: the confidence set is the smallest in this case (see Section 6.3 below), and thus the number of evaluation points is also the smallest.

Panel B illustrates the tractability of our method even when Theorem S1.1 and Corollary S1.1 are not applicable. Even with more than 300,000 inequalities, we can test whether a given $\boldsymbol{\theta}$ belongs to $\Theta_I$ in less than two seconds, and we recover the full confidence set in less than 20 hours. The only computational challenge we encounter is inadequate memory to utilize all 24 CPU cores when the number of inequalities becomes very large.[43] But even then, we can check a very large number of candidate values $\boldsymbol{\theta}$ in a reasonable amount of time, thus demonstrating that our method can be employed in a wide range of applications.

## 6.3 Simulation Results

Figures 6.1 and 6.2 depict the AS 95 percent confidence sets for $(\mathrm{E}(\nu_i), \mathrm{Var}(\nu_i))$ for the large $\mathcal{D}$ simulations. Figure 6.3 depicts the confidence set for the high-dimension $\boldsymbol{\nu}$ simulations. The axes of the figures represent the parameter space, and hence the depictions provide a sense of the absolute and relative informativeness of each confidence set. As the figures show, each confidence set includes the data generating values of the distribution of $\nu_i$ ($0.01 \times Beta(1,1)$).

---

[43]We utilize all 24 CPU cores in every simulation but one: the mixed logit simulation with $|\mathcal{D}| = 17$, which entails checking more than 300,000 inequalities, utilizes 12 CPU cores.

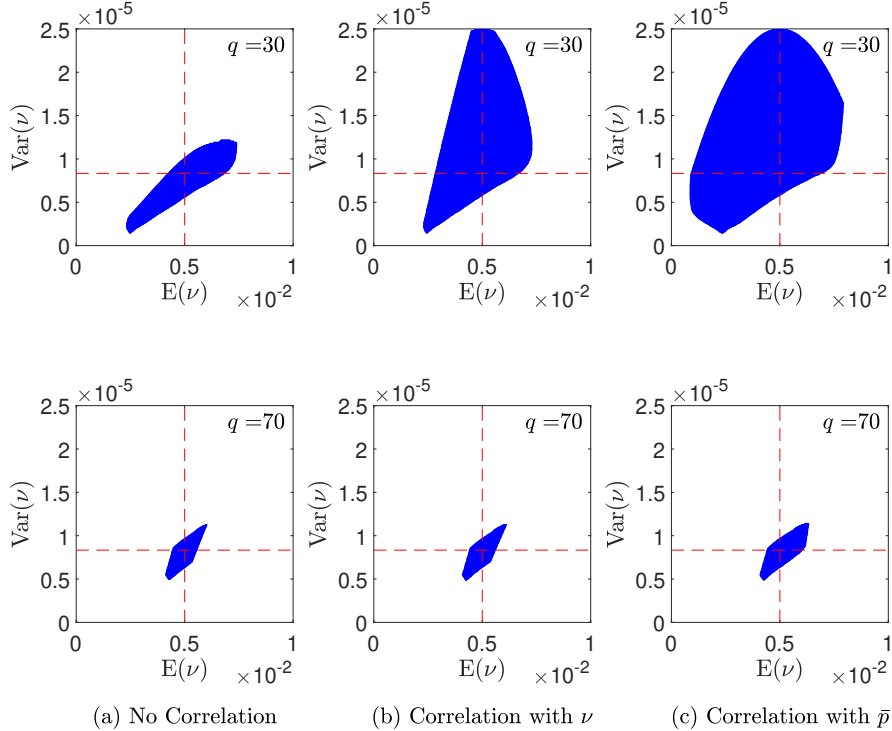

(a) No Correlation  (b) Correlation with $\nu$  (c) Correlation with $\bar{p}$

Figure 6.2: AS 95 percent confidence set – large $\mathcal{D}$ simulations, FP2.

We first comment on how the informativeness of the confidence set varies with $\kappa$, as illustrated by Figure 6.1. Note that the inequalities used in a case with a smaller $\kappa$ are a subset of those used in a case with a larger $\kappa$.[44] As a consequence, the sharp identification region $\Theta_I$ shrinks as $\kappa$ increases. However, the critical values used to account for statistical uncertainty may increase as more inequalities are used, thereby yielding an ambiguous effect on the confidence set. Nonetheless, in our simulations the confidence sets are effectively subsets of each other, and shrink substantially as $\kappa$ increases. This illustrates the important role of $\kappa$, and the fact that if $\kappa$ is very small relative to $|\mathcal{D}|$, the confidence set is substantially less informative than when $\kappa$ is closer to $|\mathcal{D}|$.

We next comment on how different forms of dependence between $C_i$, on the one hand, and $\nu_i$ or $\bar{p}_i$, on the other, impact the informativeness of the confidence set, as illustrated by Figure 6.2. Panel (a) shows that when $C_i$ is independent of $\nu_i$ and $\bar{p}_i$, the confidence set is quite informative, even when $C_i$ is relatively small ($q = 30$). Panels (b) and (c) illustrate that the confidence set expands—mainly to the "north," admitting higher values of $\mathrm{Var}(\nu_i)$—when $C_i$ is correlated with $\nu_i$ or $\bar{p}_i$, respectively, with larger expansions occurring when $C_i$ is smaller. We conjecture that this happens because the households' choices are limited by the restrictions that such dependence imposes on their choice sets. When $C_i$ is correlated with $\nu_i$, for instance, the choice sets of households with high risk aversion do not include high deductibles, and hence those households do not respond to increases in $\bar{p}_i$ beyond some level, because they cannot switch to cheaper alternatives (i.e., higher deductibles). (By comparison, when $C_i$ is independent of $\nu_i$, all households have a positive probability of drawing a choice set that includes high deductibles.) Put differently, there is not enough "price elasticity" for the econometrician to (relatively precisely) trace out the distribution of $\nu_i$. Moreover, this problem becomes more severe the smaller are the households' choice sets. A similar logic applies when $C_i$ is correlated with $\bar{p}_i$.

---

[44]To conserve space, Figure 6.1 depicts the confidence set for $\kappa = 10, 50, 90$. The plots for $\kappa = 30, 70$, which exhibit the same nesting pattern, are available from the authors upon request.

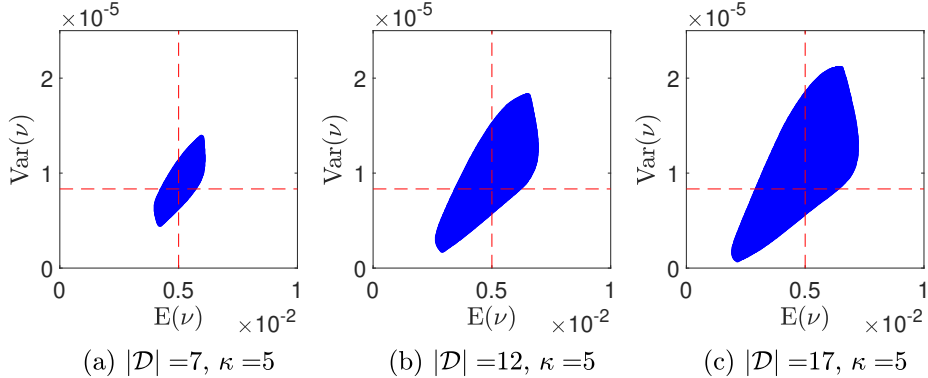

Figure 6.3: AS 95 percent confidence set – high-dimension $\boldsymbol{\nu}$ simulations.

Finally, we comment briefly on Figure 6.3. Again, the larger is $|\mathcal{D}|$ relative to $\kappa$, the larger is the confidence set. Nevertheless, despite the substantial amount of unobserved heterogeneity captured by $\boldsymbol{\nu} \in \mathbb{R}^{18}$, the confidence set remains informative.

# 7 Discussion

In what follows we provide an overview of the assumptions made in the econometrics and applied literatures on discrete choice analysis to grapple with the identification problem created by unobserved heterogeneity in choice sets.[45] We describe four prominent approaches

---

[45]Many important papers in the theory literature—including papers on revealed preference analysis under limited attention, limited consideration, and other forms of bounded rationality that manifest in unobserved heterogeneity in choice sets—also grapple with the identification problem (e.g., Masatlioglu et al. 2012; Manzini and Mariotti 2014; Caplin and Dean 2015; Lleras et al. 2017; Cattaneo et al. 2020). However, these papers generally assume rich datasets—e.g., observed choices from every possible subset of the feasible

and provide examples of recent papers that take each approach. We do not provide a comprehensive review of the literature, which is vast and spans a diverse array of fields in economics. However, our overview of the landscape enables us to situate our approach within the literature and provides context for our contributions, which we recap at the end.

The most common approach in the discrete choice literature to the identification problem created by unobserved choice sets is to assume that all choice sets comprise the feasible set or a known subset of the feasible set (Swait 2001, p. 643; Honka et al. 2017, p. 615). This is the approach taken by, for example, Berry et al. (1995) in estimating demand curves from data on U.S. auto sales; Cohen and Einav (2007) in estimating risk preferences from data on deductible choices in Israeli auto insurance; and Chiappori et al. (2019) in estimating risk preferences from betting data on U.S. horse races. We also take this approach in prior work on estimating risk preferences from data on deductible choices in U.S. auto and home insurance (Barseghyan et al. 2011, 2013, 2016).

Papers that allow for heterogeneity in choice sets take three basic approaches to identification. The first is to rely on auxiliary information about the composition or distribution of choice sets. For instance, Draganska and Klapper (2011), who study ground coffee sales, use survey data on brand awareness; De los Santos et al. (2012), who study online book purchases, use survey data on web browsing; Conlon and Mortimer (2013), who study vending machine sales, utilize periodic inventory snapshots; Honka and Chintagunta (2017), who study auto insurance purchases, use survey data on price quotes; and Honka et al. (2017), who study bank account openings, use survey data on brand awareness and search activity.[46]

The second approach is to rely on two-way exclusion restrictions—i.e., assume that certain variables impact choice sets but not preferences and vice versa. For example, Goeree (2008) assumes that media advertising affects the set of computers of which a consumer is aware (and hence her choice set) but not her preferences over computers, while computer attributes affect her preferences but not her choice set; Gaynor et al. (2016) assume that waiting times and mortality rates directly impact a patient's preferences over hospitals but not her referring physician's preferences (which determine her choice set), while distance to hospital and hospital fixed effects directly impact her referring physician's preferences (and hence her choice set) but not her preferences; and Hortaçsu et al. (2017) assume that a retail electricity customer's decision to consider alternatives to her retailer is a function of her last period

---

set—that often are not available in applied work, especially outside of the laboratory. A notable exception is Dardanoni et al. (2020), which assumes that only a single cross-section of aggregate choice shares is observed.

[46]For earlier papers, see, e.g., Roberts and Lattin (1991) and Ben-Akiva and Boccara (1995).

retailer (e.g., a bad customer service experience) but not her next period retailer, while her choice of retailer is a function of her next period retailer but not her last period retailer.[47]

The last approach is to rely primarily on restrictions to the choice set formation process. Five recent papers that exemplify this approach are Abaluck and Adams (2020), Barseghyan et al. (2020), Crawford et al. (2020), Lu (2019), and Cattaneo et al. (2020).[48]

Abaluck and Adams (2020) consider two models of choice set formation: a variant of the ASR model described above and a "default specific" model in which each agent's choice set comprises either a single, default alternative or the entire feasible set. They show that the restrictions imposed on choice probabilities by these models are sufficient for point identification of preferences and choice set probabilities due to induced asymmetries in cross-attribute responses ('Slutsky asymmetries'), assuming that choice sets and preferences are independent conditional on observables and that every alternative has a continuous attribute with large support that is additively separable in utility and shifts choice set probabilities.

Barseghyan et al. (2020) study point identification of discrete choice models with unobserved heterogeneity in preferences and choice sets. They establish conditions for point identification of the preference distribution under generic choice set formation processes. They also illustrate the tradeoff between the common exclusion restrictions and the restrictions on choice set formation required for semi-nonparametric point identification.

Crawford et al. (2020) show that with panel data (or group-homogeneous cross-section data) and preferences in the logit family, point identification of preferences is possible, without any exclusion restrictions, under the assumption that choice sets and preferences are independent conditional on observables and with restrictions on how choice sets evolve over time. These restrictions enable the construction of proper subsets of agents' true choice sets ('sufficient sets') that can be utilized to estimate the preference model.

Lu (2019) provides conditions for both partial and point identification of a random coefficient logit model. He assumes that each agent's unobserved choice set is bounded by two observed sets, her largest possible choice set (e.g., the feasible set) and her smallest possible choice set (containing a default alternative and at least one other alternative). He shows that availability of these data, together with the assumption that agents' choices obey Sen's property $\alpha$, yields moment inequalities on the choice probabilities, which he uses to obtain outer regions on the model's preference parameters.

---

[47]Heiss et al. (2019) similarly assume that a Medicare Part D insured's decision to consider alternatives to her existing drug plan is triggered by past changes in her plan's attributes (e.g., a price increase), while her plan choice is determined by current attributes of available plans. See also Ho et al. (2017).

[48]Dardanoni et al. (2020) also take this approach. However, they rule out unobserved preference heterogeneity and focus on point identification of the choice set formation model.

Cattaneo et al. (2020) propose a random attention model in which agents' preferences are homogeneous (and thus independent of choice sets) and the probability of a particular choice set does not decrease when the number of possible choice sets decreases. Within this framework, they provide revealed preference theory and testable implications for observable choice probabilities, as well as partial identification results for preference orderings.

The method that we propose and apply in this paper falls into this last category. However, it relies on fewer and weaker restrictions on the choice set formation process than any other paper in that category. Our core model imposes—and hence our main identification result requires—only one mild assumption on the choice set formation process, namely that choice sets have a known minimum size greater than one. Importantly, our core model does not assume that choice sets are independent of preferences conditional on observables (Abaluck and Adams 2020; Crawford et al. 2020; Cattaneo et al. 2020). Nor do we impose other restrictions on how agents' choice sets are formed (Abaluck and Adams 2020; Barseghyan et al. 2020) or evolve over time (Crawford et al. 2020), rely on exclusion restrictions or large support assumptions (Abaluck and Adams 2020; Barseghyan et al. 2020), require that the econometrician knows the composition of the smallest possible choice set for each agent (Abaluck and Adams 2020; Lu 2019), or assume that choice sets satisfy a monotonicity or other regularity condition (Lu 2019; Cattaneo et al. 2020).

Due to the parsimony of our method we obtain partial and not point identification of the underlying model of preferences. Nevertheless, we demonstrate that much can be learned about the distribution of preferences under our approach. Moreover, what is learned has more credibility because we avoid making a host of arbitrary or unverifiable assumptions about the choice set formation process to achieve point identification. Our primary contribution, therefore, is that we offer a new, robust, informative, and implementable method of discrete choice analysis when choice sets are unobserved. We show how one can use this method to partially identify and conduct inference on the distribution of preferences as well as the distribution of choice set size (with an additional independence assumption). Through our empirical application we also contribute new insights to the literature on risky choice.

# Appendix

## A.1 Random Closed Sets

The theory of random closed sets generally applies to the space of closed subsets of a locally compact Hausdorff second countable topological space $\mathbb{F}$. For simplicity we consider here the case $\mathbb{F} = \mathbb{R}^k$ and refer to Molchanov (2017) for the general case. Denote by $\mathcal{F}$ (respectively, $\mathcal{K}$) the collection of closed (compact) subsets of $\mathbb{R}^k$. Denote by $(\Omega, \mathfrak{F}, P)$ the nonatomic probability space on which all random variables and random sets are defined.

DEFINITION A.1 (Random Closed Set): *A map $Y : \Omega \to \mathcal{F}$ is a* random closed set *if for every compact set $K$ in $\mathbb{R}^k$, $Y^{-1}(K) = \{\omega \in \Omega : Y(\omega) \cap K \neq \varnothing\} \in \mathfrak{F}$.*

DEFINITION A.2 (Selection): *For any random set $Y$, a (measurable)* selection *of $Y$ is a random vector $y$ (taking values in $\mathbb{R}^k$) such that $y(\omega) \in Y(\omega)$, $P - a.s.$*

THEOREM A.1 (Artstein's Theorem): *A random vector $y$ and a random set $Y$ can be realized on the same probability space as random elements $y'$ and $Y'$, distributed as $y$ and $Y$, respectively, so that $P(y' \in Y') = 1$, if and only if $P(y \in K) \leq P(Y \cap K \neq \varnothing) \; \forall K \in \mathcal{K}$.*

Because in this paper the random closed set of interest $D_\kappa^*(\mathbf{x}_i, \boldsymbol{\nu}_i; \boldsymbol{\delta})$ is a subset of $\mathcal{D}$, it suffices to consider $\mathbb{F} = \mathcal{D}$; see Molchanov (2017, Example 1.1.9).

LEMMA A.1: *The set $D_\kappa^*(\mathbf{x}_i, \boldsymbol{\nu}_i; \boldsymbol{\delta})$ in equation (3.1) is a random closed set.*

*Proof.* Let $D_\kappa^* \equiv D_\kappa^*(\mathbf{x}_i, \boldsymbol{\nu}_i; \boldsymbol{\delta})$. Because $D_\kappa^*$ is a finite set, we have that $\{D_\kappa^* \cap K \neq \varnothing\} = \bigcup_{G \subseteq \mathcal{D}: |G| = \kappa} \{d_i^*(G, \mathbf{x}_i, \boldsymbol{\nu}_i; \boldsymbol{\delta}) \in K\}$. As $d_i^*(G, \mathbf{x}_i, \boldsymbol{\nu}_i; \boldsymbol{\delta})$ is a random variable, the result follows (see Molchanov and Molinari 2018, Example 1.5). □

## A.2 Proof of Theorem 3.1

Let $d^*(G, \mathbf{x}, \boldsymbol{\nu}; \boldsymbol{\delta})$ denote the model-implied optimal choice for an agent with attributes $(\mathbf{x}, \boldsymbol{\nu})$ and choice set $G$. Recall that by Assumption 2.2(II), $\Pr(C = G | \mathbf{x}, \boldsymbol{\nu}) = 0$ for all $G \subseteq \mathcal{D}$ such that $|G| < \kappa$. Then by definition the sharp identification region $\Theta_I$ is given by the set of values $\boldsymbol{\theta} \in \Theta$ for which there exists a distribution $\mathsf{F}(\cdot; \mathbf{x}, \boldsymbol{\nu})$ such that $\mathsf{F}(G; \mathbf{x}, \boldsymbol{\nu}) \geq 0$ for all $G \subseteq \mathcal{D}$, $\mathsf{F}(G; \mathbf{x}, \boldsymbol{\nu}) = 0$ if $|G| < \kappa$, $\sum_{G \subseteq \mathcal{D}} \mathsf{F}(G; \mathbf{x}, \boldsymbol{\nu}) = 1$, and for all $c \in \mathcal{D}$

$$\Pr(d = c | \mathbf{x}) = \int_{\boldsymbol{\tau} \in \mathcal{V}} \sum_{G \subseteq \mathcal{D}} \mathbf{1}(d^*(G, \mathbf{x}, \boldsymbol{\tau}; \boldsymbol{\delta}) = c) \mathsf{F}(G; \mathbf{x}, \boldsymbol{\tau}) dP(\boldsymbol{\tau}; \boldsymbol{\gamma}), \; \mathbf{x} - a.s. \quad (A.1)$$

This is because for such values $\boldsymbol{\theta} \in \Theta$, one can complete the model with a distribution $\mathsf{F}(\cdot; \mathbf{x}, \boldsymbol{\nu})$ so that the model-implied conditional distribution of optimal choices matches the distribution of observed choices. We are then left to show that this set is equal to the one in equation (3.5). Molchanov and Molinari (2018, Theorem 2.33) show that the observed vector $(d, \mathbf{x})$ is a selection of the random closed set $(D_\kappa^*(\mathbf{x}, \boldsymbol{\nu}; \boldsymbol{\delta}), \mathbf{x})$ if and only if the condition in equation (3.5) holds $\mathbf{x} - a.s.$ for all $K \subseteq \mathcal{D}$. Take a value $\boldsymbol{\theta} \in \Theta$ such that there exists a distribution $\mathsf{F}(G; \mathbf{x}, \boldsymbol{\nu})$ under which equation (A.1) holds. By definition $(d^*(G, \mathbf{x}, \boldsymbol{\nu}; \boldsymbol{\delta}), \mathbf{x})$ is a selection of $(D^*_\kappa(\mathbf{x}, \boldsymbol{\nu}; \boldsymbol{\delta}), \mathbf{x})$, and by Molchanov and Molinari (2018, Theorem 2.33) the inequality in equation (3.5) holds $\mathbf{x} - a.s.$ for all $K \subseteq \mathcal{D}$. Conversely, take a value $\boldsymbol{\theta} \in \Theta$ for which the inequalities in equation (3.5) are satisfied $\mathbf{x} - a.s.$ for all $K \subseteq \mathcal{D}$. Then, by Theorem A.1, there exists a selection $(\tilde{d}(G), \mathbf{x})$ of $(D^*_\kappa(\mathbf{x}, \boldsymbol{\nu}; \boldsymbol{\delta}), \mathbf{x})$ such that $\Pr(d = c|\mathbf{x}) = \Pr(\tilde{d}(G) = c|\mathbf{x})$, $\mathbf{x} - a.s.$, for all $c \in \mathcal{D}$ for some $G$ such that $|G| \geq \kappa$. Let $\mathsf{F}(G; \mathbf{x}, \boldsymbol{\nu})$ equal 1 for one such set $G$ with $\tilde{d}(G) = c$, and equal 0 for all other $G \subseteq \mathcal{D}$. Then equation (A.1) holds $\mathbf{x} - a.s.$ for all $c \in \mathcal{D}$. To conclude the proof, we show that if the inequalities in equation (3.5) hold for all $K \subseteq \mathcal{D} : |K| < \kappa$, then they hold for all $K \subseteq \mathcal{D}$. Recall that the set $D^*_\kappa(\mathbf{x}, \boldsymbol{\nu}; \boldsymbol{\delta})$ comprises the $|\mathcal{D}| - \kappa + 1$ best alternatives in $\mathcal{D}$. Then any $K \subseteq \mathcal{D} : |K| \geq \kappa$ includes at least the $(|\mathcal{D}| - \kappa + 1)$-th best alternative for all realizations of $\boldsymbol{\nu}$ in $\mathcal{V}$, so that $\Pr(D^*_\kappa(\mathbf{x}, \boldsymbol{\nu}; \boldsymbol{\delta}) \cap K \neq \varnothing) = 1$ and the inequality in equation (3.5) holds mechanically. $\square$

Ben-Akiva, M. and B. Boccara (1995): "Discrete Choice Models with Latent Choice Sets," *International Journal of Marketing Research*, 12, 9–24.

Ben-Akiva, M. E. (1973): "Structure of Passenger Travel Demand Models," Ph.D. Dissertation, Department of Civil Engineering, Massachusetts Institute of Technology.

Beresteanu, A., I. Molchanov, and F. Molinari (2011): "Sharp Identification Regions in Models with Convex Moment Predictions," *Econometrica*, 79, 1785–1821.

Beresteanu, A. and F. Molinari (2008): "Asymptotic Properties for a Class of Partially Identified Models," *Econometrica*, 76, 763–814.

Berry, S., J. Levinsohn, and A. Pakes (1995): "Automobile Prices in Market Equilibrium," *Econometrica*, 63, 841–890.

Bhargava, S., G. Loewenstein, and J. Sydnor (2017): "Choose to Lose: Health Plan Choices from a Menu with Dominated Options," *Quarterly Journal of Economics*, 132, 1319–1372.

Canay, I. A. and A. M. Shaikh (2017): "Practical and Theoretical Advances in Inference for Partially Identified Models," in *Advances in Economics and Econometrics: Eleventh World Congress,* Vol. 2, ed. by B. Honóre, A. Pakes, M. Piazzesi, and L. Samuelson, Cambridge: Cambridge University Press, 271–306.

Caplin, A. (2016): "Measuring and Modeling Attention," *Annual Review of Economics*, 8, 379–403.

Caplin, A. and M. Dean (2015): "Revealed Preference, Rational Inattention, and Costly Information Acquisition," *American Economic Review*, 105, 2183–2203.

Cattaneo, M. D., X. Ma, Y. Masatlioglu, and E. Suleymanov (2020): "A Random Attention Model," *Journal of Political Economy*, 128, 2796–2836.

Chamberlin, E. H. (1933): *The Theory of Monopolistic Competition: A Re-orientation of the Theory of Value*, Cambridge, MA: Harvard University Press.

Chesher, A. and A. M. Rosen (2017): "Generalized Instrumental Variable Models," *Econometrica*, 85, 959–989.

Chesher, A., A. M. Rosen, and K. Smolinski (2013): "An Instrumental Variable Model of Multiple Discrete Choice," *Quantitative Economics*, 4, 157–196.

Chiappori, P.-A., A. Gandhi, B. Salanié, and F. Salanié (2019): "From Aggregate Betting Data to Individual Risk Preferences," *Econometrica*, 87, 1–36.

Cicchetti, C. J. and J. A. Dubin (1994): "A Microeconometric Analysis of Risk Aversion and the Decision to Self-Insure," *Journal of Political Economy*, 102, 169–186.

Ciliberto, F. and E. Tamer (2009): "Market Structure and Multiple Equilibria in Airline Markets," *Econometrica*, 77, 1791–1828.

Cohen, A. and L. Einav (2007): "Estimating Risk Preferences from Deductible Choice," *American Economic Review*, 97, 745–788.

# Supplemental Material for "Heterogeneous Choice Sets and Preferences"

Levon Barseghyan  
Cornell University

Maura Coughlin  
Rice University

Francesca Molinari  
Cornell University

Joshua C. Teitelbaum  
Georgetown University

February 9, 2021

# S1 Theory

## S1.1 Unobserved Heterogeneity in Choice Sets as Additively Separable Disturbances

It is possible to represent unobserved heterogeneity in choice sets through additively separable disturbances. In a classic random utility model with $U_i(c) = W_i(c) + \epsilon_{ic}$, one may let $\epsilon_{ic} \in \{-\infty, 0\}$ for each alternative $c \in \mathcal{D}$ and allow $\epsilon_{ic}$ to be correlated with $\epsilon_{ic'}$ for any two alternatives $c, c' \in \mathcal{D}$. One would then posit that: if $\kappa = |\mathcal{D}|$ then $\epsilon_{ic} = 0$ for each alternative $c \in \mathcal{D}$; if $\kappa = |\mathcal{D}| - 1$ then $\epsilon_{ic} = -\infty$ for at most one alternative in $\mathcal{D}$ (the identity of which is left unspecified); if $\kappa = |\mathcal{D}| - 2$ then $\epsilon_{ic} = -\infty$ for at most two alternatives in $\mathcal{D}$ (the identities of which are left unspecified); and so forth. This model yields that alternative $c$ is not chosen if $\epsilon_{ic} = -\infty$, which is analogous to alternative $c$ not being chosen when it is not contained in the agent's choice set.

## S1.2 Positive Probability of Utility Ties

When utility ties are allowed, one can adapt the definition of $D^*_\kappa(\mathbf{x}_i, \boldsymbol{\nu}_i; \boldsymbol{\delta})$ as follows:

$$D^*_\kappa(\mathbf{x}_i, \boldsymbol{\nu}_i; \boldsymbol{\delta}) = \bigcup_{G \subseteq \mathcal{D}: |G| \geq \kappa} \left\{ \arg\max_{c \in G} W(\mathbf{x}_{ic}, \boldsymbol{\nu}_i; \boldsymbol{\delta}) \right\} = \bigcup_{G \subseteq \mathcal{D}: |G| = \kappa} \left\{ \arg\max_{c \in G} W(\mathbf{x}_{ic}, \boldsymbol{\nu}_i; \boldsymbol{\delta}) \right\}, \tag{S1.1}$$

where again the last equality follows from Sen's property $\alpha$, and now $\arg\max_{c \in G} W(\mathbf{x}_{ic}, \boldsymbol{\nu}_i; \boldsymbol{\delta})$ may include multiple elements of $\mathcal{D}$ due to the possibility of utility ties. The random closed set $D^*_\kappa(\mathbf{x}_i, \boldsymbol{\nu}_i; \boldsymbol{\delta})$ contains alternatives up to the $(|\mathcal{D}| - \kappa + 1)$-th best in $\mathcal{D}$, where "best" is defined with respect to $W(\mathbf{x}_{ic}, \boldsymbol{\nu}_i; \boldsymbol{\delta})$. Due to the possibility of ties, $|D^*_\kappa(\mathbf{x}_i, \boldsymbol{\nu}_i; \boldsymbol{\delta})|$ may be larger than $|\mathcal{D}| - \kappa + 1$.[1]

To see that our characterization in Theorem 3.1 applied with this new definition of $D^*_\kappa(\mathbf{x}_i, \boldsymbol{\nu}_i; \boldsymbol{\delta})$ remains sharp, note that the model-implied optimal choice for an agent with attributes $(\mathbf{x}_i, \boldsymbol{\nu}_i)$, utility parameters $\boldsymbol{\delta}$, and choice set $G$ is no longer unique. But this additional multiplicity of optimal choices is incorporated into $D^*_\kappa(\mathbf{x}_i, \boldsymbol{\nu}_i; \boldsymbol{\delta})$, and all model restrictions continue to be embedded in the requirement that $d_i \in D^*_\kappa(\mathbf{x}_i, \boldsymbol{\nu}_i; \boldsymbol{\delta})$, almost surely. The proof of Theorem 3.1 continues to apply, although at the price of additional notation (a selection mechanism that determines the probability with which each optimal choice $d^*_i(G, \mathbf{x}_i, \boldsymbol{\nu}_i; \boldsymbol{\delta}) \in \arg\max_{c \in G} W(\mathbf{x}_{ic}, \boldsymbol{\nu}_i; \boldsymbol{\delta})$ is selected when multiple alternatives are optimal for a realization $G$ of $C_i$).

---

[1] To illustrate, consider the case $|\mathcal{D}| = 5$ and $\kappa = 4$. When utility ties occur with positive probability, for a given $(\mathbf{x}, \boldsymbol{\nu}; \boldsymbol{\delta})$ it might be, for example, that three alternatives are tied as first best, and hence at least one of them is in any realization of $C_i$ and $|D^*_\kappa(\mathbf{x}_i, \boldsymbol{\nu}_i; \boldsymbol{\delta})| = 3$.

## S1.3  Computational Simplifications

We omit the subscript $i$ on random variables and random sets throughout this section.

### S1.3.1  Sufficient Collection of Test Sets $K$

Theorem 3.1 and Corollary 3.1 provide a characterization of $\Theta_I$ as the collection of $\boldsymbol{\theta} \in \Theta$ that satisfy a finite number of conditional moment inequalities, indexed by the *test sets* $K \subset \mathcal{D}$. In this subsection we provide results to reduce the collection of test sets $K$ for which to check the inequalities from all nonempty proper subsets of $\mathcal{D}$ to a smaller collection.

THEOREM S1.1: *Let the assumptions of Theorem 3.1 hold. Then the following steps yield a sufficient collection of sets $K$, denoted $\mathbb{K}$, on which to check the inequalities in equation (3.5) to verify if $\boldsymbol{\theta} \in \Theta_I$. Initialize $\mathbb{K} = \{K \subset \mathcal{D} : |K| < \kappa\}$. Then:*

*(1) For a given set $K \in \mathbb{K}$, if it holds that $\forall \boldsymbol{\nu} \in \mathcal{V}$ an element of $K$ (possibly different across values of $\boldsymbol{\nu}$) is among the $|\mathcal{D}| - \kappa + 1$ best alternatives in $\mathcal{D}$, then set $\mathbb{K} = \mathbb{K} \backslash K$;[2]*

*(q) Repeat the following step for $q = 2, \ldots, \kappa - 1$. Take any set $K \in \mathbb{K}$ such that $K = K_{q-1} \cup \{c_j\}$ for some $K_{q-1}$ with $|K_{q-1}| = q - 1$ and $\{c_j\} \in \mathbb{K}, K_{q-1} \in \mathbb{K}$ after Steps (1) and (q-1). If $\nexists \boldsymbol{\nu} \in \mathcal{V}$ such that both $c_j$ and at least one element of $K_{q-1}$ are among the $|\mathcal{D}| - \kappa + 1$ best alternatives in $\mathcal{D}$, then set $\mathbb{K} = \mathbb{K} \backslash K$.*

*If the set $D_\kappa^*$ does not depend on $\boldsymbol{\delta}$, as in our application in Sections 4–5, the collection $\mathbb{K}$ is invariant across $\boldsymbol{\theta} \in \Theta$.*

*Proof.* Step (1) follows because under the stated condition, $\Pr(D_\kappa^*(\mathbf{x}, \boldsymbol{\nu}; \boldsymbol{\delta}) \cap K \neq \varnothing) = 1$. Step (q) follows because under the stated condition, the events $\{D_\kappa^*(\mathbf{x}, \boldsymbol{\nu}; \boldsymbol{\delta}) \cap \{c_j\} \neq \varnothing\}$ and $\{D_\kappa^*(\mathbf{x}, \boldsymbol{\nu}; \boldsymbol{\delta}) \cap K_{q-1} \neq \varnothing\}$ are disjoint. This implies that the right-hand side of the inequality in equation (3.5) is additive, and therefore that inequality evaluated at $K$ is implied by the ones evaluated at $\{c_j\}$ and at $K_{q-1}$. □

Depending on the structure of the realizations of the random set $D_\kappa^*(\mathbf{x}, \boldsymbol{\nu}; \boldsymbol{\delta})$, Theorem S1.1 can be further simplified. The following corollary provides an example.

COROLLARY S1.1: *Let Assumptions 2.1 and 2.2 hold. Suppose all possible realizations of $D_\kappa^*(\mathbf{x}, \boldsymbol{\nu}; \boldsymbol{\delta})$ are given by adjacent elements of $\mathcal{D}$, as $\{c_j, c_{j+1}, \ldots, c_{j+|\mathcal{D}|-\kappa}\}$, for $j = 1, \ldots, \kappa$.*

---

[2] Here the notation $\mathbb{K} \backslash K$ indicates that the set $K$ is removed from the collection of sets $\mathbb{K}$. In practice, one can implement this step first on sets $K : |K| = 1$, and for $K$ that satisfies the condition remove from $\mathbb{K}$ all sets $K' \supseteq K$. Then repeat the procedure for the remaining sets $K : |K| = 2$, and so forth.

Then the collection of test sets $\mathbb{K}$ in Theorem S1.1 can be initialized to

$$\mathbb{K} = \Big\{ \{c_1\}, \{c_1, c_2\}, \{c_1, c_2, c_3\}, \cdots, \{c_1, c_2, \ldots, c_{\kappa-1}\},$$
$$\{c_{|\mathcal{D}|}\}, \{c_{|\mathcal{D}|}, c_{|\mathcal{D}|-1}\}, \{c_{|\mathcal{D}|}, c_{|\mathcal{D}|-1}, c_{|\mathcal{D}|-2}\}, \cdots, \{c_{|\mathcal{D}|}, c_{|\mathcal{D}|-1}, \ldots, c_{|\mathcal{D}|-\kappa+2}\} \Big\}, \quad \text{(S1.2)}$$

which contains $2(\kappa - 1)$ elements.

*Proof.* We first establish that if the inequalities in equation (3.5) are satisfied for sets of size $|K| = m$, $m = 1, \ldots, \kappa - 1$, comprised of adjacent alternatives (with respect to $|\mathcal{D}|$), then they are satisfied for all $K \subset \mathcal{D}$.

Let the inequality in equation (3.5) be satisfied for $K_1 = \{c_j, c_{j+1}, \ldots, c_p\}$, for $K_2 = \{c_q, c_{q+1}, \ldots, c_t\}$, with $p < q - 1$ so that $K_1 \cap K_2 = \varnothing$, and for $K = K_1 \cup \{c_{p+1}, \ldots, c_{q-1}\} \cup K_2$ (the set that comprises all adjacent alternatives between $c_j$ and $c_t$). We then show that the inequality for $K_1 \cup K_2$ is redundant. The same argument generalizes to sets comprised of the union of disjoint collections of adjacent alternatives.

Consider first the case that $q - p \geqslant |\mathcal{D}| - \kappa + 1$. Then $D_\kappa^*(\mathbf{x}, \boldsymbol{\nu}; \boldsymbol{\delta})$ cannot intersect both $K_1$ and $K_2$, and hence

$$P(D_\kappa^*(\mathbf{x}, \boldsymbol{\nu}; \boldsymbol{\delta}) \cap (K_1 \cup K_2) \neq \varnothing; \boldsymbol{\gamma}) = P(D_\kappa^*(\mathbf{x}, \boldsymbol{\nu}; \boldsymbol{\delta}) \cap K_1 \neq \varnothing; \boldsymbol{\gamma}) + P(D_\kappa^*(\mathbf{x}, \boldsymbol{\nu}; \boldsymbol{\delta}) \cap K_2 \neq \varnothing; \boldsymbol{\gamma})$$

and the result follows.

Consider next the case that $q - p < |\mathcal{D}| - \kappa + 1$. We claim that in this case

$$D_\kappa^*(\mathbf{x}, \boldsymbol{\nu}; \boldsymbol{\delta}) \cap K \backslash (K_1 \cup K_2) \neq \varnothing \Rightarrow D_\kappa^*(\mathbf{x}, \boldsymbol{\nu}; \boldsymbol{\delta}) \cap (K_1 \cup K_2) \neq \varnothing. \quad \text{(S1.3)}$$

To establish this claim, take $c_s \in \{c_{p+1}, \ldots, c_{q-1}\} \equiv K \backslash (K_1 \cup K_2)$. Suppose $c_s \in D_\kappa^*(\mathbf{x}, \boldsymbol{\nu}; \boldsymbol{\delta})$. Then either $c_p \in D_\kappa^*(\mathbf{x}, \boldsymbol{\nu}; \boldsymbol{\delta})$ or $c_q \in D_\kappa^*(\mathbf{x}, \boldsymbol{\nu}; \boldsymbol{\delta})$, because $|D_\kappa^*(\mathbf{x}, \boldsymbol{\nu}; \boldsymbol{\delta})| = |\mathcal{D}| - \kappa + 1$. The claim follows because $K_1 \cup K_2 \subset K$, and hence $\Pr(d \in K_1 \cup K_2 | \mathbf{x}) \leqslant \Pr(d \in K | \mathbf{x})$, while $P(D_\kappa^*(\mathbf{x}, \boldsymbol{\nu}; \boldsymbol{\delta}) \cap (K_1 \cup K_2) \neq \varnothing; \boldsymbol{\gamma}) = P(D_\kappa^*(\mathbf{x}, \boldsymbol{\nu}; \boldsymbol{\delta}) \cap K \neq \varnothing; \boldsymbol{\gamma})$ due to equation (S1.3).

Finally, we show that it suffices to verify equation (3.5) for the sets $K \in \mathbb{K}$ as specified in equation (S1.2). Consider first the case where $|\mathcal{D}| - \kappa + 1 > \kappa - 1$. Then for all $1 < p < q < \kappa$ and $K = \{c_p, c_{p+1}, \ldots, c_q\}$, it holds that $|K| < \kappa - 1$ and, denoting $K^c = \mathcal{D} \backslash K$,

$$P(D_\kappa^*(\mathbf{x}, \boldsymbol{\nu}; \boldsymbol{\delta}) \cap K \neq \varnothing; \boldsymbol{\gamma}) = 1 - P(D_\kappa^*(\mathbf{x}, \boldsymbol{\nu}; \boldsymbol{\delta}) \subset K^c; \boldsymbol{\gamma})$$
$$= 1 - P(D_\kappa^*(\mathbf{x}, \boldsymbol{\nu}; \boldsymbol{\delta}) \subset \{c_1, \ldots, c_{p-1}\}; \boldsymbol{\gamma}) - P(D_\kappa^*(\mathbf{x}, \boldsymbol{\nu}; \boldsymbol{\delta}) \subset \{c_{q+1}, \ldots, c_\mathcal{D}\}; \boldsymbol{\gamma})$$
$$= 1 - P(D_\kappa^*(\mathbf{x}, \boldsymbol{\nu}; \boldsymbol{\delta}) \subset \{c_{q+1}, \ldots, c_\mathcal{D}\}; \boldsymbol{\gamma}), \quad \text{(S1.4)}$$

where the first equality follows by definition, the second follows because $D_\kappa^*(\mathbf{x}, \boldsymbol{\nu}; \boldsymbol{\delta})$ is comprised of $|\mathcal{D}| - \kappa + 1$ adjacent alternatives, and the last follows because $P(D_\kappa^*(\mathbf{x}, \boldsymbol{\nu}; \boldsymbol{\delta}) \subset \{c_1, \ldots, c_{p-1}\}; \boldsymbol{\gamma}) = 0$ as $|\{c_1, \ldots, c_{p-1}\}| < \kappa - 1 < |\mathcal{D}| - \kappa + 1$. On the other hand,

$$\Pr(d \in \{c_p, \ldots, c_q\}) \leqslant \Pr(d \in \{c_1, \ldots, c_q\}),$$

and hence if equation (3.5) is satisfied for $K = \{c_1, \ldots, c_q\}$, it is also satisfied for $K = \{c_p, c_{p+1}, \ldots, c_q\}$ for all $1 < p < q < \kappa$. A similar reasoning, with appropriate modifications, holds for sets $K = \{c_{|\mathcal{D}|-q+1}, c_{p+1}, \ldots, c_{|\mathcal{D}|-p+1}\}$.

When $|\mathcal{D}| - \kappa + 1 \leqslant \kappa - 1$, equation (S1.4) continues to hold as stated whenever $p < |\mathcal{D}| - \kappa + 1$. If $p > |\mathcal{D}| - \kappa + 1$, the result follows by the additivity in the second line of equation (S1.4) and the additivity of probabilities, because

$$\Pr(d \in K|\mathbf{x}) \leqslant P(D_\kappa^*(\mathbf{x}, \boldsymbol{\nu}; \boldsymbol{\delta}) \cap K \neq \varnothing; \boldsymbol{\gamma}) \Leftrightarrow \Pr(d \in K^c|\mathbf{x}) \geqslant P(D_\kappa^*(\mathbf{x}, \boldsymbol{\nu}; \boldsymbol{\delta}) \subset K^c; \boldsymbol{\gamma}).$$

Hence, the inequality for $K = \{c_p, \ldots, c_q\}$ is implied whenever it is satisfied for $K = \{c_1, \ldots, c_p\}$ and $K = \{c_q, \ldots, c_{|\mathcal{D}|}\}$. $\square$

The following claim establishes that Corollary S1.1 applies when $\boldsymbol{\nu} \in \mathbb{R}$ and the alternatives in the feasible set are vertically differentiated.

CLAIM S1.1: *Let Assumptions 2.1 and 2.2 hold. Let $\mathcal{D} = \{c_1, \ldots, c_{|\mathcal{D}|}\}$ and $\boldsymbol{\nu} = \nu \in \mathbb{R}$. Suppose that: (I) for every pair of alternatives $c_j, c_k \in \mathcal{D}$, $j < k$, and given any $\mathbf{x} \in \mathcal{X}$, there exists a unique threshold $\bar{\nu}_{j,k}(\mathbf{x})$ such that for all $\nu > \bar{\nu}_{j,k}(\mathbf{x})$ alternative $c_j$ has greater utility than alternative $c_k$ and for all $\nu < \bar{\nu}_{j,k}(\mathbf{x})$ alternative $c_k$ has greater utility than alternative $c_j$; and (II) for every alternative $c_j \in \mathcal{D}$ and given any $\mathbf{x} \in \mathcal{X}$, there exists a $\nu \in \mathbb{R}$ such that $c_j$ is the first best in $\mathcal{D}$. Then, given any $(\mathbf{x}, \nu) \in \mathcal{X} \times \mathbb{R}$ and any $\kappa \geqslant 2$, the set $D_\kappa^*(\mathbf{x}, \boldsymbol{\nu}; \boldsymbol{\delta})$ comprises adjacent elements of $\mathcal{D}$, as $\{c_j, c_{j+1}, \ldots, c_{j+|\mathcal{D}|-\kappa}\}$, for $j = 1, \ldots, \kappa$.*

*Proof.* The proof builds on Fact 4 in Barseghyan et al. (2020). Let $|\mathcal{D}| \geqslant 3$ (otherwise the claim holds trivially). Take any $\mathbf{x} \in \mathcal{X}$ and any three alternatives $c_j, c_{j+1}, c_{j+2} \in \mathcal{D}$. Conditions (I) and (II) imply that $\bar{\nu}_{j,j+1}(\mathbf{x}) > \bar{\nu}_{j,j+2}(\mathbf{x}) > \bar{\nu}_{j+1,j+2}(\mathbf{x})$. (In particular, $\bar{\nu}_{j+1,j+2}(\mathbf{x}) > \bar{\nu}_{j,j+2}(\mathbf{x}) > \bar{\nu}_{j,j+1}(\mathbf{x})$ violates condition (II) because $c_{j+1}$ is not first best for any $\nu \in \mathbb{R}$, and every other permutation violates condition (I) due to the transitivity of utility). In other words, the alternatives are *vertically differentiated* in that $c_j$ is first best for all $\nu > \bar{\nu}_{j,j+1}(\mathbf{x})$; $c_{j+1}$ is first best for all $\nu \in (\bar{\nu}_{j+1,j+2}(\mathbf{x}), \bar{\nu}_{j,j+1}(\mathbf{x}))$; and $c_{j+2}$ is first best for all $\nu < \bar{\nu}_{j+1,j+2}(\mathbf{x})$. Consequently, for all $\nu \in \mathbb{R}$, the only possible strict utility rankings of the three alternatives are: $U(c_j) > U(c_{j+1}) > U(c_{j+2})$ (when $\nu > \bar{\nu}_{j,j+1}(\mathbf{x})$);

$U(c_{j+1}) > U(c_j) > U(c_{j+2})$ (when $\bar{\nu}_{j,j+1}(\mathbf{x}) > \nu > \bar{\nu}_{j,j+2}(\mathbf{x})$); $U(c_{j+1}) > U(c_{j+2}) > U(c_j)$ (when $\bar{\nu}_{j,j+2}(\mathbf{x}) > \nu > \bar{\nu}_{j+1,j+2}(\mathbf{x})$); and $U(c_{j+2}) > U(c_{j+1}) > U(c_j)$ (when $\nu < \bar{\nu}_{j+1,j+2}(\mathbf{x})$). Thus, alternative $c_{j+1}$ is never the third best among the three alternatives. This implies that if $c_j$ and $c_{j+2}$ both have greater utility than a fourth alternative $c_m$, $m \notin \{j, j+1, j+2\}$, then $c_{j+1}$ also has greater utility than $c_m$. It follows that for any $(\mathbf{x}, \nu) \in \mathcal{X} \times \mathbb{R}$, the set $D^*_\kappa(\mathbf{x}_i, \nu_i; \boldsymbol{\delta})$ comprises adjacent elements of $\mathcal{D}$, as $\{c_j, c_{j+1}, \ldots, c_{j+|\mathcal{D}|-\kappa}\}$, for $j = 1, \ldots, \kappa$. □

When Assumption 3.1 is maintained, the logic of Theorem S1.1 can be used to obtain a collection of sufficient test sets $K$ on which to verify the inequalities in (3.7), by applying its Steps 2.1-2.($\kappa - 1$) to the random sets $D^*_q(\mathbf{x}, \nu; \boldsymbol{\delta})$, $q = \kappa, \ldots, |\mathcal{D}|$. Further simplifications are possible when interest centers on specific projections of $\Theta_I$, using the fact that $D^*_{q+1}(\mathbf{x}_i, \nu_i; \boldsymbol{\delta}) \subset D^*_q(\mathbf{x}_i, \nu_i; \boldsymbol{\delta})$ for all $q \geq \kappa$. As discussed following Corollary 3.1, when Assumption 3.1 is maintained the projection of $\Theta_I$ on $[\boldsymbol{\delta}; \boldsymbol{\gamma}]$ is obtained by setting $\pi_\kappa(\mathbf{x}; \boldsymbol{\eta}) = 1$ and $\pi_q(\mathbf{x}; \boldsymbol{\eta}) = 0$, $q = \kappa + 1, \ldots, |\mathcal{D}|$. Hence, Steps 2.1-2.($\kappa - 1$) in Theorem S1.1 applied only to $D^*_\kappa(\mathbf{x}, \nu; \boldsymbol{\delta})$ deliver the sufficient collection of sets $K$ on which to verify (3.7) to obtain the sharp identification region for $[\boldsymbol{\delta}; \boldsymbol{\gamma}]$. On the other hand, the projection of $\Theta_I$ on $\pi_q(\mathbf{x}; \boldsymbol{\eta})$, $q = \kappa + 1, \ldots, |\mathcal{D}|$, is obtained by setting $\pi_l(\mathbf{x}; \boldsymbol{\eta}) = 0$ for all $l \notin \{q, \kappa\}$, and that on $\pi_\kappa(\mathbf{x}; \boldsymbol{\eta})$ by setting $\pi_l(\mathbf{x}; \boldsymbol{\eta}) = 0$ for all $l = \kappa + 2, \ldots, |\mathcal{D}|$. Hence, Steps 2.1-2.($\kappa - 1$) in Theorem S1.1 applied, respectively, to only $D^*_\kappa(\mathbf{x}, \nu; \boldsymbol{\delta})$ and $D^*_q(\mathbf{x}, \nu; \boldsymbol{\delta})$ deliver the sufficient collection of sets $K$ on which to verify (3.7) to obtain the sharp identification region for $\pi_q$, $q = \kappa + 1, \ldots, |\mathcal{D}|$, and applied only to $D^*_\kappa(\mathbf{x}, \nu; \boldsymbol{\delta})$ and $D^*_{\kappa+1}(\mathbf{x}, \nu; \boldsymbol{\delta})$ deliver the sufficient collection of sets $K$ on which to verify (3.7) to obtain the sharp identification region for $\pi_\kappa$.

The two corollaries that follow illustrate the specific adaptations of Theorem S1.1 that we use in our application in Sections 4–5. Proofs are omitted because the corollaries follow immediately from Theorem S1.1.

COROLLARY S1.2: *Let $\mathcal{D} = \{c_1, c_2, c_3, c_4, c_5\}$ and $\kappa = 3$. Suppose that all assumptions in Corollary 3.1 hold and that $\boldsymbol{\nu} = \nu \in \mathbb{R}$ with support $[0, \bar{\nu}]$, $\bar{\nu} < \infty$. Then the following steps yield a sufficient collection of sets $K$, denoted $\mathbb{K}$, on which to check the inequalities in equation (3.7) to obtain sharp bounds on $\pi_5$. Initialize $\mathbb{K} = \{K : K \subset \mathcal{D}\}$. Then:*

1. *For any set $K = \{c_j, c_k\} \subset \mathcal{D}$, if $\nexists \nu \in [0, \bar{\nu}]$ such that both $c_j$ and $c_k$ are among the best 3 alternatives in $\mathcal{D}$, then set $\mathbb{K} = \mathbb{K} \backslash \{c_j, c_k\}$;*

2. *Set $\mathbb{K} = \mathbb{K} \backslash \{c_j, c_k, c_l\}$ for all $j, k, l \in \{1, 2, 3, 4, 5\}$.*

COROLLARY S1.3: *Let $\mathcal{D} = \{c_1, c_2, c_3, c_4, c_5\}$ and $\kappa = 3$. Suppose that all assumptions in Corollary 3.1 hold and that $\boldsymbol{\nu} = \nu \in \mathbb{R}$ with support $[0, \bar{\nu}]$, $\bar{\nu} < \infty$. Then the following* *steps yield a sufficient collection of sets $K$, denoted $\mathbb{K}$, on which to check the inequalities in equation (3.7) to obtain sharp bounds on $\pi_4$. Initialize $\mathbb{K} = \{K : K \subset \mathcal{D}\}$. Then:*

1. *For any set $K = \{c_j, c_k\} \subset \mathcal{D}$, if $\nexists \nu \in [0, \bar{\nu}]$ such that both $c_j$ and $c_k$ are among the best 3 alternatives in $\mathcal{D}$, then set $\mathbb{K} = \mathbb{K} \backslash \{\{c_j, c_k\}, \{\mathcal{D} \backslash \{c_j, c_k\}\}\}$;*

2. *For any set $K = \{c_j, c_k, c_l\} \subset \mathcal{D}$ such that $\{c_j, c_k\} \in \mathbb{K}$ after Step 1, if $\nexists \nu \in [0, \bar{\nu}]$ such that both $c_l$ and at least one element of $\{c_j, c_k\}$ are among the best 3 alternatives in $\mathcal{D}$, then set $\mathbb{K} = \mathbb{K} \backslash \{c_j, c_k, c_l\}$;*

3. *For any set $K \in \mathbb{K}$, if $\forall \nu \in [0, \bar{\nu}]$ one element of $K$, possibly different across values of $\nu$, is among the best 2 alternatives in $\mathcal{D}$, then set $\mathbb{K} = \mathbb{K} \backslash K$.*

In our application in Sections 4–5, the number of inequalities obtained through application of the foregoing results (taking into account the 65 hypercubes on $(\mu, \bar{p})$) is $6 \times 65 = 390$ for the sharp identification region of $\boldsymbol{\gamma}$; $17 \times 65 = 1,105$ for the sharp identification region of $\pi_5$; and $15 \times 65 = 975$ for the sharp identification region of $\pi_4$.

## S1.4 An Equivalent Characterization Based on Convex Optimization

The characterization in Theorem 3.1 can equivalently be written in terms of a convex optimization problem.

COROLLARY S1.4: *Let Assumptions 2.1 and 2.2 hold and let $\Theta = \Delta \times \Gamma$. Then*

$$\Theta_I = \left\{ \theta \in \Theta : \max_{\mathbf{u} \in \mathbb{R}^{|\mathcal{D}|} : ||\mathbf{u}|| \leqslant 1} \left[ \mathbf{u}^\top \mathbf{p}(\mathbf{x}) - \int_{\boldsymbol{\tau} \in \mathcal{V}} \max_{d^* \in D^*_\kappa(\mathbf{x}, \boldsymbol{\tau}; \boldsymbol{\delta})} \left( \mathbf{u}^\top \mathbf{q}^{d^*} \right) dP(\boldsymbol{\tau}; \boldsymbol{\gamma}) \right] = 0, \mathbf{x} - a.s. \right\}, \tag{S1.5}$$

*where $\mathbf{p}(\mathbf{x}) = [\Pr(d = c_1|\mathbf{x}) \ldots \Pr(d = c_{|\mathcal{D}|}|\mathbf{x})]^\top$ and, for a given $d^* \in D^*_\kappa(\mathbf{x}, \boldsymbol{\nu}; \boldsymbol{\delta})$, $\mathbf{q}^{d^*} = [\mathbf{1}(d^* = c_1) \ldots \mathbf{1}(d^* = c_{|\mathcal{D}|})]^\top$.*

*Proof.* We establish the equivalence between equations (3.5) in the paper and (S1.5) here.[3] Due to the positive homogeneity in $\mathbf{u}$ of $\mathbf{u}^\top \mathbf{p}(\mathbf{x}) - \int_{\boldsymbol{\tau} \in \mathcal{V}} \max_{d^* \in D^*_\kappa(\mathbf{x}, \boldsymbol{\tau}; \boldsymbol{\delta})} \mathbf{u}^\top \mathbf{q}^{d^*} dP(\boldsymbol{\tau}; \boldsymbol{\gamma})$, we have that

$$\mathbf{u}^\top \mathbf{p}(\mathbf{x}) - \int_{\boldsymbol{\tau} \in \mathcal{V}} \max_{d^* \in D^*_\kappa(\mathbf{x}, \boldsymbol{\tau}; \boldsymbol{\delta})} \mathbf{u}^\top \mathbf{q}^{d^*} dP(\boldsymbol{\tau}; \boldsymbol{\gamma}) \leqslant 0 \tag{S1.6}$$

holds for all $\mathbf{u} : ||\mathbf{u}|| \leqslant 1$ if and only if expression (S1.6) holds for all $\mathbf{u} \in \mathbb{R}^{|\mathcal{D}|}$. Consider the specific subset of vectors $\mathsf{U} = \{\mathbf{u} \in \mathbb{R}^{|\mathcal{D}|} : u_j \in \{0, 1\}, j = 1, \ldots, |\mathcal{D}|\}$. Each vector $\mathbf{u} \in \mathsf{U}$ uniquely corresponds to a subset $K_{\mathbf{u}} = \{c_1 u_1, \ldots, c_{|\mathcal{D}|} u_{|\mathcal{D}|}\}$. For a given $\mathbf{u}$, $\mathbf{u}^\top \mathbf{q}^{d^*} = 1$ if $d^* \in K_{\mathbf{u}}$ and $\mathbf{u}^\top \mathbf{q}^{d^*} = 0$ otherwise. Hence, the corresponding inequality in (S1.6) reduces to

$$\Pr(d \in K_{\mathbf{u}} | \mathbf{x}) = \mathbf{u}^\top \mathbf{p}(\mathbf{x}) \leqslant \mathrm{E}\left[\max_{d^* \in D_\kappa^*(\mathbf{x}, \boldsymbol{\tau}; \boldsymbol{\delta})} \mathbf{u}^\top \mathbf{q}^{d^*} \Big| \mathbf{x}; \boldsymbol{\gamma}\right] = P(D_\kappa^*(\mathbf{x}, \boldsymbol{\nu}; \boldsymbol{\delta}) \cap K_{\mathbf{u}} \neq \varnothing; \boldsymbol{\gamma}).$$

---
[3]The argument of proof goes through similar steps as in Molchanov and Molinari (2018, Theorem 3.28).

It then follows that any $\theta$ in the set defined in equation (S1.5) belongs to the set defined in equation (3.5) because $\{K : K \subseteq \mathcal{D}\} = \{K_{\mathbf{u}} : \mathbf{u} \in \mathsf{U}\}$.

Conversely, take a $\theta$ in the set defined by equation (3.5). Then, by Theorem A.1, there exists a selection $d^*$ of $D_\kappa^*(\mathbf{x}, \boldsymbol{\nu}; \boldsymbol{\delta})$ such that for all $c \in \mathcal{D}$ and $\mathbf{x} - a.s.$, $\Pr(d = c | \mathbf{x}_i) = \Pr(d^* = c | \mathbf{x}_i)$. Hence, $\theta$ belongs to the set defined in equation (S1.5). □

As the set $D_\kappa^*(\mathbf{x}, \boldsymbol{\nu}; \boldsymbol{\delta})$ is comprised of the $|\mathcal{D}| - \kappa + 1$ best alternatives in $\mathcal{D}$, it can have only a finite number of realizations, as discussed in Section 3.4, which we denote $D^1, \ldots, D^h$. Hence, the characterization in equation (S1.5) can be rewritten as

$$\Theta_I = \left\{\theta \in \Theta : \max_{\mathbf{u} \in \mathbb{R}^{|\mathcal{D}|} : \|\mathbf{u}\| \leqslant 1} \left[\mathbf{u}^\top \mathbf{p}(\mathbf{x}) - \sum_{j=1}^h \left(\max_{d^* \in D^j} \mathbf{u}^\top \mathbf{q}^{d^*}\right) P(D_\kappa^*(\mathbf{x}, \boldsymbol{\nu}; \boldsymbol{\delta}) = D^j; \boldsymbol{\gamma})\right] = 0, \mathbf{x} - a.s.\right\}.$$

This means that to determine whether a given $\theta \in \Theta$ belongs to $\Theta_I$, it suffices to maximize an easy-to-compute superlinear, hence concave, function over a convex set, and check if the resulting objective value vanishes. Several efficient algorithms in convex programming are available to solve this problem; see, for example, the Matlab software for disciplined convex programming CVX (Grant and Boyd 2010).

## S1.5 Additively Separable Extreme Value Type 1 Unobserved Heterogeneity

We now explain how to compute $P(D_\kappa^*(\mathbf{x}, \boldsymbol{\nu}; \boldsymbol{\delta}) \cap K \neq \varnothing; \boldsymbol{\gamma})$ when $\boldsymbol{\nu} = (\boldsymbol{v}, (\epsilon_c, c \in \mathcal{D}))$ and $W(\mathbf{x}_c, \boldsymbol{\nu}; \boldsymbol{\delta}) = \omega(\mathbf{x}_c, \boldsymbol{v}; \boldsymbol{\delta}) + \epsilon_c$, with $\epsilon_c$ independently and identically distributed Extreme Value Type 1 *and* independent of $\boldsymbol{v}$, as in a mixed logit (McFadden and Train 2000).

Given a realization $G$ of the choice set and $\tilde{c} \in G$ (and no utility ties), we have

$$\Pr(d^*(G, \mathbf{x}, \boldsymbol{\nu}; \boldsymbol{\delta}) = \tilde{c} | \mathbf{x}, \boldsymbol{v}) = \Pr(W(\mathbf{x}_{\tilde{c}}, \boldsymbol{\nu}; \boldsymbol{\delta}) \geqslant W(\mathbf{x}_c, \boldsymbol{\nu}; \boldsymbol{\delta}) \; \forall c \in G | \boldsymbol{v})$$
$$= \frac{\exp(\omega(\mathbf{x}_{\tilde{c}}, \boldsymbol{v}; \boldsymbol{\delta}))}{\sum_{c \in G} \exp(\omega(\mathbf{x}_c, \boldsymbol{v}; \boldsymbol{\delta}))}. \tag{S1.7}$$

Conditional on $\boldsymbol{v}$, one can leverage the closed-form expressions in equation (S1.7) to compute $P(D_\kappa^*(\mathbf{x}, \boldsymbol{\nu}; \boldsymbol{\delta}) \cap K \neq \varnothing; \boldsymbol{\gamma})$ so that numerical integration is needed *only* for $\boldsymbol{v}$. The same result applies, with $q$ replacing $\kappa$, to compute $P(D_q^*(\mathbf{x}, \boldsymbol{\nu}; \boldsymbol{\delta}) \cap K \neq \varnothing; \boldsymbol{\gamma})$ in Corollary 3.1.

THEOREM S1.2: *Suppose that $\boldsymbol{\nu} = (\boldsymbol{v}, (\epsilon_c, c \in \mathcal{D}))$ and $W(\mathbf{x}_c, \boldsymbol{\nu}; \boldsymbol{\delta}) = \omega(\mathbf{x}_c, \boldsymbol{v}; \boldsymbol{\delta}) + \epsilon_c$, with $\epsilon_c$ independently and identically distributed Extreme Value Type 1* and *independent of $\boldsymbol{v}$. Conditional on $\boldsymbol{v}$, any $P(D_\kappa^*(\mathbf{x}, \boldsymbol{\nu}; \boldsymbol{\delta}) \cap K \neq \varnothing | \boldsymbol{v}; \boldsymbol{\gamma})$ can be computed as a linear combination over different sets $G$ of expression* (S1.7). *Hence, any $P(D_\kappa^*(\mathbf{x}, \boldsymbol{\nu}; \boldsymbol{\delta}) \cap K \neq \varnothing; \boldsymbol{\gamma})$ can be computed as an integral with respect to the distribution of $\boldsymbol{v}$ of linear combinations over different sets $G$ of expression* (S1.7).

To prove this theorem, we first establish two auxiliary results. The first one states that the probability of at least one alternative in $K$ being preferred to all alternatives in $\mathcal{D}\backslash K$ is the sum over all elements of $K$ that each is first best in $\mathcal{D}$.

CLAIM S1.2: *Conditional on $\boldsymbol{v}$, the probability that at least one alternative in a set $K \subset \mathcal{D}$ is better than all alternatives in the set $\mathcal{D}\backslash K$ is given by*

$$\Pr(\vee_{c' \in K} W(\mathbf{x}_{c'}, \boldsymbol{\nu}; \boldsymbol{\delta}) > W(\mathbf{x}_c, \boldsymbol{\nu}; \boldsymbol{\delta}) \ \forall c \in \mathcal{D}\backslash K | \boldsymbol{v}) = \sum_{c' \in K} \frac{\exp(\omega(\mathbf{x}_{c'}, \boldsymbol{v}; \boldsymbol{\delta}))}{\sum_{c \in \mathcal{D}} \exp(\omega(\mathbf{x}_c, \boldsymbol{v}; \boldsymbol{\delta}))}.$$

*Proof of Claim S1.2.* We first establish equivalence of the following events:

$$\{\exists c' \in K \ s.t. \ W(\mathbf{x}_{c'}, \boldsymbol{\nu}; \boldsymbol{\delta}) > W(\mathbf{x}_c, \boldsymbol{\nu}; \boldsymbol{\delta}); \ \forall c \in \mathcal{D}\backslash K\}$$
$$\iff \cup_{c' \in K} \{W(\mathbf{x}_{c'}, \boldsymbol{\nu}; \boldsymbol{\delta}) > W(\mathbf{x}_c, \boldsymbol{\nu}; \boldsymbol{\delta}), \forall c \in \mathcal{D}\backslash c'\}. \quad (S1.8)$$

The right-to-left implication in (S1.8) is immediate. The left-to-right implication can be established by contradiction, observing that the complement of the event in the right-hand side of (S1.8) is that there exists a $c \in \mathcal{D}\backslash K$ that is preferred to all other alternatives. The result then follows because the events in the right-hand side of (S1.8) are disjoint. □

Next, recall that, as discussed in Section 3.4, the set $D_\kappa^*(\mathbf{x}, \boldsymbol{\nu}; \boldsymbol{\delta})$ can only take on a finite number of realizations, denoted $D^1, \ldots, D^h$, with $|D^j| = |\mathcal{D}| - \kappa + 1$ for all $j = 1, \ldots, h$. We show how to compute the probability of any of these realizations.

CLAIM S1.3: *For each $j = 1, \ldots, h$, $P(D_\kappa^*(\mathbf{x}, \boldsymbol{\nu}; \boldsymbol{\delta}) = D^j | \boldsymbol{v}; \boldsymbol{\gamma})$ can be computed as a linear combination of expression* (S1.7) *for different sets $G$.*

*Proof of Claim S1.3.* Note that

$$P(D_\kappa^*(\mathbf{x}, \boldsymbol{\nu}; \boldsymbol{\delta}) = D^j | \boldsymbol{v}; \boldsymbol{\gamma}) = P(W(\mathbf{x}_{c'}, \boldsymbol{\nu}; \boldsymbol{\delta}) > W(\mathbf{x}_c, \boldsymbol{\nu}; \boldsymbol{\delta}), \ \forall c' \in D^j, \forall c \in \mathcal{D}\backslash D^j | \boldsymbol{v}; \boldsymbol{\gamma}).$$

Given this, the proof proceeds sequentially. Suppose $|D_\kappa^*(\mathbf{x}, \boldsymbol{\nu}; \boldsymbol{\delta})| = 1$. Then the result follows immediately (with $G = \mathcal{D}$). Suppose $|D_\kappa^*(\mathbf{x}, \boldsymbol{\nu}; \boldsymbol{\delta})| = 2$. Then we have $D^j = \{c', c''\}$

for some $c', c'' \in \mathcal{D}$, and

$$P(\{W(\mathbf{x}_{c'}, \boldsymbol{\nu}; \boldsymbol{\delta}) > W(\mathbf{x}_c, \boldsymbol{\nu}; \boldsymbol{\delta})\} \cap \{W(\mathbf{x}_{c''}, \boldsymbol{\nu}; \boldsymbol{\delta}) > W(\mathbf{x}_c, \boldsymbol{\nu}; \boldsymbol{\delta})\} \; \forall c \in \mathcal{D} \backslash D^j | \boldsymbol{\upsilon}; \boldsymbol{\gamma})$$
$$= P(W(\mathbf{x}_{c'}, \boldsymbol{\nu}; \boldsymbol{\delta}) > W(\mathbf{x}_c, \boldsymbol{\nu}; \boldsymbol{\delta}) \; \forall c \in \mathcal{D} \backslash D^j | \boldsymbol{\upsilon}; \boldsymbol{\gamma}) + P(W(\mathbf{x}_{c''}, \boldsymbol{\nu}; \boldsymbol{\delta}) > W(\mathbf{x}_c, \boldsymbol{\nu}; \boldsymbol{\delta} | \boldsymbol{\upsilon}; \boldsymbol{\gamma}) \; \forall c \in \mathcal{D} \backslash D^j)$$
$$- P(\{W(\mathbf{x}_{c'}, \boldsymbol{\nu}; \boldsymbol{\delta}) > W(\mathbf{x}_c, \boldsymbol{\nu}; \boldsymbol{\delta})\} \cup \{W(\mathbf{x}_{c''}, \boldsymbol{\nu}; \boldsymbol{\delta}) > W(\mathbf{x}_c, \boldsymbol{\nu}; \boldsymbol{\delta})\} \; \forall c \in \mathcal{D} \backslash D^j | \boldsymbol{\upsilon}; \boldsymbol{\gamma}).$$

The first term in this expression can be computed by applying equation (S1.7) with $G = \mathcal{D} \backslash c''$; the second term can be computed by applying equation (S1.7) with $G = \mathcal{D} \backslash c'$; the last term, by Claim S1.2, can be computed as the sum over $\tilde{c} \in D^j$ of equation (S1.7) with $G = \mathcal{D}$.

For $|D_\kappa^*(\mathbf{x}, \boldsymbol{\nu}; \boldsymbol{\delta})| \geqslant 3$ one can proceed iteratively using the inclusion/exclusion formula and applying Claim S1.2. $\square$

With these results in hand, we prove Theorem S1.2.

*Proof of Theorem S1.2.* By Claim S1.3 we can compute $P(D_\kappa^*(\mathbf{x}, \boldsymbol{\nu}; \boldsymbol{\delta}) = D^j | \boldsymbol{\upsilon}; \boldsymbol{\gamma})$ for each $D^j$ such that $|D^j| = |\mathcal{D}| - \kappa + 1$ as a linear combination of expression (S1.7) with different sets $G$. To obtain the result in Theorem S1.2, for each set $K$ one can simply sum $P(D_\kappa^*(\mathbf{x}, \boldsymbol{\nu}; \boldsymbol{\delta}) = D^j | \boldsymbol{\upsilon}; \boldsymbol{\gamma})$ over the sets $D^j$ such that $D^j \cap K \neq \varnothing$. $\square$

## S2    Additional Details on Statistical Inference

As explained in Section 5, we base our confidence sets for the vector $\boldsymbol{\theta}$ on the Kolmogorov-Smirnov test statistic suggested by Andrews and Shi (2013, equation (3.7) on p. 618) [hereafter, AS], which in our framework simplifies to

$$T_n(\boldsymbol{\theta}) = n \max_{j=1,\dots,J; K \in \mathbb{K}} \max\left\{\frac{\bar{m}_{n,K,j}(\boldsymbol{\theta})}{\hat{\sigma}_{n,K,j}(\boldsymbol{\theta})}, 0\right\}^2$$

where $\bar{m}_{n,K,j}(\boldsymbol{\theta})$ and $\hat{\sigma}_{n,K,j}(\boldsymbol{\theta})$ are defined in Section 5. Our application of the method proposed by AS computes bootstrap-based critical values to obtain a confidence set

$$CS = \{\boldsymbol{\theta} \in \Theta : T_n(\boldsymbol{\theta}) \leqslant \hat{c}_{n,1-\alpha+\xi}(\boldsymbol{\theta}) + \xi\}$$

where $\xi > 0$ is an arbitrarily small constant which we set equal to $10^{-6}$ as suggested by AS (p. 625). In practice, we evaluate $T_n(\boldsymbol{\theta})$ and the bootstrap-based critical value $\hat{c}_{n,1-\alpha+\xi}(\boldsymbol{\theta})$ on a grid of values of $\boldsymbol{\theta}$ designed to give good coverage of the $(\mathrm{E}(\nu), \mathrm{Var}(\nu))$-space to obtain a precise description of the confidence set for this pair of parameters. To explain how this grid is constructed, we note that given the assumption that $\nu_i \sim Beta(\gamma_1, \gamma_2)$ with support $[0, 0.03]$, $E(\nu) \in 0.03 \times (0, 1]$ and $\text{Var}(\nu) \in 0.0009 \times (0, 0.25]$. We therefore obtain a grid of values over $(\gamma_1, \gamma_2)$ comprised of 665,603 points, such that the associated grid on $(E(\nu), \text{Var}(\nu))$ has first coordinate in $0.03 \times [0.0005, 0.9995]$ with step size $0.03 \times 0.0005$, and second coordinate in $0.0009 \times (0.0005, 0.25]$ with step size $0.0009 \times 0.0005$.[4] The approximation of $\hat{c}_{n,1-\alpha+\xi}(\boldsymbol{\theta})$ is based on the bootstrap procedure detailed in AS (Section 9) and uses 1,000 bootstrap replications.[5] The procedure takes as inputs a *GMS function* $\varphi$, a *GMS sequence* $\tau_n$ such that $\tau_n \to \infty$ as $n \to \infty$, and a *non-decreasing sequence of positive constants* $\beta_n$ such that $\beta_n/\tau_n \to 0$ as $n \to \infty$, which together are used to determine which moment inequalities are sufficiently close to binding to contribute to the limiting distribution of $T_n(\boldsymbol{\theta})$. We use the GMS function proposed by AS (equation (4.10) on p. 627):[6]

$$\varphi_{K,j}(\boldsymbol{\theta}) = \begin{cases} 0 & \text{if } \tau_n^{-1}\sqrt{n}\bar{m}_{n,K,j}(\boldsymbol{\theta})/\hat{\sigma}_{n,K,j}(\boldsymbol{\theta}) \geqslant -1 \\ -\beta_n & \text{otherwise,} \end{cases}$$

and we set $\tau_n = (0.3 \ln n)^{1/2}$ and $\beta_n = (0.4 \ln n / \ln \ln n)^{1/2}$ as recommended by AS (p. 643).

Similar to AS, the KMS procedure takes as inputs a GMS function $\varphi$ and a GMS sequence $\tau_n$.[7] To simplify computations, we use the *hard threshold* GMS function:[8]

$$\varphi_{K,j}(\boldsymbol{\theta}) = \begin{cases} 0 & \text{if } \tau_n^{-1}\sqrt{n}\bar{m}_{n,K,j}(\boldsymbol{\theta})/\hat{\sigma}_{n,K,j}(\boldsymbol{\theta}) \geqslant -1 \\ -\infty & \text{otherwise.} \end{cases}$$

The procedure also requires a regularization parameter $\rho \geqslant 0$, which (like $\varphi$ and $\tau_n$) enters the calibration of $\hat{c}^f_{n,1-\alpha}$ and introduces a conservative distortion that is required to obtain uniform coverage of projections. The smaller is the value of $\rho$, the larger is the conservative distortion, but the higher is the confidence that the critical value is uniformly valid in situations where the local geometry of $\Theta_I$ makes inference especially challenging. For a discussion, see KMS (Section 2.2). We choose the value of $\rho$ as follows. We begin with the recommendation in KMS (Section 2.4). To further guard against possible irregularities in the local geometry of $\Theta_I$, we reduce the resulting value of $\rho$ by 20 percent.

---

[4] To obtain confidence intervals on $\pi_5$, $\pi_4$, and $\pi_3$, we first evaluate $T_n(\boldsymbol{\theta})$ on a coarser grid and compare it with the AS critical value. For each $\pi_q$, $q = 3, 4, 5$, we then refine the grid around the extreme values of $\pi_q$ that are not rejected, for a final step size of 0.01 on $\pi_q$ and 0.05 on each component of $(\gamma_1, \gamma_2)$.

[5] Compared to the description in AS (Section 9), note that our moment inequalities are of the $\leqslant$ form, whereas AS's are of the $\geqslant$ form.

[6] AS label the GMS sequence $\kappa_n$, but we use $\tau_n$ to avoid confusion with our use of $\kappa$ for the (known and fixed) minimum choice set size in Assumption 2.2.

[7] Our findings based on the AS and KMS methods are robust to the choice of tuning parameters, as indicated by results available from the authors upon request.

[8] This function was proposed by Andrews and Soares (2010) and labeled $\varphi^{(1)}$ on p. 131 of their article.

# S3  Additional Results

## S3.1  Claim Probabilities

The claim probabilities originate from Barseghyan et al. (2018). We estimate the households' claim probabilities using the company's claims data. We assume that household $i$'s auto collision claims in year $t$ follow a Poisson distribution with mean $\lambda_{it}$. We also assume that the household's deductible choice does not influence its claim rates $\lambda_{it}$ (Assumption 4.1(II)). We treat the household's claim rate as a latent random variable and assume that $\ln \lambda_{it} = \mathbf{X}'_{it}\boldsymbol{\beta} + \varepsilon_i$, where $\mathbf{X}_{it}$ is a vector of observables and $\exp(\varepsilon_i)$ follows a Gamma distribution with unit mean and variance $\phi$. We perform a Poisson panel regression with random effects to obtain maximum likelihood estimates of $\boldsymbol{\beta}$ and $\phi$. In an effort to obtain the most precise estimates, we use the full set of auto collision claims data, which comprises 1,349,853 household-year records. For each household, we calculate a fitted claim rate $\widehat{\lambda}_i$ conditional on the household's observables at the time of first purchase and its subsequent claims experience. More specifically, $\widehat{\lambda}_i = \exp(\mathbf{X}'_i\widehat{\boldsymbol{\beta}}) \operatorname{E}(\exp(\varepsilon_i)|\mathbf{Y}_i)$, where $\mathbf{Y}_i$ records household $i$'s claims experience after purchasing the policy and $\operatorname{E}(\exp(\varepsilon_i)|\mathbf{Y}_i)$ is calculated using the maximum likelihood estimate of $\phi$. In principle, a household may experience one or more claims during the policy period. We assume that households disregard the possibility of experiencing more than one claim (Assumption 4.1(I)). Given this, we transform $\widehat{\lambda}_i$ into a claim probability $\mu_i \equiv 1 - \exp(-\widehat{\lambda}_i)$, which follows from the Poisson probability mass function, and round it to the nearest half percentage point. We treat $\mu_i$ as data.

## S3.2  Deductible Choices

Table S3.1 reports the sample distribution of deductible choices by octiles of base price $\bar{p}_i$ and claim probability $\mu_i$. The octiles are the hypercubes referenced in Sections 5 and S2 (other than the one that contains all households).

## S3.3  Subgroup Results

Figure S3.1 depicts the AS 95 percent confidence set for $(\operatorname{E}(\nu_i), \operatorname{Var}(\nu_i))$ for population subgroups based on gender, age, and insurance score of the principal driver. In addition, Table S3.2 reports (i) the KMS 95 percent confidence interval for the mean of $\nu_i$ and (ii) 95 percent confidence intervals for the 25th and 75th percentiles of $\nu_i$ based on projections of the AS confidence set. For the mean, we report the actual confidence interval as well as the risk premium, for a lottery that yields a loss of $1000 with probability 10 percent, implied

Table S3.1: Deductible Choices by Octiles of $\bar{p}$ and $\mu$

| $\bar{p}$ octile | $\mu$ octile | Obs. | Percent choosing deductible | | | | |
|---|---|---|---|---|---|---|---|
| | | | $100 | $200 | $250 | $500 | $1000 |
| 1 | 1 | 2,756 | 3.3 | 31.2 | 18.9 | 43.8 | 2.9 |
| 1 | 2 | 2,901 | 3.6 | 31.8 | 18.7 | 43.6 | 2.2 |
| 1 | 3 | 2,661 | 2.9 | 32.1 | 20.0 | 43.6 | 1.5 |
| 1 | 4 | 2,113 | 3.4 | 34.2 | 20.6 | 40.8 | 1.0 |
| 1 | 5 | 2,116 | 3.9 | 32.1 | 20.2 | 42.2 | 1.5 |
| 1 | 6 | 1,630 | 4.2 | 34.5 | 21.9 | 38.9 | 0.6 |
| 1 | 7 | 1,233 | 4.4 | 34.1 | 22.8 | 38.7 | 0.0 |
| 1 | 8 | 660 | 5.0 | 39.4 | 25.6 | 30.0 | 0.0 |
| 2 | 1 | 1,949 | 1.0 | 20.8 | 17.0 | 57.1 | 4.0 |
| 2 | 2 | 1,944 | 2.0 | 22.3 | 16.9 | 56.4 | 2.5 |
| 2 | 3 | 1,543 | 1.9 | 25.7 | 19.1 | 50.7 | 2.6 |
| 2 | 4 | 2,152 | 2.0 | 23.1 | 18.5 | 54.4 | 2.0 |
| 2 | 5 | 1,320 | 2.3 | 26.7 | 18.0 | 50.8 | 2.2 |
| 2 | 6 | 1,979 | 1.6 | 25.6 | 20.1 | 51.1 | 1.6 |
| 2 | 7 | 1,584 | 1.8 | 26.5 | 22.6 | 47.9 | 1.3 |
| 2 | 8 | 1,151 | 2.0 | 26.5 | 22.7 | 48.7 | 0.2 |
| 3 | 1 | 1,362 | 0.7 | 20.4 | 14.3 | 59.8 | 4.7 |
| 3 | 2 | 1,914 | 0.8 | 18.5 | 14.6 | 62.1 | 3.9 |
| 3 | 3 | 2,127 | 0.8 | 19.8 | 16.1 | 60.0 | 3.2 |
| 3 | 4 | 1,518 | 1.3 | 20.3 | 17.7 | 59.4 | 1.4 |
| 3 | 5 | 2,255 | 1.0 | 19.9 | 17.6 | 59.4 | 2.1 |
| 3 | 6 | 1,773 | 0.8 | 19.9 | 18.4 | 59.1 | 1.9 |
| 3 | 7 | 1,729 | 1.2 | 21.1 | 20.0 | 56.7 | 1.1 |
| 3 | 8 | 1,602 | 1.2 | 20.7 | 22.2 | 54.9 | 0.9 |
| 4 | 1 | 1,340 | 0.7 | 12.7 | 13.7 | 67.5 | 5.3 |
| 4 | 2 | 1,458 | 0.8 | 14.1 | 15.2 | 65.8 | 4.3 |
| 4 | 3 | 1,632 | 0.7 | 15.1 | 15.4 | 66.1 | 2.8 |
| 4 | 4 | 1,595 | 0.6 | 14.7 | 16.6 | 64.8 | 3.3 |
| 4 | 5 | 1,606 | 0.8 | 14.3 | 17.1 | 65.4 | 2.5 |
| 4 | 6 | 1,705 | 0.6 | 16.1 | 15.2 | 65.5 | 2.6 |
| 4 | 7 | 1,974 | 0.7 | 15.4 | 17.0 | 65.5 | 1.5 |
| 4 | 8 | 1,914 | 1.0 | 17.3 | 17.7 | 62.8 | 1.2 |
| 5 | 1 | 1,126 | 0.4 | 11.4 | 12.6 | 70.5 | 5.2 |
| 5 | 2 | 1,547 | 0.1 | 11.8 | 11.9 | 71.7 | 4.5 |
| 5 | 3 | 1,609 | 0.5 | 10.4 | 13.0 | 71.6 | 4.5 |
| 5 | 4 | 1,522 | 0.5 | 10.6 | 14.5 | 71.4 | 3.0 |
| 5 | 5 | 2,066 | 0.7 | 10.8 | 12.8 | 72.1 | 3.5 |
| 5 | 6 | 1,697 | 0.6 | 12.5 | 14.7 | 69.2 | 2.9 |
| 5 | 7 | 1,801 | 0.2 | 12.2 | 14.6 | 70.9 | 2.2 |
| 5 | 8 | 2,128 | 0.5 | 11.9 | 17.1 | 68.8 | 1.6 |
| 6 | 1 | 1,303 | 0.3 | 6.7 | 9.1 | 78.3 | 5.6 |
| 6 | 2 | 1,403 | 0.2 | 6.9 | 11.4 | 75.5 | 6.0 |
| 6 | 3 | 1,326 | 0.5 | 7.3 | 11.2 | 76.8 | 4.2 |
| 6 | 4 | 1,784 | 0.3 | 8.1 | 11.2 | 76.2 | 4.2 |
| 6 | 5 | 1,589 | 0.2 | 7.9 | 9.8 | 78.0 | 4.1 |
| 6 | 6 | 1,725 | 0.5 | 8.9 | 12.0 | 74.7 | 3.9 |
| 6 | 7 | 2,061 | 0.1 | 7.3 | 11.2 | 78.4 | 3.1 |
| 6 | 8 | 2,363 | 0.1 | 9.0 | 12.3 | 76.3 | 2.2 |
| 7 | 1 | 1,521 | 0.3 | 5.2 | 6.9 | 81.1 | 6.5 |
| 7 | 2 | 1,351 | 0.1 | 5.6 | 7.5 | 80.1 | 6.7 |
| 7 | 3 | 1,665 | 0.2 | 4.1 | 8.6 | 80.2 | 6.8 |
| 7 | 4 | 1,646 | 0.1 | 5.0 | 6.7 | 81.7 | 6.4 |
| 7 | 5 | 1,726 | 0.1 | 5.0 | 7.4 | 82.6 | 5.0 |
| 7 | 6 | 1,865 | 0.1 | 4.9 | 7.9 | 82.5 | 4.6 |
| 7 | 7 | 2,045 | 0.1 | 5.7 | 7.6 | 82.4 | 4.2 |
| 7 | 8 | 2,452 | 0.2 | 5.4 | 9.1 | 81.0 | 4.4 |
| 8 | 1 | 2,636 | 0.0 | 1.3 | 2.5 | 74.2 | 21.9 |
| 8 | 2 | 1,553 | 0.1 | 1.5 | 1.8 | 80.3 | 16.4 |
| 8 | 3 | 1,463 | 0.0 | 1.6 | 3.1 | 82.8 | 12.4 |
| 8 | 4 | 1,568 | 0.0 | 1.4 | 2.7 | 80.2 | 15.6 |
| 8 | 5 | 1,384 | 0.0 | 1.8 | 2.0 | 80.6 | 15.6 |
| 8 | 6 | 1,570 | 0.1 | 2.0 | 3.0 | 78.9 | 16.1 |
| 8 | 7 | 1,501 | 0.0 | 1.2 | 2.5 | 82.7 | 13.7 |
| 8 | 8 | 1,698 | 0.1 | 2.1 | 3.3 | 81.0 | 13.5 |

Notes: Analysis sample (111,890 households).

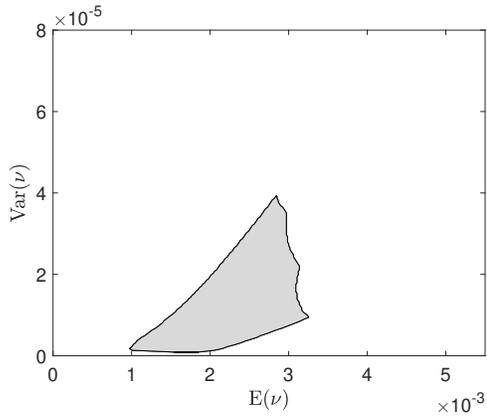
(a) Male

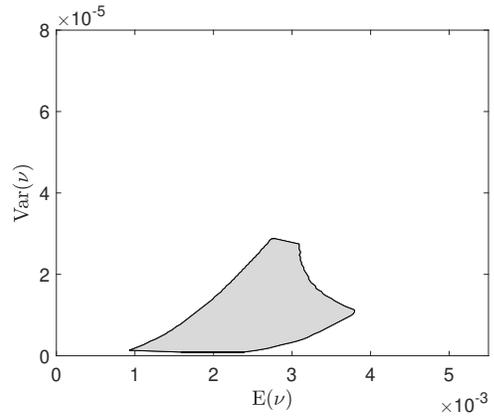
(b) Female

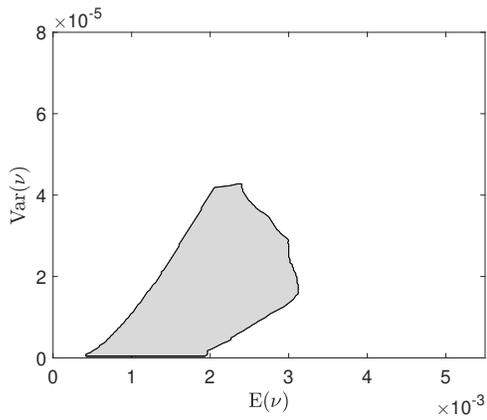
(c) Young

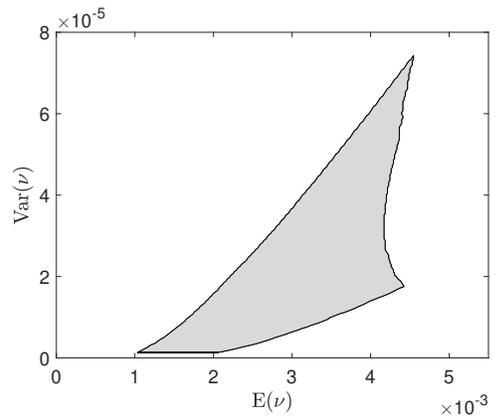
(d) Old

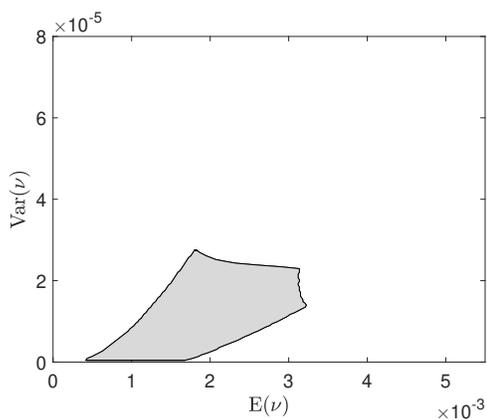
(e) Low insurance score

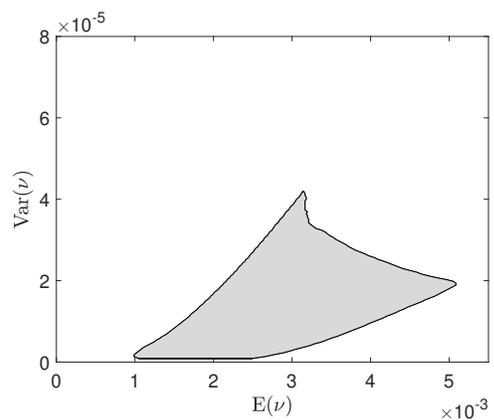
(f) High insurance score

Figure S3.1: AS 95 percent confidence sets for $(\mathrm{E}(\nu), \mathrm{Var}(\nu))$.

Table S3.2: Distribution of Absolute Risk Aversion

|  | Mean | | Implied risk premium | | | | | |
|---|---|---|---|---|---|---|---|---|
|  | | | Mean | | 25th pctl. | | 75th pctl. | |
|  | LB | UB | LB | UB | LB | UB | LB | UB |
| Male | 0.00104 | 0.00321 | $ 61 | $279 | $ 0 | $ 73 | $ 76 | $426 |
| Female | 0.00101 | 0.00377 | $ 59 | $339 | $ 0 | $117 | $ 81 | $485 |
| Young | 0.00044 | 0.00306 | $ 22 | $263 | $ 0 | $ 95 | $ 0 | $407 |
| Old | 0.00107 | 0.00432 | $ 63 | $393 | $ 0 | $ 73 | $ 95 | $548 |
| Low insurance score | 0.00042 | 0.00315 | $ 21 | $273 | $ 0 | $ 73 | $ 7 | $425 |
| High insurance score | 0.00102 | 0.00501 | $ 60 | $452 | $ 0 | $127 | $ 85 | $591 |

Notes: 95 percent confidence intervals. LB = lower bound. UB = upper bound. Implied risk premia for a lottery that yields a loss of $1000 with probability 10 percent.

by each bound. For the percentiles, we report only the implied risk premia. For the most part, the subgroup results are comparable to the results for all households. The notable exceptions are the lower bounds on the mean for households with young principal drivers and households with low insurance scores. These lower bounds are on the order of $4 \cdot 10^{-4}$ (which implies a risk premium of about $20), whereas the corresponding lower bounds for the other subgroups and the population are on the order of $10^{-3}$ (which implies a risk premium of about $60).[9] Strikingly, the lower bounds on the 75th percentile for these two subgroups correspond to risk premia of 17 cents and $7, respectively.

Table S3.3 reports KMS 95 percent confidence intervals for $\pi_5$, $\pi_4$, and $\pi_3$ for the same population subgroups. The interesting quantities are the upper bounds on $\pi_5$ and $\pi_4$. The former is the maximum fraction of households whose deductible choices can be rationalized with full size choice sets, while the latter is the maximum fraction of households whose deductible choices can be rationalized with full-1 choice sets.[10] We find, inter alia, that: (i) at least 70 percent of households with female principal drivers require limited choice sets to explain their deductible choices, whereas at least 74 percent of households with male principal drivers require limited choice sets; (ii) at least 73 percent of households with old principal drivers require limited choice sets to explain their deductible choices, whereas at least 75 percent of households with young principal drivers require limited choice sets; and (iii) at least 67 percent of households with low insurance scores require limited choice sets to explain their deductible choices, whereas at least 73 percent of households with high insurance scores require limited choice sets.[11]

---

[9]Because the subgroups all have different confidence sets (as well as different sample sizes), it is possible that a result for all households is not a weighted average of the corresponding results within a subgroup.

[10]With $\kappa = 3$, the lower bounds on $\pi_5$ and $\pi_4$ are zero, the lower bound on $\pi_3$ is one minus the upper

Table S3.3: Distribution of Choice Set Size

|  | $\pi_5$ (full) | | $\pi_4$ (full-1) | | $\pi_3$ (full-2) | |
| --- | --- | --- | --- | --- | --- | --- |
|  | LB | UB | LB | UB | LB | UB |
| Male | 0.00 | 0.26 | 0.00 | 0.85 | 0.15 | 1.00 |
| Female | 0.00 | 0.30 | 0.00 | 0.90 | 0.10 | 1.00 |
| Young | 0.00 | 0.25 | 0.00 | 1.00 | 0.00 | 1.00 |
| Old | 0.00 | 0.27 | 0.00 | 0.96 | 0.04 | 1.00 |
| Low insurance score | 0.00 | 0.33 | 0.00 | 1.00 | 0.00 | 1.00 |
| High insurance score | 0.00 | 0.27 | 0.00 | 1.00 | 0.00 | 1.00 |

Notes: KMS 95 percent confidence intervals. LB = lower bound. UB = upper bound.

## S3.4 Admissible Probability Density Functions

Figure S3.2 depicts a 95 percent confidence set for an outer region of admissible probability density functions of $\nu_i$ for all households. To construct the outer region (shaded in grey), we find at each point on a grid of 101 values of $\nu_i$ the minimum and maximum values of all probability density functions implied by values of $\boldsymbol{\theta}$ in the AS 95 percent confidence set. This gives us 101 points on the lower and upper envelopes of admissible probability density functions. In other words, we obtain pointwise sharp lower and upper bounds on the set of admissible probability density functions. Although the bounds are pointwise sharp, the region is labeled an outer region because not all probability density functions in it are consistent with the distribution of observed choices. The figure also superimposes the predicted density functions of $\nu_i$ based on point estimates obtained under the UR and ASR models. The UR predicted density function does not lie entirely inside the confidence set, whereas the AR predicted density function does (although we note that this does not necessarily imply that the true choice formation process is an ASR process).

## S3.5 Suboptimal Choices

As we state in Section 5.2.1, with full size choice sets, our model cannot explain the frequency of the $200 deductible in our data. The reason is that, with full size choice sets, our model satisfies the following conditional rank order property, which is a generalization of the rank order property established by Manski (1975) for random utility models that are linear in the nonrandom parameters and feature an additive i.i.d. disturbance in the utility function.

---

bound on $\pi_4$, and the upper bound on $\pi_3$ is one.

[11]Because the subgroups all have different confidence sets (as well as different sample sizes), it is possible that the upper bound on $\pi_5$ for all households is not a weighted average of the upper bounds on $\pi_5$ within a subgroup. The same is true for the upper bound on $\pi_4$ (and, therefore, for the lower bound on $\pi_3$).

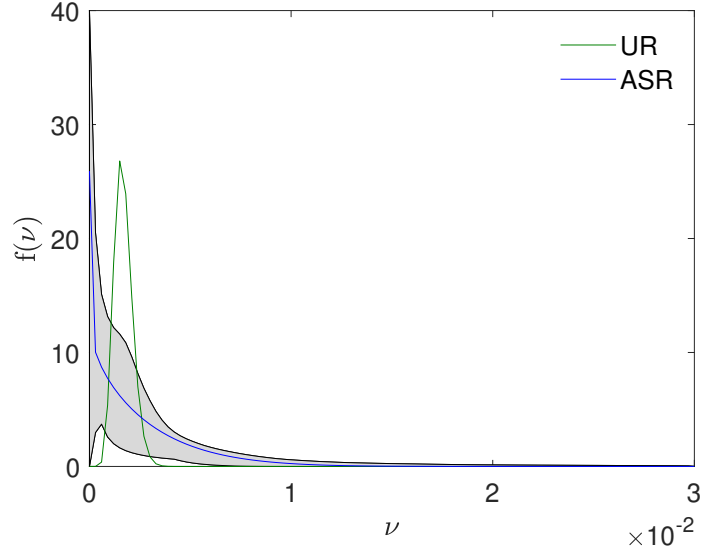

Figure S3.2: Confidence set for outer region of admissible probability density functions of $\nu$.

Notes: The figure depicts a 95 percent confidence set for an outer region of admissible probability density functions of $\nu_i$. It also superimposes the implied probability density functions of $\nu_i$ based on point estimates obtained under the UR and ASR models.

PROPERTY S3.1 (Conditional Rank Order Property): *For all $c, c' \in \mathcal{D}$, $\Pr(d = c'|\mathbf{x}, \boldsymbol{\nu}) \geq \Pr(d = c|\mathbf{x}, \boldsymbol{\nu})$ if and only if $W(\mathbf{x}_{c'}, \boldsymbol{\nu}; \boldsymbol{\delta}) \geq W(\mathbf{x}_c, \boldsymbol{\nu}; \boldsymbol{\delta})$, $(\mathbf{x}, \boldsymbol{\nu}) - a.s.$*

Indeed, *any* model that satisfies an analogous property is incapable of explaining the relative frequency of \$200 in the distribution of observed deductible choices.[12] This includes, inter alia, the conditional logit model (McFadden 1974), the mixed logit model (McFadden 1974; McFadden and Train 2000), the multinomial probit model (e.g., Hausman and Wise 1978), and semiparametric models such as the one in Manski (1975). At the same time, not all choice set formation processes can explain the relative frequency of \$200 in our data. For instance, UR cannot but ASR can.

CLAIM S3.1: *Take the model in Section 2. Suppose for a given $c \in \mathcal{D}$ there exist $a, b \in \mathcal{D}$, $a \neq b \neq c$, such that for each $\boldsymbol{\nu} \in \mathcal{V}$, $W(\mathbf{x}_a, \boldsymbol{\nu}; \boldsymbol{\delta}) > W(\mathbf{x}_c, \boldsymbol{\nu}; \boldsymbol{\delta})$ or $W(\mathbf{x}_b, \boldsymbol{\nu}; \boldsymbol{\delta}) > W(\mathbf{x}_c, \boldsymbol{\nu}; \boldsymbol{\delta})$. Then for any distribution of $\boldsymbol{\nu}$ with support $\mathcal{V}$:*

(I) *Property S3.1 implies $\Pr(d = a|\mathbf{x}) + \Pr(d = b|\mathbf{x}) > \Pr(d = c|\mathbf{x})$, $\mathbf{x} - a.s.$*

(II) *Under UR, $\Pr(d = a|\mathbf{x}) + \Pr(d = b|\mathbf{x}) > \Pr(d = c|\mathbf{x})$, $\mathbf{x} - a.s.$*

(III) *Under ASR, $\Pr(d = a|\mathbf{x}) + \Pr(d = b|\mathbf{x}) < \Pr(d = c|\mathbf{x})$ is possible.*

---

[12]In the case of a model with additively separable noise where $\boldsymbol{\nu} = (\boldsymbol{v}, (\epsilon_c, c \in \mathcal{D}))$ and $W(\mathbf{x}_c, \boldsymbol{\nu}; \boldsymbol{\delta}) = \omega(\mathbf{x}_c, \boldsymbol{v}; \boldsymbol{\delta}) + \epsilon_c$, the analogous property is: For all $c, c' \in \mathcal{D}$, $\Pr(d = c'|\mathbf{x}, \boldsymbol{v}) \geq \Pr(d = c|\mathbf{x}, \boldsymbol{v})$ if and only if $\omega(\mathbf{x}_{c'}, \boldsymbol{v}; \boldsymbol{\delta}) \geq \omega(\mathbf{x}_c, \boldsymbol{v}; \boldsymbol{\delta})$, $(\mathbf{x}, \boldsymbol{v}) - a.s.$

*Proof.* The implication in Claim S3.1(I) follows from Property S3.1 by integrating with respect to the distribution of $\boldsymbol{\nu}$.

Claim S3.1(II) follows from the fact that the UR model satisfies Property S3.1. Suppose alternative $c'$ is preferred to alternative $c$. Alternative $c'$ may be chosen from choice sets that contain both $c'$ and $c$ and from choice sets that contain $c'$ but not $c$. However, alternative $c$ may be chosen only from choice sets that contain $c$ but not $c'$. Because all choice sets, conditional on the draw of $|C|$, are equiprobable, $c'$ is chosen more frequently than $c$.

We can establish Claim S3.1(III) with a trivial example. Suppose $\varphi(a) = \varphi(b) = 0$ and $\varphi(c) = 1$. Then $\Pr(d = a|\mathbf{x}) = \Pr(d = b|\mathbf{x}) = 0$ and $\Pr(d = c|\mathbf{x}) > 0$ provided there exists a positive measure of values $\boldsymbol{\nu} \in \mathcal{V}$ such that $W(\mathbf{x}_c, \boldsymbol{\nu}; \boldsymbol{\delta}) > W(\mathbf{x}_{c'}, \boldsymbol{\nu}; \boldsymbol{\delta})$ for all $c' \in \mathcal{D}\backslash\{a, b\}$, $c' \neq c$. More generally, $\Pr(d = a|\mathbf{x}) + \Pr(d = b|\mathbf{x}) < \Pr(d = c|\mathbf{x})$ is possible provided $\varphi(a)$ and $\varphi(b)$ are sufficiently low, $\varphi(c)$ is sufficiently high, and $c$ is the first best alternative in $\mathcal{D}\backslash\{a, b\}$ for some positive measure of values $\boldsymbol{\nu} \in \mathcal{V}$. $\square$

We emphasize that Claim S3.1 does not rely on Assumption 3.1 or the assumptions of the empirical model in Section 4.1. It thus exemplifies a new approach to testing assumptions on choice set formation in any random utility model under weak restrictions on the utility function and without parametric restrictions on the distribution of preferences or choice sets.

An analogous claim holds in the case of a model with additively separable disturbances, such as the mixed logit model in Section 5.1.1, for any distribution of $\boldsymbol{v}$ with support $\Upsilon$, where the predicate is: Suppose for a given $c \in \mathcal{D}$ there exist $a, b \in \mathcal{D}$, $a \neq b \neq c$, such that for each $\boldsymbol{v} \in \Upsilon$, $\omega(\mathbf{x}_a, \boldsymbol{v}; \boldsymbol{\delta}) > \omega(\mathbf{x}_c, \boldsymbol{v}; \boldsymbol{\delta})$ or $\omega(\mathbf{x}_b, \boldsymbol{v}; \boldsymbol{\delta}) > \omega(\mathbf{x}_c, \boldsymbol{v}; \boldsymbol{\delta})$.